\def\CHfour{{\fam0 CH_4}}
\def\CHthree{{\fam0 CH_3}}
\def\CN{{\fam0 CN}}
\def\HCN{{\fam0 HCN}}
\def\CHtwoOH{{\fam0 CH_2OH}}
\def\CH{{\fam0 CH}}
\def\COtwo{{\fam0 CO_2}}
\def\CO{{\fam0 CO}}
\def\CtwoHtwo{{\fam0 C_2H_2}}
\def\CtwoHthree{{\fam0 C_2H_3}}
\def\CtwoHfour{{\fam0 C_2H_4}}
\def\CtwoHfive{{\fam0 C_2H_5}}
\def\CtwoHsix{{\fam0 C_2H_6}}
\def\C{{\fam0 C}}
\def\threeCHtwo{{\fam0 ^{3}CH_2}}

\def\HtwoO{{\fam0 H_2O}}
\def\Htwo{{\fam0 H_2}}
\def\H{{\fam0 H}}
\def\M{{\fam0 M}}
\def\NO{{\fam0 NO}}
\def\NH{{\fam0 NH}}

\def\NHthree{{\fam0 NH_3}}
\def\Net{{\fam0 Net:}}
\def\N{{\fam0 N}}
\def\Ntwo{{\fam0 N_2}}

\def\OH{{\fam0 OH}}

\def\O{{\fam0 O}}

\def\cmmtwo{{\fam0\,cm^{-2}}}

\def\cmtwo{{\fam0 cm^2}}

\def\scinot#1.{\hbox{$\times 10^{#1}$}}
\def\smone{{\fam0\,s^{-1}}}

\def\ten#1.{\hbox{$10^{#1}$}}

\def\deg{\ifmmode^\circ\else$\null^\circ$\fi}
\def\spose#1{\hbox to 0pt{#1\hss}}
\def\lta{\mathrel{\spose{\lower 3pt\hbox{$\mathchar "218$}}\raise 2.0pt\hbox{$\mathchar"13C$}}}
\def\gta{\mathrel{\spose{\lower 3pt\hbox{$\mathchar "218$}}\raise 2.0pt\hbox{$\mathchar"13E$}}}
\def\lrarrow{\mathrel{\spose{\lower 1pt\hbox{$\rightarrow$}}\raise 3.0pt\hbox{$\leftarrow$}}}

\documentclass{emulateapj}

\slugcomment{submitted to Astrophys. J.}

\shorttitle{Chemical Consequences of the C/O Ratio on Hot Jupiters}
\shortauthors{Moses et al.}

\begin{document}


\title{Chemical Consequences of the C/O Ratio on Hot Jupiters: Examples From
WASP-12\lowercase{b}, CoRoT-2\lowercase{b}, XO-1\lowercase{b}, and HD~189733\lowercase{b}}

\author{J. I. Moses}
\affil{Space Science Institute, 4750 Walnut Street, Suite 205, Boulder, CO, 80301, USA}
\email{jmoses@spacescience.org}

\author{N. Madhusudhan}
\affil{Department of Physics and Department of Astronomy, Yale University, New Haven, CT, 06520-8101, USA}

\author{C. Visscher}
\affil{Southwest Research Institute, Boulder, CO, 80302, USA}

\and

\author{R. S. Freedman}
\affil{SETI Institute, Mountain View, CA, 94043 and NASA Ames Research Center, Moffett Field, CA, 94035, USA}

%
%

\begin{abstract}
Motivated by recent spectroscopic evidence for carbon-rich atmospheres on some transiting 
exoplanets, we investigate the influence of the C/O ratio on the chemistry, composition, and 
spectra of extrasolar giant planets both from a thermochemical-equilibrium perspective 
and from consideration of disequilibrium processes like photochemistry and transport-induced 
quenching.  We find that although CO is predicted to be a major atmospheric constituent on hot 
Jupiters for all C/O ratios, other oxygen-bearing molecules like $\HtwoO$ and $\COtwo$ are much 
more abundant when C/O $<$ 1, whereas CH$_4$, HCN, and $\CtwoHtwo$ gain significantly in 
abundance when C/O $>$ 1.  Other notable species like N$_2$ and $\NHthree$ that do not 
contain carbon or oxygen are relatively unaffected by the C/O ratio.  Disequilibrium processes 
tend to enhance the abundance of CH$_4$, NH$_3$, HCN, and $\CtwoHtwo$ over a wide range of C/O 
ratios.  We compare the results of our models with secondary-eclipse photometric data from the 
{\it Spitzer\ Space\ Telescope\/} and conclude that 
(1) disequilibrium models with C/O $\sim$ 1 are consistent with spectra of WASP-12b, XO-1b, 
and CoRoT-2b, confirming the possible carbon-rich nature of these planets, (2) spectra from 
HD 189733b are consistent with C/O $\lta$ 1, but as the assumed metallicity is increased above 
solar, the required C/O ratio must increase toward 1 to prevent too much $\HtwoO$ absorption, 
(3) species like HCN can have a significant influence on spectral behavior in the 3.6 and 
8.0 $\mu$m {\it Spitzer\/} channels, potentially providing even more opacity than CH$_4$ when 
C/O $>$ 1, and (4) the very high CO$_2$ abundance inferred for HD 189733b from near-infrared 
observations cannot be explained through equilibrium or disequilibrium chemistry in a 
hydrogen-dominated atmosphere.  We discuss possible formation mechanisms for carbon-rich 
hot Jupiters, including scenarios in which the accretion of CO-rich, H$_2$O-poor gas dominates
the atmospheric envelope, and scenarios in which the planets accrete carbon-rich solids while 
migrating through disk regions inward of the snow line.  The C/O ratio and bulk 
atmospheric metallicity provide important clues regarding the formation and evolution of the 
giant planets.
\end{abstract}


\keywords{planetary systems --- 
planets and satellites: atmospheres --- 
planets and satellites: composition --- 
planets and satellites: individual (HD 189733b, WASP-12b, XO-1b, CoRoT-2b) --- 
stars: individual (HD 189733, WASP-12, XO-1, CoRoT-2)}

\section{Introduction}

Transit and eclipse observations provide a unique means with which to characterize 
the physical and chemical properties of ``hot Jupiter''  exoplanet atmospheres 
\citep[e.g.,][]{seager10}.  
Of particular importance in terms of furthering our understanding of planetary 
formation and evolution are the constraints on atmospheric composition provided by 
multi-wavelength photometric and spectral observations of the star-planet system 
as the transiting planet passes in front of and behind its host star as seen from 
an observer's perspective.  
Transit and eclipse observations to date have been used to infer the presence of the 
molecules H$_2$O, CO, CO$_2$, and CH$_4$ in the tropospheres and/or stratospheres of extrasolar 
giant planets 
\citep[e.g.,][]{charb05,charb08,burrows05,burrows07,burrows08,fort07,tinetti07,tinetti10,barman07,barman08,grill08,swain08a,swain08b,swain09hd189,swain09hd209,swain10,beaulieu08,beaulieu10,snellen10,madhu09,madhu10inv,madhu11gj436,madhu11wasp12b,madhu12coratio,desert09,knut11,line11opti,lee12,waldmann12}. 

Despite this remarkable feat, several factors can complicate the compositional analyses 
from transit and eclipse observations, and the results regarding molecular abundances 
are typically model dependent.  For example, in addition to being sensitive to the atmospheric 
composition, the transit and eclipse signatures are strongly affected by the atmospheric thermal 
structure, the presence of clouds and hazes, the efficiency of atmospheric dynamics, and the 
atmospheric mean molecular mass, bulk metallicity, and elemental ratios \citep[e.g.,][]{seager10}.  
The complexity of the problem is best attacked through a 
combination of sophisticated theoretical modeling and careful data analysis 
\citep[e.g., see][and references therein]{marley07,burrowsort10,showman10,seager10,winn10}, 
but degeneracies between solutions, and resulting uncertainties in composition, will continue 
to plague the determination of constituent abundances until higher-spectral-resolution observations 
become available to more unambiguously identify atmospheric species.

The largest source of degeneracy arises from uncertainties in the atmospheric thermal structure 
and composition.  To get around this problem, one can attempt to derive simultaneous constraints 
on both the composition and thermal structure through statistical treatments that systematically 
compare millions of forward models with the observed photometric and spectral data \citep{madhu09} 
or through parameter estimation and retrieval methods 
\citep{madhu10inv,madhu11gj436,madhu11wasp12b,lee12,line11opti}. 
Potential drawbacks to these techniques are the necessary treatment of a three-dimensional (3D) system 
as a one-dimensional (1D) problem and the adoption of some built-in assumptions, such as the 
presence and extinction characteristics of any clouds/hazes, or the selection of a limited number 
of molecules that are presumed to contribute to the observed behavior.  Moreover, the best 
mathematical solution may not necessarily make physical sense.  As an example, \citet{madhu09}, 
\citet{lee12}, and \citet{line11opti} all derive an unexpectedly large CO$_2$ abundance for HD 
189733b based on the dayside eclipse {\it Hubble\ Space\ Telescope\/} (HST) NICMOS data of 
\citet{swain09hd189} that so far defies theoretical explanation, even when disequilibrium processes 
are considered \citep[see][]{moses11,line10}.  It is not clear at the moment whether the theoretical 
models are at fault (e.g., they could be missing some key mechanisms that enhance photochemical 
production of CO$_2$) or whether there are some problems with the data and/or analysis (e.g.,
underestimated systematic error bars for the data, or missing opacity sources for the 
spectral models).  Although these techniques are powerful, it is always a good idea to follow 
up such studies with physically based forward models to confirm that the derived characteristics 
can be explained theoretically.

One interesting and surprising result from these and other observational analyses is the 
suggestion of a low derived water abundance for several of the close-in transiting exoplanets 
examined to date, in comparison with CO and/or CH$_4$, or in comparison with expectations based 
on solar-composition atmospheres in chemical equilibrium.  The H$_2$O abundance has been 
suggested to be anomalously low on WASP-12b (\citealt{madhu11wasp12b}, see also \citealt{cowan12}), 
and this conclusion seems firm based on model comparisons with the {\it Spitzer\ Space\ Telescope\/} 
dayside eclipse data of \citet{campo11}.  Water has also been reported to be low on HD 189733b 
\citep{grill07,swain09hd189,madhu09,desert09,sing09,line11opti,lee12}, on HD 209458b 
\citep{seager05,richardson07,swain09hd209,madhu09,madhu10inv}, and on XO-1b and CoRoT-2b 
\citep{madhu12coratio}.
The interpretation for these latter planets is not definitive given uncertainties in the thermal 
structure and other model parameters, and given that solutions with solar-like water abundances 
have been found  that provide acceptable fits to the transit and eclipse data for these planets 
\citep[e.g.,][]{burrows05,burrows06,burrows08,fort05tres1,fort08unified,fort07,barman05,barman07,barman08,tinetti07,tinetti10,grill08,charb08,swain08b,machalek08,madhu09,madhu10inv,beaulieu10,knut12}. 
However, because the fit is often improved for the case of lower water abundances
\citep{madhu09,madhu10inv,madhu11wasp12b,madhu12coratio}, the suggestion of lower-than-expected 
H$_2$O abundance should not be dismissed out of hand.  

Water is thermodynamically stable under a wide variety of temperature and pressure conditions 
expected for hot-Jupiter atmospheres \citep[e.g.,][]{burrows99,lodders02,hubeny07,sharp07}, and
due to the large cosmic abundance of oxygen, H$_2$O is expected to be one of the most 
abundant molecules on these transiting planets if one assumes solar-like elemental abundances. 
Although H$_2$O can be destroyed by ultraviolet photons, photochemical models predict that 
water will be effectively recycled in hot-Jupiter atmospheres 
\citep{moses11,line10,line11gj436}, keeping H$_2$O abundances close to 
thermochemical-equilibrium predictions.
Extinction from clouds and hazes could help mask the water absorption signatures, 
but in that case other molecular bands would be affected as well, making the spectrum appear more 
like a blackbody \citep[e.g.,][]{liou02}.
If the interpretation is robust, one possible explanation could be that the atmospheres are relatively 
enriched in carbon, with C/O ratios $>$ 1, which leads to small H$_2$O/CO and H$_2$O/CH$_4$ 
ratios in hot-Jupiter atmospheres
\citep{seager05,kuchner05,fort05,lodders10,madhu11wasp12b,madhu11carbrich,madhu12coratio,koppa12}.

The influence of elemental abundances such as the bulk C/O and C/N ratios on exoplanet 
composition and spectra has not been as well studied as some of the other factors that 
can affect the observations 
--- investigators tend to simply assume solar-like elemental ratios in their models.  This 
neglect is unfortunate, as giant planets can conceivably have elemental compositions different 
from their host stars, and the atmospheric C/O ratio can provide a particularly important clue 
for unraveling the formation and evolutionary history of the planets 
\citep[e.g.,][]{lunine04,owen06,johnson09,mousis11,madhu11carbrich,oberg11}.  
The C/O ratio in a giant planet's atmosphere will reflect not only the composition of the 
circumstellar disk in which the planet formed, but physical properties of the disk 
(such as the total disk mass and temperature structure), the planet's formation location within the 
disk (especially in relation to various condensation fronts), the rate of planet formation, 
and the evolutionary history of the planet and disk, including migration history and relative 
contribution of gas versus solid planetesimals in delivering heavy elements to the atmosphere.  

Regardless of whether the planet formed and remained beyond what was once a putative 
``snow line'' in the disk, like our own solar-system giant planets, or whether it 
migrated from behind the snow line or from elsewhere in the disk to its present short-period 
orbital positions, like the close-in transiting hot Jupiters, the C/O ratio can help unravel 
the conditions in the protoplanetary disk and the timing, history, and mechanism of the planet's 
formation and evolution 
\citep[e.g.,][]{larimer75,larimer79,lattimer78,lunine85,stevenson88,prinn89,lodders93,lodders95,lodders99,niemann98,cyr98,cyr99,owen99,atreya99,krot00,gaidos00,gautier01,hersant04,hersant08,lodders04,lodders10,lunine04,wong04,cuzzi04,cuzzi05,alibert05,kuchner05,ciesla06,owen06,guillot06,wooden08,dodson09,mousis09hd189,mousis09min,mousis11,bond10,madhu11carbrich,madhu12coratio,najita11,oberg11,ebel11,koppa12}.

The atmospheric C/O ratio itself can have a strong influence on the relative abundance of the 
spectroscopically important H$_2$O, CH$_4$, CO, CO$_2$, C$_2$H$_2$, and HCN molecules
\citep{lodders02,seager05,kuchner05,zahnle09soot,madhu10inv,lodders10,line10,madhu11wasp12b,madhu12coratio,koppa12}.
A photospheric value of C/O $\approx$ 0.55 seems to be the consensus for our Sun, despite some 
disagreement over the exact absolute C/H and O/H ratios 
\citep{allende01,allende02,asplund05,asplund09,grevesse07,grevesse10,caffau08,caffau10,lodders09}, but
possible departures from the canonical solar value or from the host-star's value need to be considered 
in atmospheric models.  

This need is spotlighted by the low derived H$_2$O abundances for some exoplanets mentioned above,
and in particular, by the recent inference of a carbon-rich atmosphere for WASP-12b 
\citep{madhu11wasp12b}.  In fact, \citet{madhu12coratio} and \citet{madhu11carbrich} suggest 
that super-solar C/O ratios could potentially explain not only the unusual eclipse photometric 
band ratios of several transiting exoplanets (e.g., WASP-12b, WASP-14b, WASP-19b, WASP-33b, XO-1b, 
CoRoT-2b), but also the lack of an apparent thermal inversion in some highly irradiated hot Jupiters, 
due to the fact that TiO --- one candidate absorber that could cause a stratospheric thermal inversion
\citep{hubeny03,fort06hd149,burrows06} --- does not form in significant quantities when 
atmospheric C/O ratios $\gta$ 1.  To provide a good fit to the secondary-eclipse data, the 
carbon-rich models identified by \citet{madhu11wasp12b} and \citet{madhu12coratio} all require 
specific constraints on species abundances and thermal profiles, and it remains to be demonstrated 
that the derived abundances make sense from a disequilibrium-chemistry standpoint.

We therefore develop equilibrium and disequilibrium models to better quantify the effects of 
the C/O ratio on exoplanet atmospheric composition in a general sense, as well as to specifically 
check whether the molecular abundances derived from analyses such as 
\citet{madhu09,madhu10inv,madhu11gj436}, \citet{madhu11wasp12b}, \citet{madhu12coratio}, 
\citet{lee12}, and \citet{line11opti} are physically and chemically plausible for assumptions 
of either equilibrium or disequilibrium conditions.  We will also check whether additional 
molecules that are typically not considered in spectral models could be contributing to the 
observed behavior of transiting hot Jupiters.  Our disequilibrium models include the effects 
of thermochemical and photochemical kinetics and vertical transport in controlling the vertical 
profiles for the major neutral carbon, nitrogen, and oxygen species through the entire relevant 
atmospheric column, from the equilibrium-dominated deep troposphere up through the quench regions 
for the major species, and on up to the photochemistry-dominated upper stratosphere 
\citep{viss10co,moses10,moses11,viss11}.  
Differences that occur as a result of the C/O ratio will be highlighted.

\section{Models}

We restrict our study to C-, N-, and O-bearing species (along with H and He) because these are 
the most cosmically abundant of the reactive heavy elements, because they will be 
incorporated into the most abundant and 
most spectrally significant molecules in the infrared photospheres of hot Jupiters, and because 
their kinetic behavior is better understood than that of other elements like sulfur, phosphorus, 
alkalis, halides, or metals.   The chemistry of other elements will be interesting in 
its own right, but given the lower cosmic abundances of these elements, their coupled chemistry 
with C-, N-, and O-bearing compounds is not expected to have a significant impact on the molecular 
abundances predicted here.  One exception is the loss of $\sim$20\% of the oxygen if the 
planet is cool enough for silicate condensation to occur \citep{lodders10}, and we account for 
that loss in our model. 

We focus on transiting planets that are hot enough that CO rather than CH$_4$ is expected to be 
the dominant carrier of atmospheric carbon because the composition of these warmer planets is 
expected to be more sensitive to changes in the C/O ratio \citep{madhu12coratio}.  We ignore 
planets with well-defined stratospheric thermal inversions because we assume (perhaps naively; 
see \citealt{knut10}) that TiO is responsible for upper-atmospheric heating 
\citep[e.g.,][]{hubeny03,fort06hd149,fort08unified,burrows06,burrows07,burrows08}, and 
thus these planets may be expected to have solar-like C/O ratios \citep{madhu11carbrich}.  The
disequilibrium chemistry of such planets has already been explored
\citep[e.g.,][]{moses11,line10,zahnle09sulf,liang04,liang03}.  

For our equilibrium calculations, we use the NASA CEA code developed by \citet{gordon94}, with 
thermodynamic parameters adapted from \citet{gurvich94}, the JANAF tables \citep{chase98}, 
\citet{burcat05}, and other literature sources.  For the disequilibrium calculations, we use
the Caltech/JPL 1D pho{\-}to{\-}chemistry/diffu{\-}sion code KINETICS \citep{allen81}
to solve the coupled continuity equations that control the vertical distributions of 
tropospheric and stratospheric species in the case of incident UV irradiation  
\citep[e.g.,][]{yung84,glad96,moses00a,moses00b,moses05,moses10,moses11,liang03,liang04,line10,line11gj436,viss10co,viss11}.  

\subsection{Photochemical and Thermochemical Kinetics and Transport Model\label{pchemmod}}

When transport time scales are shorter than chemical kinetics time scales, the atmospheric 
composition can depart from strict thermodynamic equilibrium 
\citep[e.g.,][]{prinn77,lewis84,viss11}, making thermochemical equilibrium assumptions 
no longer valid for spectroscopic modeling.  This so-called ``transport-induced quenching'' 
disequilibrium mechanism will affect the observed composition.  The consequences for 
extrasolar giant planets have been explored by \citet{lodders02}, \citet{coop06}, \citet{viss06}, 
\citet{fort06a}, \citet{hubeny07}, \citet{line10,line11gj436}, \citet{moses11}, \citet{viss11}, 
and \citet{madhu11gj436}.  The atmospheric composition can also be altered by 
the intense ultraviolet radiation from the host star.  Absorption of UV photons by 
atmospheric species can lead to dissociation of the molecules, and the resulting species 
can react to form new disequilibrium constituents.  Models of the photochemistry of 
transiting hot Jupiters have been presented by \citet{liang03,liang04}, \citet{yelle04}, 
\citet{garcia07}, \citet{koskinen07}, \citet{zahnle09soot,zahnle09sulf}, 
\citet{line10,line11gj436}, \citet{moses11}, \citet{miller-ricci12}, \citet{koppa12}, and 
\citet{venot12}.  The 
bottom line from the most recent models is that photochemical processes will affect the composition of 
the middle atmospheres (i.e., the infrared ``photospheres'') of all but the hottest of the 
hot Jupiters, leading to the removal of some expected species like CH$_4$ and NH$_3$ from 
high altitudes and to the production of some other interesting spectrally active constituents 
like atomic C, O, N, and molecular HCN, $\CtwoHtwo$, and in some instances, $\COtwo$.

\begin{figure}
\includegraphics[angle=-90,scale=0.37]{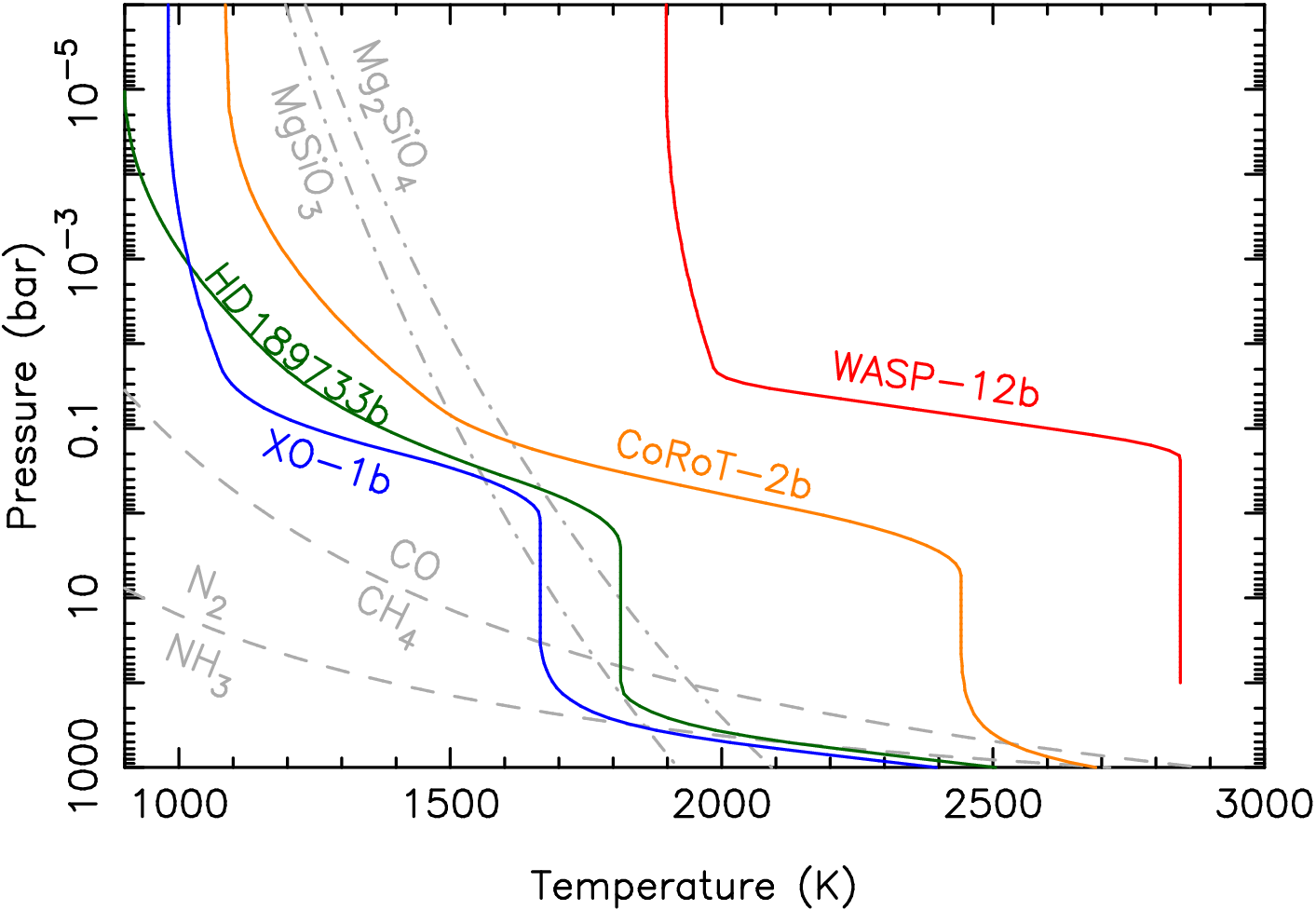}
\caption{Temperature profiles (solid lines, as labeled) adopted for our dayside atmospheres, 
selected from models presented in \citet{madhu09}, \citet{madhu11wasp12b}, and 
\citet{madhu12coratio}.  The gray dot-dashed lines illustrate the condensation curves for 
enstatite (MgSiO$_3$) and forsterite (Mg$_2$SiO$_4$), and the gray dashed lines indicate 
equal-abundance curves for N$_2$-NH$_3$ and CO-CH$_4$, all for assumed solar-composition, 
solar-metallicity atmospheres.  The more reduced components (CH$_4$ and NH$_3$) dominate 
to the left of the gray-dashed curves, and the more oxidized species (CO and N$_2$) dominate 
to the right.  Note that because our solutions do not necessarily favor solar-composition 
atmospheres, the gray curves will need to be shifted around in response to specific model 
assumptions; however, CO and N$_2$ are expected to be the dominant carbon and nitrogen 
components in the photospheres (a few bar to $\sim$0.1 mbar) of all these planets for all 
our assumed bulk elemental compositions and metallicities.  A color version of this figure 
is available in the online journal.\label{figtemp}}
\end{figure}

For our disequilibrium models, 
vertical transport is assumed to occur via molecular and eddy diffusion. The eddy diffusion 
coefficient $K_{zz}$, which is a free parameter in our models, provides a measure of the strength 
of atmospheric mixing via convection, atmospheric waves, and eddies of all scales.  
Although the vertical winds derived from general-circulation models (GCMs) such as those 
of \citet{showman09} can be used as a guide 
for defining $K_{zz}$ \citep[see][]{moses11}, we simply assume an altitude-independent 
(constant) value for $K_{zz}$ for the models presented here. We allow the solutions for the vertical 
species profiles to reach a steady state.  The models all have 198 vertical grid points spaced 
equally in log(pressure), typically ranging from the deep troposphere ($\sim$1 kbar) to 
well into the thermosphere (pressures $\lta$10$^{-10}$ bar).  The lower boundary is chosen at 
a hot-enough temperature (typically $>$ 2400 K) to encompass the quench levels for the various 
major disequilibrium species \citep[e.g.,][]{moses11,viss11}, and the upper boundary is chosen
at a high-enough altitude such that the atmosphere is optically thin to the UV radiation 
that dissociates the key photochemically active species in the model.  Note, however, that we 
do not consider ion chemistry or other processes that are important in the thermosphere and 
upper stratosphere, and our results at pressures less than $\sim$1 $\mu$bar should not be taken 
seriously.  For discussions of thermospheric composition and chemistry, see \citet{yelle04}, 
\citet{garcia07}, and \citet{koskinen10}.  Plane-parallel geometry is assumed for the radiative 
calculations in the photochemical model, and multiple Rayleigh scattering by H$_2$ and He is 
considered using a Feautrier radiative-transfer method \citep{michel92}.

The thermal structure is an input to the KINETICS code; that is, temperatures are not 
calculated self consistently.  We use specific profiles from the secondary-eclipse 
analyses of \citet{madhu09,madhu10inv}, \citet{madhu11wasp12b}, and \citet{madhu12coratio} 
to define the thermal structure (see Fig.~\ref{figtemp}).  The dayside hemisphere of the 
planet contributes to the secondary-eclipse observations, and in fact the hottest regions 
on the dayside will disproportionately dominate the planetary flux; therefore, our adopted profiles 
are typically hotter than those derived from 1D radiative-equilibrium models with efficient 
heat redistribution \citep[e.g.,][]{burrows06,burrows08,fort06hd149,fort10,barman08}. In 
principle, it would seem better to use results from the 3D GCMs to define our 
thermal structure; however, 3D models are not available for most of the planets we are 
investigating, and the complexity of the real atmospheres 
will make it difficult to realistically simulate all the chemical and physical mechanisms 
that affect the spectrally active molecules and other opacity sources in the GCMs.  We 
therefore rely on the statistical models listed above, which at least satisfy energy 
balance and provide an adequate reproduction of the observed eclipse data, to define
our thermal profiles.  Since the quenching of the nitrogen species occurs at high 
temperatures and pressures deeper in the planet's troposphere than is accounted for with 
the thermal profiles of \citet{madhu09,madhu10inv}, \citet{madhu11wasp12b}, and 
\citet{madhu12coratio}, we extend these profiles to deeper pressures assuming a 
somewhat arbitrary adiabat based on the 1D models of \citet{fort08unified}.  At the 
low-pressure end, we extend the profiles upward nearly isothermally (see Fig.~\ref{figtemp}).
The real upper atmospheres will transition into a high-temperature thermosphere, but \citet{moses11} 
and \citet{line11gj436} demonstrate that the adopted thermospheric structure has little 
effect on the results for the observable portions of the troposphere and stratosphere.  
Assuming these fixed temperature-pressure profiles, we then use the hydrostatic 
equilibrium equation to define the altitude grid.

The reaction list is taken from \citet{moses11}, and the reader can refer to that paper 
for further details.  The reactions are fully reversed assuming thermodynamic reversibility, 
which allows the model atmosphere to reach thermochemical equilibrium when kinetic reaction 
time scales are much shorter than transport time scales.  Our nominal 1$\times$ solar composition 
atmospheres assume the protosolar composition from Table 10 of \citet{lodders09}; however, 
because our model does not include the rock-forming elements, we assume that 21\% of the oxygen 
is tied up with condensed silicates and metals and is therefore not available for gas-phase 
chemistry \citep[e.g.,][]{viss05,lodders10}.  Our C/O ratios therefore refer strictly to the 
gas-phase C/O ratio after any silicate and metal-oxide condensation has occurred and not to the 
bulk planetary C/O ratio.  Note that when we alter the C/O ratio above solar, the oxygen abundance 
is kept fixed while the carbon abundance is adjusted.  Thermochemical equilibrium abundances 
from the CEA code are used to define the initial species profiles, and zero flux boundary 
conditions are adopted at the top and bottom boundaries.  Again, as discussed in \citet{moses11}, 
we expect the atmospheres of these planets to be actively escaping, but our upper boundary 
condition has no effect on the stratospheric and tropospheric profiles unless the hydrodynamic 
wind extends down below the thermosphere, at which point our hydrostatic models will be no longer 
valid.

For the incident stellar ultraviolet flux that drives the photochemistry, we use HST/STIS
spectra from epsilon Eri (a K$2\, $V star) and Chi$^1$ Ori (a G$0\, $V star) from 
the CoolCAT database \citep{ayers05} for the 1150--2830 \AA\ region to mimic spectra from
HD 189733 and WASP-12, respectively, after adjusting for the appropriate distance scalings.  
For wavelengths above and below this range, we scale the solar spectrum to match 
the expected magnitude of the flux for the appropriate stellar type and orbital distance.  
For XO-1b and CoRoT-2b, we simply adopt the solar spectrum, adjusted to the appropriate 
orbital distance.  The stellar zenith angle is fixed at 48\deg, which is appropriate for 
secondary-eclipse conditions \citep{moses11}.

\subsection{Spectral Models\label{specmod}}

Given the atmospheric thermal and chemical profiles, as described above for each planet under 
consideration (namely XO-1b, HD 189733b, CoRoT-2b, and WASP-12b), we use a radiative transfer 
code to generate a corresponding model spectrum for comparison with observations.  We compute 
emergent thermal spectra using the exoplanetary atmospheric modeling program of \citet{madhu09}. 
The code computes 1D line-by-line radiative transfer in a plane-parallel atmosphere under the 
assumptions of local thermodynamic equilibrium (LTE) and hydrostatic equilibrium.  Line-by-line 
molecular absorption due to H$_2$O, CO, CH$_4$, CO$_2$, NH$_3$, C$_2$H$_2$, HCN, and H$_2$-H$_2$ 
collision-induced absorption (CIA) is considered. The molecular linelists for H$_2$O, CO, CH$_4$, 
and NH$_3$ are obtained from \citet{freedman08} and references therein, and the HITRAN database 
\citep{rothman05,rothman09}.  The linelists for CO$_2$, $\CtwoHtwo$, and HCN are obtained from 
\citet{wattson86}, \citet{rothman05}, and \citet{harris08}, respectively.  The CIA opacities are 
obtained from \citet{borysow97} and \citet{borysow02}. The emergent spectrum is calculated 
assuming the planet is at secondary eclipse, i.e. that the integrated flux from the entire 
dayside atmosphere of the is observed. In order to compute planet-star flux ratios, a Kurucz 
model spectrum \citep{castelli04} is adopted for the given stellar parameters.  While comparing 
our models with observations of a given planet, we integrate the model spectra over the 
instrumental bandpasses for which data are available for that system and compute a goodness 
of fit \citep[e.g.,][]{madhu09,madhu11wasp12b}.

\section{Results}

Assuming a ``nominal'' temperature-pressure profile selected for each of the exoplanets 
under consideration (see Fig.~\ref{figtemp}), we investigate how the atmospheric chemistry 
changes for different assumptions of the atmospheric C/O ratio and metallicity.  We first 
calculate thermochemical equilibrium constituent abundances as a function of pressure 
for C/O ratios ranging from 0.1 to 1.9, for a variety of assumed bulk atmospheric 
metallicities.  Then, we select one or more representative models with different 
metallicities and C/O ratios to investigate possible disequilibrium chemistry 
in more detail with our thermochemical and photochemical kinetics and transport 
models.  Disequilibrium-model solutions will be presented for carbon-to-oxygen ratios 
of both solar-like (C/O $\sim$ 0.5) and carbon-rich (C/O $\gta$ 1) atmospheres.  We then 
create synthetic spectra from the model results and compare these with the secondary-eclipse 
data to determine if we can find disequilibrium model solutions that provide an acceptable 
fit to the data.  Our resulting models will be non-unique.
Note that because haze extinction is more likely 
to affect the transit observations (due to the long slant path being able to amplify extinction 
from layers that would otherwise be optically thin in the vertical), we restrict our data 
comparisons to the secondary-eclipse observations.  In addition, because {\it Spitzer\/} 
photometric data are available for all the planets in question, we focus on the {\it Spitzer\/} 
dataset.

For any given metallicity, thermal structure, and finite list of molecules that contribute to 
the spectral signatures, the abundances of the major gas-phase opacity sources required to 
fit the secondary-eclipse data are well constrained, and there 
will be a correspondingly narrow range of C/O ratios (if any) that will produce these 
desired abundances.  However, the best-fit C/O ratio will change for different assumptions 
about the thermal structure and metallicity.  Although we examine a few different scenarios 
in this study, we do not explore the full range of possible parameter space.  Our main goal 
is to illustrate how the C/O ratio affects the composition of highly irradiated hot Jupiters.  
Our purpose in developing individual models with acceptable spectral fits is to determine 
whether atmospheres with chemical abundances that are theoretically consistent with 
``reasonable'' metallicities and thermal structures can be derived that reproduce the 
observed secondary-eclipse data, and in particular, whether models with high C/O ratios are 
reasonable for the exoplanets that have been reported to have relatively low H$_2$O abundances.  
Although we focus on four specific transiting planets in this paper (XO-1b, HD 189733b, 
CoRoT-2b, WASP-12b), the planets we have selected span a wide range of apparent dayside 
temperatures from $\sim$1300-2900 \citep{cowanagol11} and can thus serve as proxies for 
a variety of hot Jupiters.

\subsection{HD 189733b Results\label{hd189sect}}

The moderate-temperature exoplanet HD 189733b, discovered by \citet{bouchy05}, resides in 
the upper part of the O1 regime of the \citet{madhu12coratio} classification scheme, in which 
the C/O ratio is less than unity (as is favored from statistical analyses of the HD 189733b 
observational data; \citealt{madhu09}), and temperatures are warm enough that CO dominates 
over CH$_4$ in the observable portion of the atmosphere (see Fig.~\ref{figtemp}), yet temperatures 
are cool enough that Ti will be tied up in condensates rather than as gas-phase TiO.  
Various analyses suggest that HD 189733b does not have a significant stratospheric thermal 
inversion \citep{fort08unified,burrows08,barman08,charb08,grill08,swain09hd189,madhu09,showman09}.

\begin{figure*}
\begin{tabular}{ll}
{\includegraphics[angle=0,clip=t,scale=0.37]{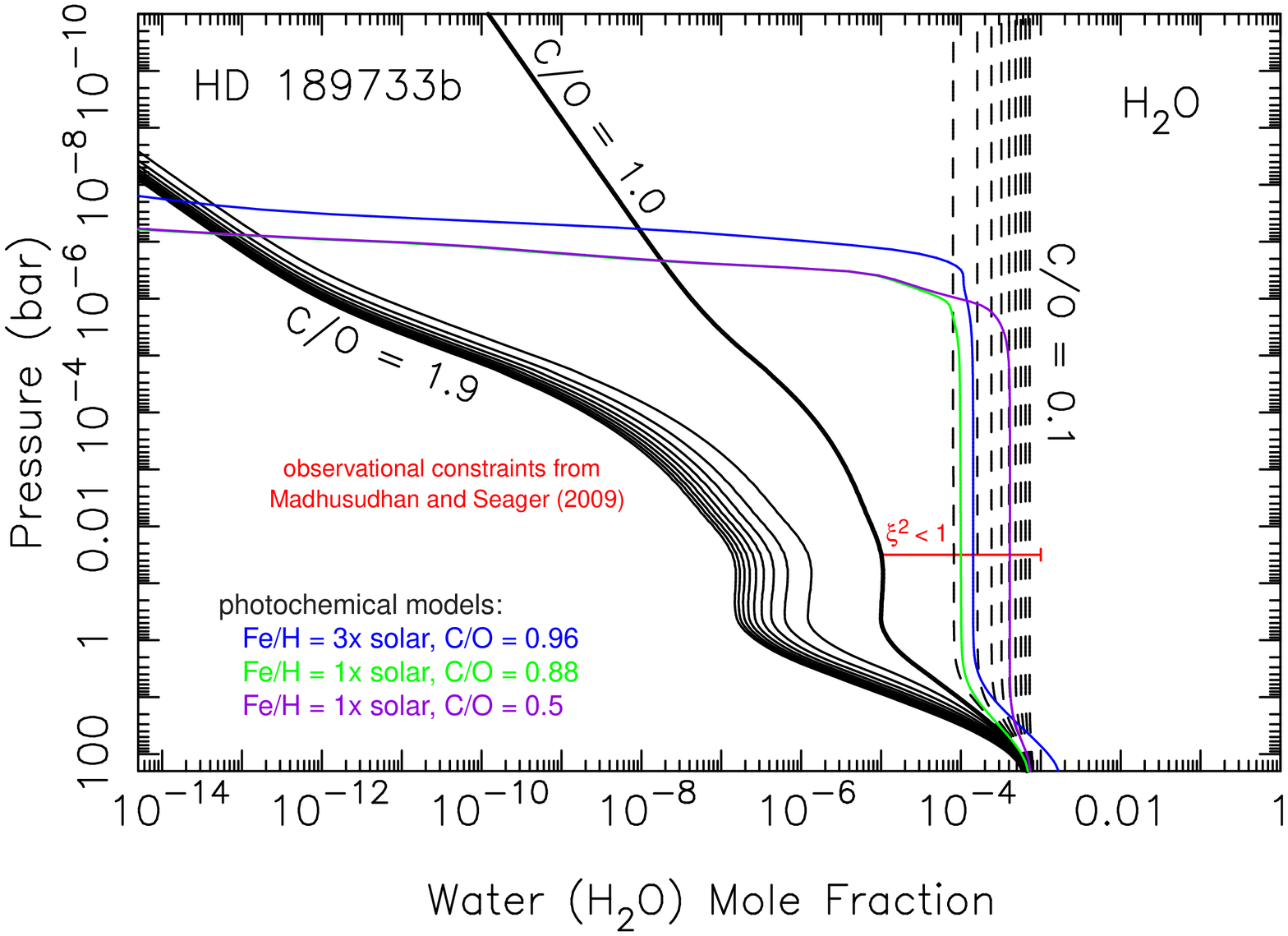}}
&
{\includegraphics[angle=0,clip=t,scale=0.37]{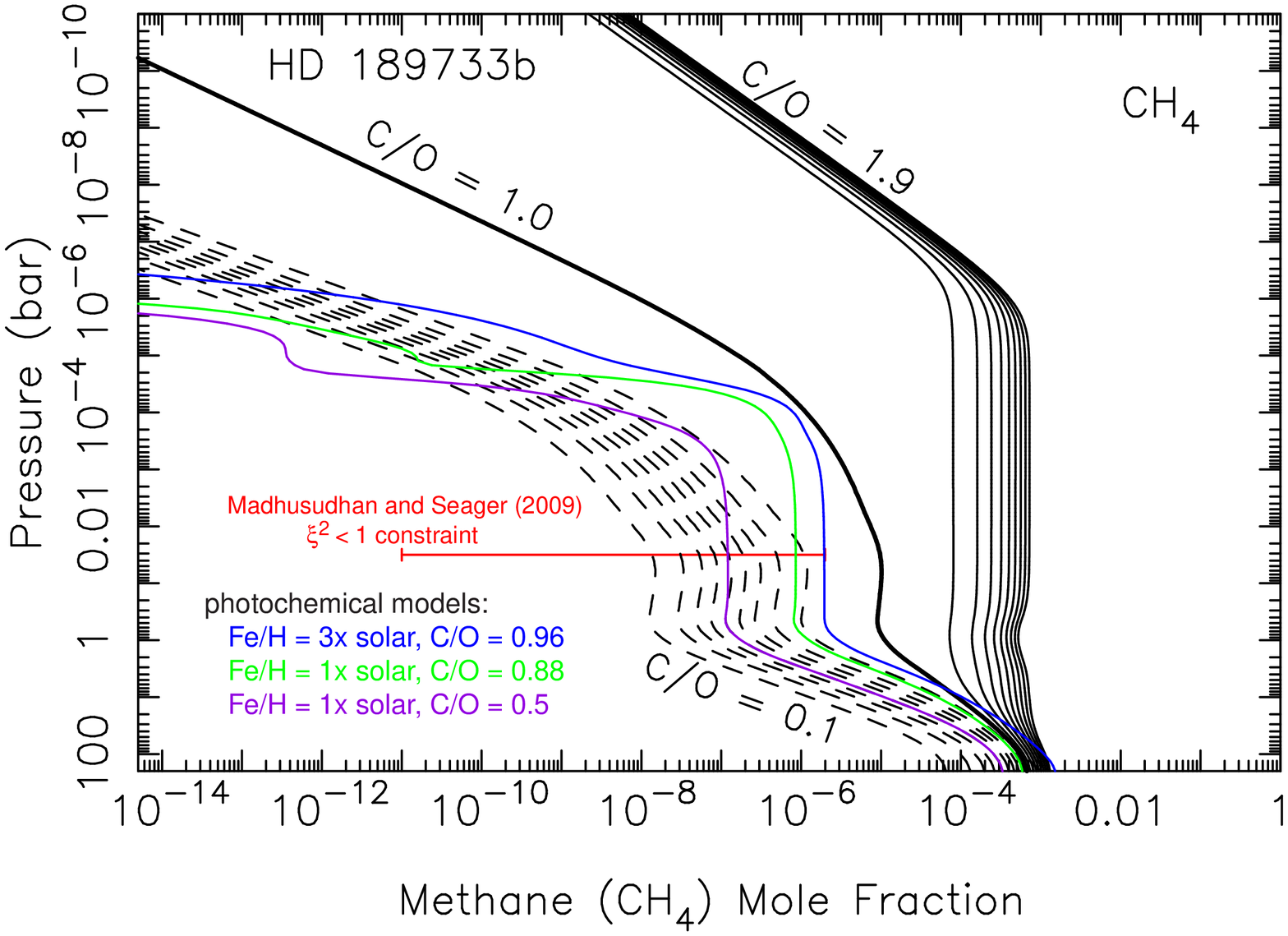}}
\\
{\includegraphics[angle=0,clip=t,scale=0.37]{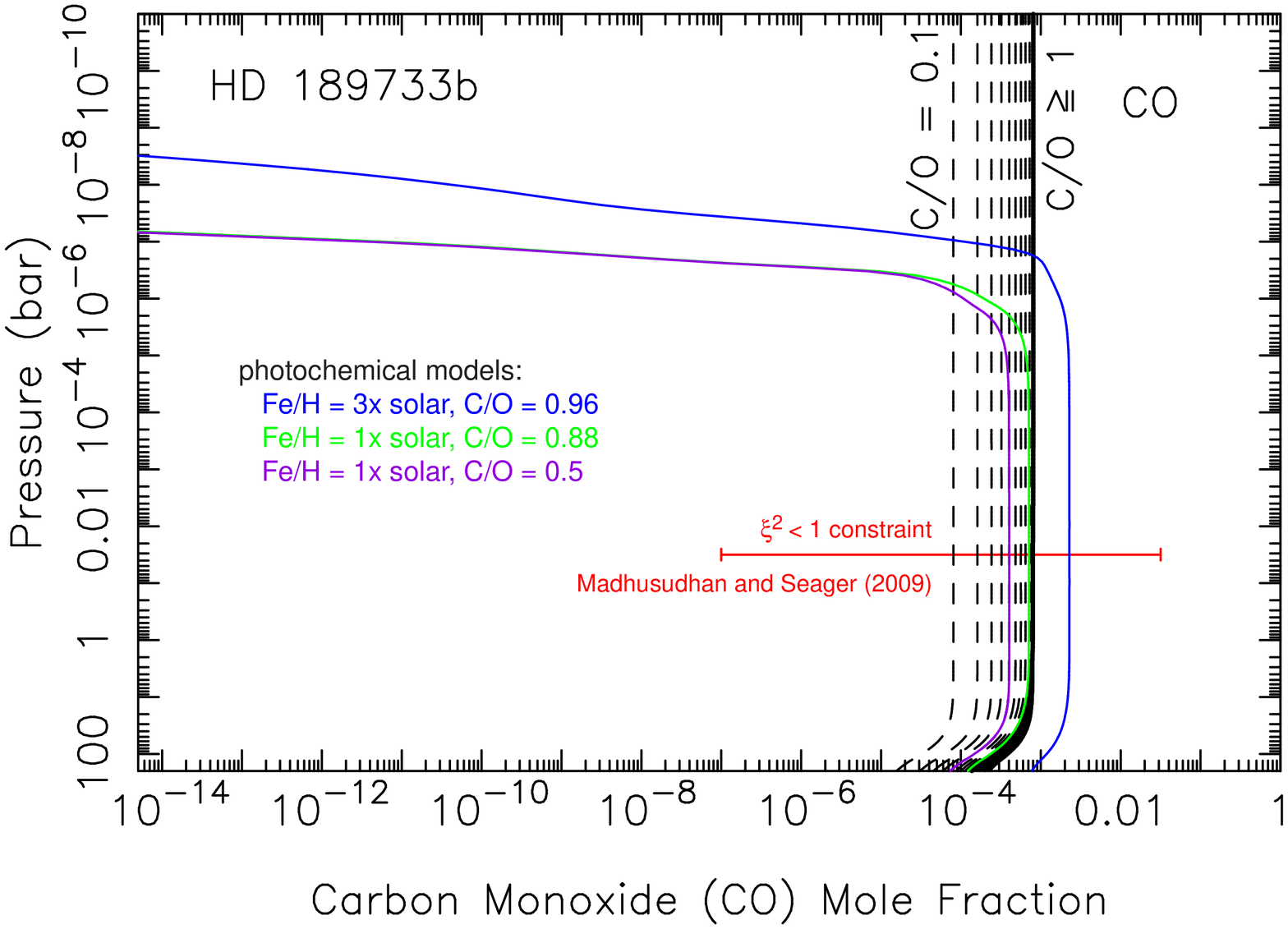}}
&
{\includegraphics[angle=0,clip=t,scale=0.37]{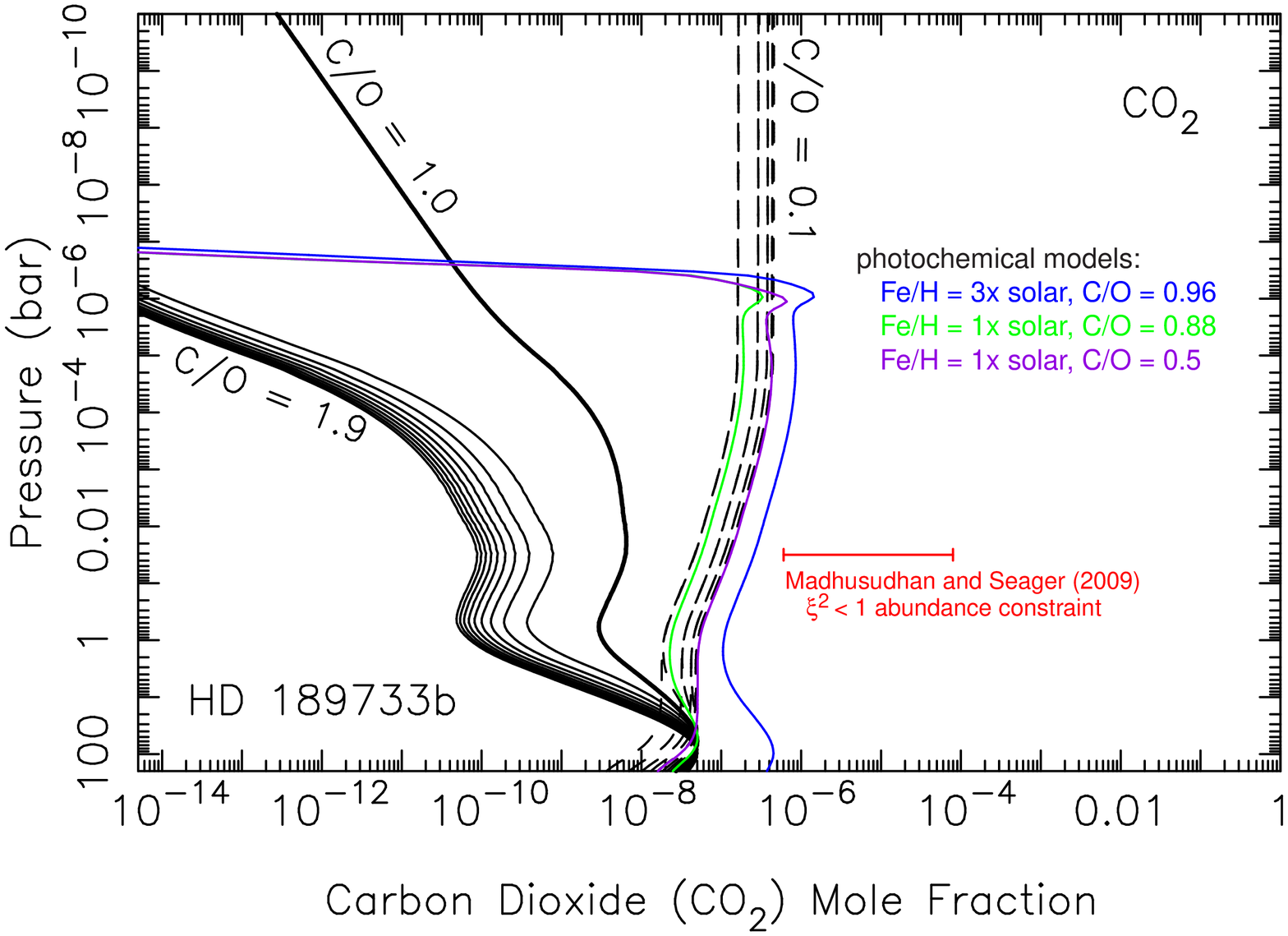}}
\\
{\includegraphics[angle=0,clip=t,scale=0.37]{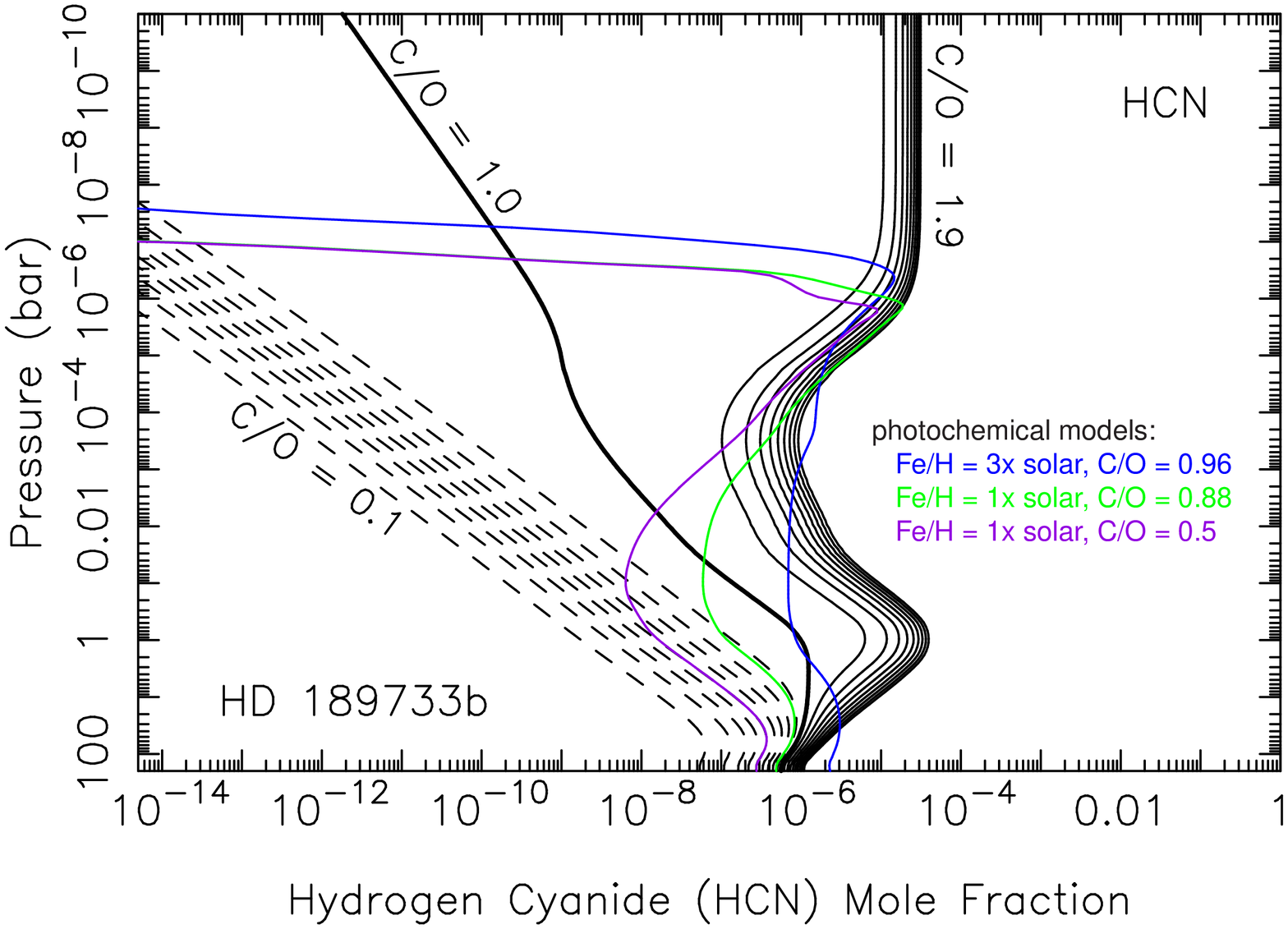}}
&
{\includegraphics[angle=0,clip=t,scale=0.37]{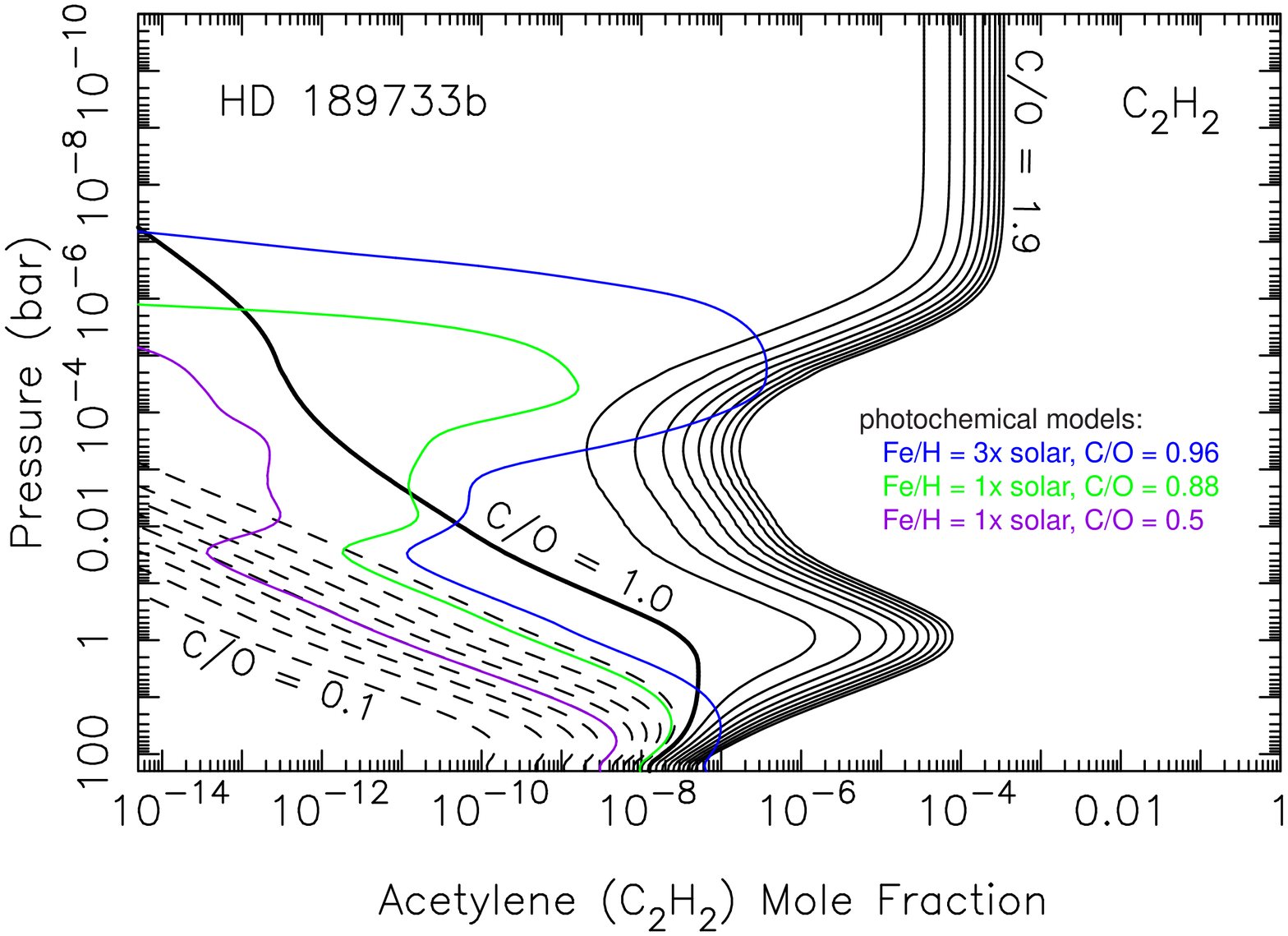}}
\\
\end{tabular}
\caption{Mixing-ratio profiles for H$_2$O, CH$_4$, CO, CO$_2$, HCN, and C$_2$H$_2$ on HD 189733b 
from our thermochemical-equilibrium models with assumed 1$\times$ solar metallicity but different 
assumed C/O ratios (ranging from 0.1 to 1.9, incrementing by 0.1 --- dashed black lines for 
C/O $<$ 1, solid lines for C/O $\ge$ 1).  The colored lines represent disequilibrium-chemistry 
results for a model with 1$\times$ solar metallicity and a C/O ratio of 0.5 (purple), a model 
with 1$\times$ solar metallicity and a C/O ratio of 0.88 (green), and a model with 3$\times$ 
solar metallicity and a C/O ratio of 0.96 (blue).  All disequilibrium models have an assumed 
moderately-low eddy diffusion coefficient of $K_{zz}$ = 10$^8$ $\cmtwo$ $\smone$.  The horizontal 
line segments (red) represent the abundance constraints derived by \citet{madhu09} from {\it 
Spitzer\/} IRAC and MIPS observations \citep{knut07,knut09hd189,charb08}.  Note that 
\citet{madhu09} did not consider HCN and C$_2$H$_2$ in their analysis, so abundance constraints 
for these molecules are not available.  
A color version of this figure is presented in the online 
journal.\label{hd189equil1x}}
\end{figure*}

\subsubsection{HD 189733b Chemistry \label{hd189chemsect}}

Figure~\ref{hd189equil1x} illustrates how the thermochemical e{\-}quil{\-}i{\-}bri{\-}um 
composition of HD 189733b varies as a function of the atmospheric C/O ratio for our assumed nominal 
dayside thermal structure and an assumed 1$\times$ solar metallicity.  For a solar-like 
C/O ratio of $\sim$0.5, CO and H$_2$O are the most abundant heavy molecules, CO$_2$ is 
much less abundant at the $\lta \,$1 ppm level, CH$_4$ and NH$_3$ are well below
ppm levels in thermochemical equilibrium at pressures less than 1 bar, and HCN and 
$\CtwoHtwo$ are trace constituents in equilibrium such that they would have no influence on 
the eclipse spectra.  Although transport-induced quenching could lead to CH$_4$ (and NH$_3$) 
mole fractions greater than 1 ppb at pressures less than 1 bar \citep[see][]{moses11,viss11}, 
the CH$_4$ mole fraction is never likely to rival that of CO for solar-like C/O ratios, unless 
effective eddy diffusion coefficients are greater than 10$^{11}$ cm$^2$ $\smone$ in the CO-CH$_4$ 
quench region \citep{moses11,viss11}.  The relatively low derived CH$_4$ abundances from the 
transit and eclipse observations \citep{swain08b,swain09hd189,madhu09,lee12,line11opti} suggest 
that atmospheric mixing is not that vigorous on HD 189733b \citep[see][]{viss11}, and CO mole 
fractions are likely greater than those of CH$_4$.

On the other hand, when the C/O ratio is greater than unity, Fig.~\ref{hd189equil1x} shows that the 
equilibrium methane abundance can approach or even exceed that of CO; other oxygen-bearing 
species like H$_2$O and CO$_2$ become much less abundant, and previously unimportant 
carbon-bearing species like HCN and C$_2$H$_2$ become abundant enough to influence spectra.  
At dayside photospheric temperatures and pressures ($\sim$1800-900 K, $\sim$1-to-10$^{-4}$ bar), 
CO is a major constituent for all assumed C/O ratios.  When C $\approx$ O, available carbon and 
oxygen will bond together to predominantly form CO; when excess O is available (i.e., C $\lta$ O), 
the extra O goes into the formation of H$_2$O, and to a lesser extent CO$_2$; when excess C is 
available (i.e., C $\gta$ O), the extra C goes into the formation of CH$_4$, and to a lesser 
extent HCN and C$_2$H$_2$.  When the atmospheric C/O ratio is very near unity, small changes 
in the C/O ratio lead to large changes in the species abundances.

\begin{figure}
\includegraphics[angle=-90,scale=0.37]{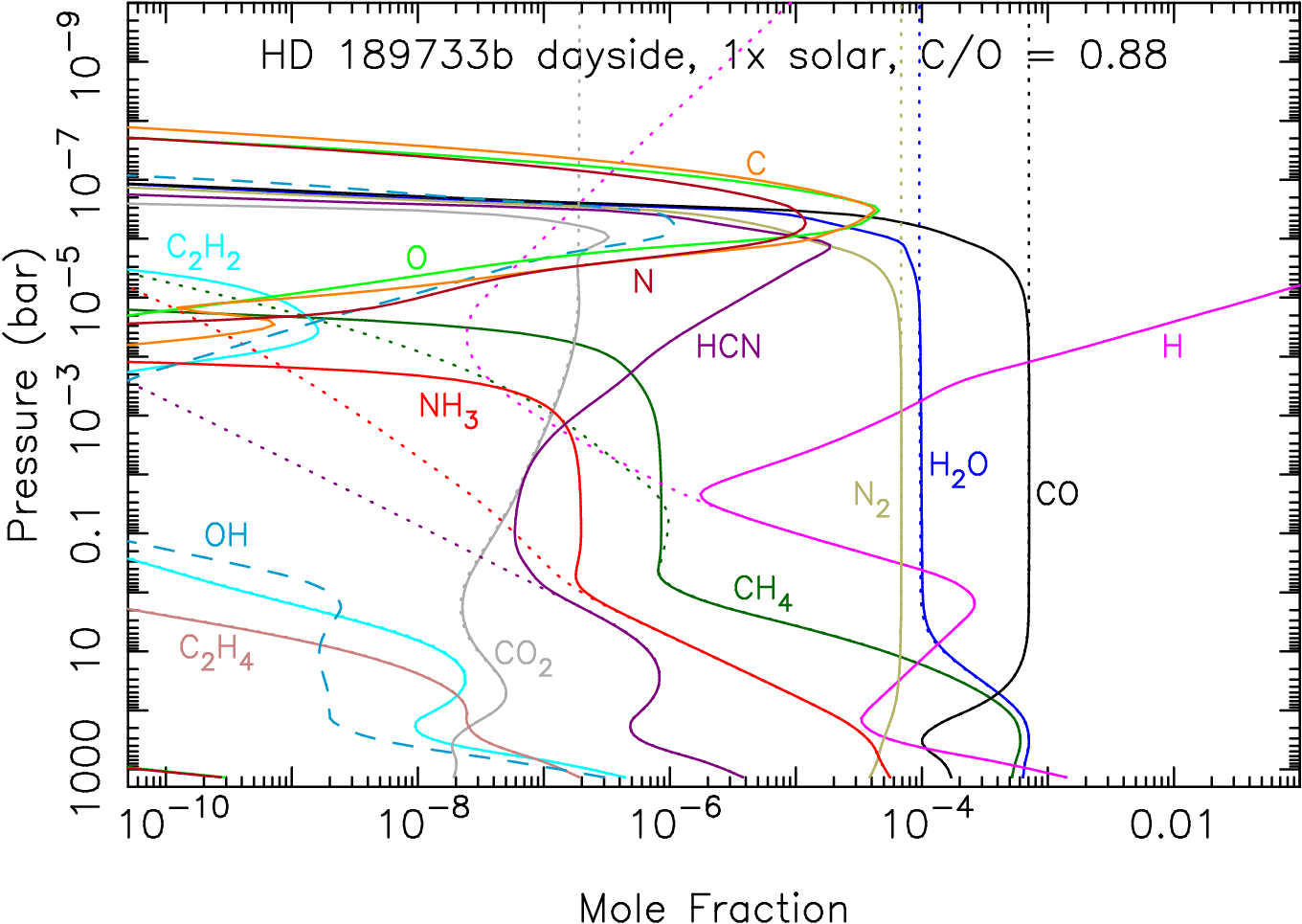}
\caption{Mole-fraction profiles for several important species (as labeled) in a 
thermochemical and photochemical kinetics and transport model for HD 189733b.  Model assumptions 
include 1$\times$ solar metallicity, a C/O ratio of 0.88, and $K_{zz}$ = 10$^8$ $\cmtwo$ 
$\smone$ (see also the green curves in Fig.~\ref{hd189equil1x}).  The dotted lines show the 
thermochemical-equilibrium abundances for the species, using the same color coding.  
A color version of this figure is available in the online journal.\label{hd189pchem}}
\end{figure}

Working solely from the \citet{madhu09} statistical constraints on the ensemble of 
models that provide the best fit to the HD 189733b {\it Spitzer\/} secondary-eclipse data of
\citet{knut07,knut09hd189} and \citet{charb08} (see the red horizontal segments in 
Fig.~\ref{hd189equil1x}),
it is clear that equilibrium models with C/O ratios $<$ 1 will be required to provide a decent fit 
to the observations.  Here we use the $\xi ^2$ $<$ 1 constraints from \citet{madhu09}, where $\xi ^2$ = 
$\chi ^2$/$N_{obs}$, with $N_{obs}$ = 6 being the number of photometric data points available for the 
goodness-of-fit calculations.  The CH$_4$ and CO$_2$ constraints 
seem to favor low C/O ratios (perhaps even C/O $<$ 0.1), whereas the H$_2$O constraints can 
be satisfied with higher C/O ratios.  In reality, the water abundance has the biggest 
influence on the spectral behavior.  For our nominal temperature profile and an assumed 
1$\times$ solar metallicity, H$_2$O mole fractions that fall near $\sim$1$\scinot-4.$ 
provide a significantly better fit to the data than mole fractions even a factor of a few 
away from the middle of the H$_2$O-constraint range provided by the \citet{madhu09} $\xi ^2$ 
$<$ 1 criterion.  It turns out that for a given temperature profile and metallicity, there 
is a very narrow range of C/O ratios that provides the necessary best-fit H$_2$O mole fraction.  
For our assumed conditions, a C/O ratio near 0.9 provides the desired equilibrium water 
abundance of $\sim$1$\scinot-4.$.  On the other hand, a C/O ratio of $\sim$0.9 also implies a 
CH$_4$ abundance at the extreme high end of what is allowable by \citet{madhu09}, and implies 
a very low CO$_2$ abundance that appears to fall well outside the $\xi ^2$ $<$ 1 range for CO$_2$.  
Keep in mind, however, that the $\COtwo$ abundance is essentially unconstrained at the $\xi ^2$ $<$ 2 
level \citep{madhu09}, and the degeneracy between the contributions of CO and CO$_2$ in the IRAC 
channels could be complicating the identification and quantification of the $\COtwo$ abundance.  

Can disequilibrium chemistry help improve the comparison with observational constraints?
Figure \ref{hd189equil1x} also shows results from a disequilibrium model with an assumed 1$\times$ 
solar metallicity and atmospheric C/O ratio of 0.88 (the green curve).  For this model, also shown 
in Fig.~\ref{hd189pchem}, the H$_2$O abundance essentially 
follows the chemical equilibrium curve until it reaches high-enough altitudes that molecular 
diffusion begins to control the profile, at altitudes well above the infrared photosphere 
(located at a few bars to $\sim$0.1 mbar).  Photochemistry and transport-induced quenching 
therefore do not change the favored C/O ratio of $\sim$0.9 for a 1$\times$-solar-metallicity 
atmosphere and our nominal temperature profile.  We also find that photochemical production of 
CO$_2$ from CO and H$_2$O chemistry results in little net additional CO$_2$, except at very high 
altitudes, because the CO$_2$ is able to recycle back to CO and H$_2$O \citep[see][]{moses11}.  
In this regard, the results from our fully-reversed reaction mechanism are consistent with those 
of \citet{moses11}, who find that CO$_2$ remains in equilibrium with H$_2$O and CO throughout 
much of the dayside atmospheric column, in contrast to the results of \citet{line10}, in which 
an incompletely reversed reaction mechanism allows CO$_2$ abundances to be enhanced by a factor 
of $\sim$2 due to photochemistry.  Even with a factor-of-two photochemical enhancement, the 
CO$_2$ mole fraction would not be large enough in the C/O = 0.88 case to fall within the desired 
\citet{madhu09} $\xi ^2$ $<$ 1 range for CO$_2$.  For the case of methane, transport-induced 
quenching slightly enhances the mole fraction as compared with equilibrium in the $\sim$1-to-10$^{-5}$ 
bar region, and only above the infrared photosphere do disequilibrium processes reduce the CH$_4$ 
abundance; thus, disequilibrium chemistry does not improve the CH$_4$ abundance comparisons with 
regard to the \citet{madhu09} $\xi ^2$ $<$ 1 range for methane.  
Given that water dominates the spectral behavior in the mid-infrared, it may not matter that 
the other species do not fall precisely within the \citet{madhu09} desired ranges 
(see Section~\ref{hd189specsect} below).  

The disequilibrium chemistry of HD 189733b with a solar-like atmospheric C/O ratio is discussed 
extensively in \citet{moses11}, and we will not repeat that discussion here \citep[see also][]{line10}.  
The main reaction mechanisms controlling the composition are very similar in an 
atmosphere with C/O = 0.88 versus 0.5, but the resulting abundances can vary significantly between 
the two cases because of the different equilibrium stabilities.  For instance, the water abundance 
is higher in the C/O = 0.5 disequilibrium model than in the C/O = 0.88 model purely because there is 
more oxygen available once carbon sequesters an amount of oxygen need to form CO, and because 
photochemical processes do not efficiently remove $\HtwoO$ and convert it to other oxygen-bearing 
species.   The dominant $\HtwoO$ recycling mechanism operating in our models is scheme (10) in 
\citet{moses11}.  Because our nominal thermal structure is warmer than that assumed in 
\citet{moses11} (and our adopted $K_{zz}$ is smaller), our quenched abundances of CH$_4$ and NH$_3$ 
are smaller than in \citet{moses11}.  Note from Fig.~\ref{hd189equil1x} that CH$_4$ quenches at 
about the same pressure level for both the C/O = 0.88 model (green) and C/O = 0.5 model (purple).  The 
quench point is under 
the control of reaction scheme (2) of \citet{moses11}, with OH + $\CHthree$ $\rightarrow$ $\CHtwoOH$ 
+ H being the rate-limiting step for the quenching of CH$_4$ $\rightarrow$ CO conversion under 
the thermal-structure conditions of our nominal model \citep[see also][]{viss11}.  Because the 
equilibrium $\CHfour$ abundance is larger at that quench point in the C/O = 0.88 model, the 
disequilibrium quenched methane abundance is larger with C/O = 0.88 than with C/O = 0.5.  The larger 
quenched methane mole fraction then enhances the effectiveness of mechanisms that convert 
$\CHfour$ to hydrocarbons like $\CtwoHtwo$ (see scheme (12) of \citealt{moses11}), and the total 
column abundance of $\CtwoHtwo$ through the middle atmosphere is significantly greater for the 
C/O = 0.88 model.

A similar result occurs for the HCN abundance at low altitudes, but not at high altitudes.  Hydrogen 
cyanide can continue to remain in equilibrium with CO, $\CHfour$, and $\NHthree$ in the lower 
and middle stratosphere after these 
molecules quench.  The abundance of molecules that do not contain carbon and oxygen, like $\NHthree$ 
and $\Ntwo$, tend to be relatively unaffected by the C/O ratio.  However, because the CO and 
quenched $\CHfour$ abundances are larger for greater C/O ratios, processes that kinetically convert 
CO and $\NHthree$ into HCN (see scheme (8) from \citealt{moses11}) or that convert $\CHfour$ and 
$\NHthree$ into HCN (see schemes (7) and (14) in \citealt{moses11}) are more effective with high 
C/O ratios than low C/O ratios.  That leads to a much larger HCN abundance at low- and 
mid-stratospheric altitudes above the quench point with the C/O = 0.88 model as compared with the 
C/O = 0.5 model.  However, the upper atmospheric peak abundance of HCN is similar in both models due 
to the fact that CO and $\Ntwo$ photochemistry are responsible for the high-altitude production 
(e.g., see the first scheme in section 3.5 of \citealt{moses11}).  High-altitude HCN production 
occurs through the reaction N + OH $\rightarrow$ NO + H, followed by NO + C $\rightarrow$ CN + O, 
followed by CN + $\Htwo$ $\rightarrow$ HCN + H.  The N and C required for this mechanism are 
derived from $\Ntwo$ and CO photolysis.  Both these molecules become optically thick to the 
photolyzing UV radiation (and therefore self shield) at high altitudes, so this mechanism only 
works within a limited altitude region, but the HCN production rates are substantial.  Because the 
CO and $\Ntwo$ column abundances at optical depths less than a few are the key defining values 
for this process to work (which does not change between models), the total production rates from 
this process are similar in our C/O = 0.5 and C/O = 0.88 models.  Hydrogen cyanide is therefore a 
key photochemical product regardless of the C/O ratio, although the total column above $\sim$1 bar 
will be larger for higher C/O ratios due to the thermochemical kinetics schemes (7), (8), and (14) 
mentioned in \citet{moses11}.  

In both models, CO and CO$_2$ track equilibrium profiles because kinetic recycling is efficient, 
and there are fewer ``leaks'' into or out of the cycles compared with CH$_4$, NH$_3$, HCN, and 
C$_2$H$_x$ hydrocarbons.  The CO recycling is discussed in \citet{moses11}.  Carbon dioxide is 
recycled through the forward-reverse reaction pair OH + CO $\lrarrow$ $\COtwo$ + H, which is 
effective throughout the stratosphere for both models.  Some net photochemical production for 
CO$_2$ does occur at high altitudes, but it does not change the total column abundance in the 
photosphere.

\begin{figure}
\includegraphics[angle=0,scale=0.48]{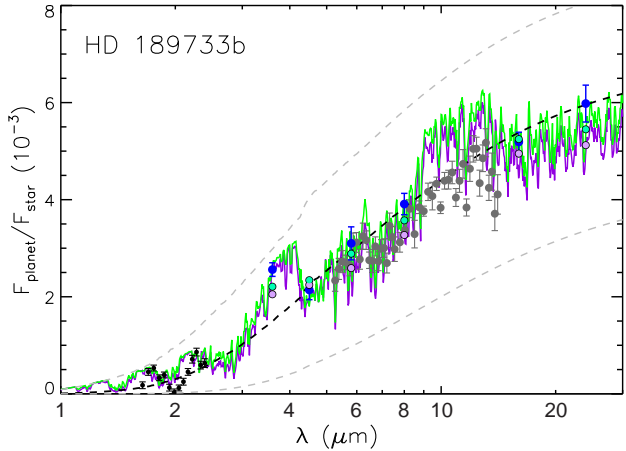}
\caption{Secondary-eclipse spectra for HD 189733b (in terms of the ratio of the flux 
of the planet to that of the host star).  The solid lines represent synthetic spectra 
from the results of our disequilibrium chemistry models with 1$\times$ solar metallicity and 
$K_{zz}$ = 10$^8$ $\cmtwo$ $\smone$, for C/O ratios of either 0.88 (green) or 0.5 (purple).
The blue circles with error bars represent the {\it Spitzer\/} IRAC and MIPS data from 
\citet{knut07,knut09hd189} and \citet{charb08}.  The small black circles with error bars 
represent the HST/NICMOS data of \citet{swain09hd189}, and the gray circles with error bars 
represent the {\it Spitzer\/} IRS data of \citet{grill08}.  The circles without error bars 
represent our C/O = 0.88 (green) or C/O = 0.5 (purple) model results convolved over the 
{\it Spitzer\/} IRAC and MIPS instrument response functions.  Note the increased 
absorption for the purple C/O = 0.5 model in all the {\it Spitzer\/} channels; the C/O = 0.88 
model fits the MIPS and IRAC data better than the C/O = 0.5 model for our particular assumed 
metallicity and thermal structure.  The dashed lines are blackbody curves for assumed 
temperatures of 900 K (bottom), 1400 K (middle), and 1800 K (top).  A color version of this 
figure is available in the online journal.\label{spechd189}}
\end{figure}

\subsubsection{HD 189733b Spectra\label{hd189specsect}}

Figure \ref{spechd189} shows synthetic spectra from our disequilibrium models, in comparison with the 
{\it Spitzer\/} IRAC and MIPS secondary-eclipse data for HD 189733b.  The plot shows spectra
from disequilibrium models with assumed C/O ratios of 0.88 and 0.5 (the green and purple curves 
from Fig.~\ref{hd189equil1x}).  These two chemical models have 
identical assumptions other than the C/O ratio.  As is obvious from the plots, the model with 
C/O = 0.88 provides a statistically better fit to the data than the more solar-like C/O = 0.5 
model for our assumed 1$\times$-solar-metallicity atmosphere with our adopted nominal thermal 
profile (i.e., $\xi ^2$ $<$ 1.9 for the C/O = 0.88 model versus $\xi ^2$ $<$ 5.0 for the C/O = 0.5 
model).  It is also apparent from the plot that the fit could have been improved 
with somewhat less methane (note the excess model absorption at 3.6 and 8.0 microns, where 
CH$_4$ contributes opacity) and perhaps more CO$_2$ (note the fit at 4.5 microns, where CO and 
CO$_2$ contribute opacity), as was suggested by the \citet{madhu09} $\xi ^2$ $<$ 1 ranges.  
However, the fit from the disequilibrium chemical model with C/O = 0.88 is within 1.4-sigma 
in all the {\it Spitzer\/} bands, on average, implying that theoretically plausible models with 
carbon-enhanced atmospheres (compared with the solar C/O ratio) can be consistent with the 
HD 189733b {\it Spitzer\/} secondary-eclipse observations.  

\begin{figure*}
\includegraphics[clip=t,scale=0.87]{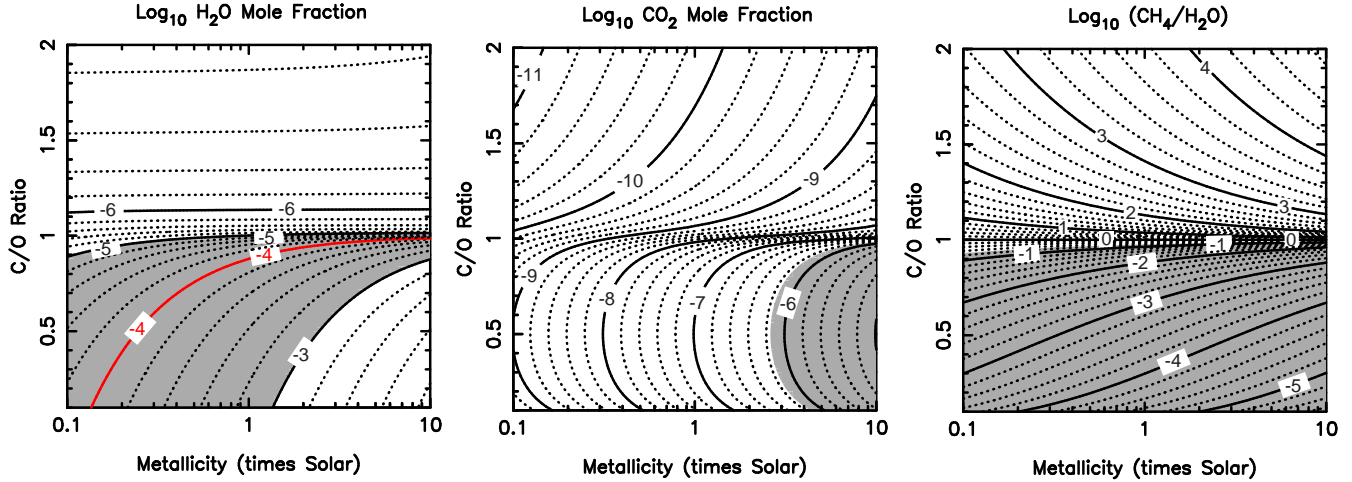}
\caption{The mole fraction of $\HtwoO$ (Left), $\COtwo$ (Middle), and the CH$_4$/H$_2$O ratio 
(Right) as a function of the C/O ratio and metallicity, assuming thermochemical-equilibrium 
conditions at 1345 K and 80 mbar (i.e., within the photosphere of our nominal model for HD 189733b).
The shaded regions represent the $\xi ^2$ $<$ 1 observational constraints from \citet{madhu09}.  
The CH$_4$/H$_2$O ratio is very sensitive to the C/O ratio, and the CO$_2$ and H$_2$O mole fractions
are very sensitive to both the atmospheric metallicity and C/O ratio.  The planetary 
spectrum itself is very sensitive to the water abundance, and the red curve in the figure on the 
left shows our best-fit $\HtwoO$ model fraction for our nominal thermal structure.  The 
observations indicate C/O $<$ 1 for HD 189733b, but not all of the observationally 
allowable phase space is chemically plausible.  The higher the atmospheric metallicity, the closer
to unity the C/O ratio must be to provide the best fit to the {\it Spitzer\/} 
secondary-eclipse spectra.  A color version of this figure is available in the online 
journal.\label{boxratios}}
\end{figure*}

It must be kept in mind that the fit is not unique.  More solar-like C/O ratios may also be 
consistent with the data, but in that case, Fig.~\ref{spechd189} demonstrates that the 
atmosphere must be much warmer than we are assuming in our nominal model, as our current 
combination of thermal structure and its resulting chemically-constrained water abundance 
in the C/O = 0.5 model leads to too much absorption in many of the {\it Spitzer\/} bandpasses.  
However, our temperatures are already at the maximum of what has been predicted from 
self-consistent 1D models with no energy redistribution from the dayside (\citealt{barman08}, 
\citealt{burrows08}; see also \citealt{fort06hd149}) or from the dayside-average atmosphere 
in 3D models \citep{showman09}, so there is a limit to how low the C/O ratio can go to fit 
the data while still maintaining a plausible thermal structure.

The solution also depends on assumptions about atmospheric metallicity.  For a fixed thermal 
structure, different metallicities require a different C/O ratio to maintain the desired 
H$_2$O mole fraction to compare well with observations (see Fig.~\ref{boxratios}).  For example, 
if the atmospheric metallicity were 3$\times$ solar, the C/O ratio would need to be $\sim$0.96 
for the H$_2$O mole fraction to be near 10$^{-4}$ in equilibrium in the photospheric region 
of HD 189733b for our nominal thermal structure (see also the disequilibrium model represented 
by the blue curve in Fig.~\ref{hd189equil1x}).  The greater the metallicity, the closer to unity 
the C/O ratio must be to remain consistent with the spectral analysis of \citet{madhu09} (i.e., 
super-solar metallicities tend to require super-solar C/O ratios to remain consistent with 
observations, for atmospheric temperatures at or below those represented by our nominal 
thermal profile).  For subsolar metallicities, more solar-like C/O ratios could produce the 
desired H$_2$O abundance.  

As was shown by \citet{madhu12coratio}, the relative abundance of methane and water is a sensitive 
indicator of the C/O ratio in hydrogen-dominated atmospheres, particularly when photospheric 
temperatures are warm enough that CO dominates over methane as the major carbon carrier.  A 
CH$_4$/H$_2$O ratio less than unity indicates a C/O ratio less than 1 (see Fig.~\ref{boxratios}).  
The \citet{madhu09} analysis firmly suggests a C/O ratio $<$ 1 for HD 189733b; however, just how 
far below unity the C/O ratio can be to remain consistent with observations depends on metallicity.  
Figure \ref{boxratios} demonstrates that while CH$_4$/H$_2$O ratios $<$ 10$^{-5}$ are allowable 
by the \citet{madhu09} spectral analysis, such values are not physically plausible unless the 
C/O ratio is strongly subsolar (i.e., $<$ 0.1) and/or metallicities are greatly supersolar.

If the CO$_2$ abundance were well constrained from observations, one could also constrain the 
overall atmospheric metallicity due to the well-known sensitivity of CO$_2$ to metallicity 
(Fig.~\ref{boxratios}; see also \citealt{lodders02}, \citealt{fort05tres1,fort06hd149,fort08colors}, 
\citealt{burrows06,burrows08}, \citealt{zahnle09soot}, \citealt{line10}, \citealt{moses11}).  
However, using the 4.5-$\mu$m {\it Spitzer\/} channel to constrain the CO$_2$ abundance 
could be problematic due to the additional contribution of CO opacity at these wavelengths 
\citep{fort10,madhu09}.  For example, although our 1$\times$ solar, C/O = 0.88, disequilibrium 
model has a CO$_2$ abundance that falls outside the \citet{madhu09} $\xi ^2$ $<$ 1 constraint, 
the model spectrum has a band-averaged 4.5-$\mu$m flux that falls within 1-sigma of the observed 
flux, and solar metallicities are still plausible for HD 189733b, at least as far as the {\it 
Spitzer\/} secondary-eclipse data are concerned.  In fact, Fig.~\ref{boxratios} demonstrates 
that there is not much overlap in the solutions that are consistent with the \citet{madhu09} 
CO$_2$ and $\HtwoO$ $\xi ^2$ $<$ 1 constraints in terms of metallicity vs.~C/O ratio phase space, 
and we consider the $\HtwoO$ abundance to be the more reliable constraint, as water has a larger 
effect on the spectrum.

The bottom line from our modeling is that although the {\it Spitzer}/IRAC secondary-eclipse 
observations indicate that the C/O ratio is likely less than unity on 
HD 189733b \citep[see][]{madhu09}, atmospheric models with C/O ratios moderately enhanced over 
the solar value of $\sim$0.55 provide better fits to the {\it Spitzer\/} IRAC and MIPS 
secondary-eclipse data than models with solar-like C/O ratios, given our assumed nominal thermal 
profile, unless the atmosphere has a subsolar metallicity.  More specifically, the flux in 
the longer-wavelength {\it Spitzer\/} bandpasses is very sensitive to the water abundance, and 
the relatively low H$_2$O abundance that \citet{madhu09} infer from the secondary-eclipse 
observations implies low metallicities, high C/O ratios, and/or high dayside temperatures 
(i.e., very inefficient heat redistribution) on HD 189733b \citep[see 
also][]{barman08,burrows08}.  

Disequilibrium processes like photochemistry and transport-induced 
quenching do not change this conclusion because water is efficiently recycled in the middle 
atmosphere through photochemical and kinetics processes (see \citealt{moses11} for more details), 
although disequilibrium chemistry is found to affect the profiles of other spectrally important 
molecules such as CH$_4$, NH$_3$, and HCN (see Fig.~\ref{hd189pchem} and \citealt{line10}, 
\citealt{moses11}).  We reproduce {\it Spitzer\/} secondary-eclipse photometric data well for 
models with metallicities between 1-5$\times$ solar and C/O ratios between 0.88-1.0.  It should 
be kept in mind, however, that our fits are non-unique.  The C/O ratio that provides the best 
fit to the observations will depend on both the metallicity and thermal structure, with higher 
metallicities and/or lower temperatures requiring higher C/O ratios to remain consistent with 
the data.  

\begin{figure*}
\begin{tabular}{ll}
{\includegraphics[angle=0,clip=t,scale=0.54]{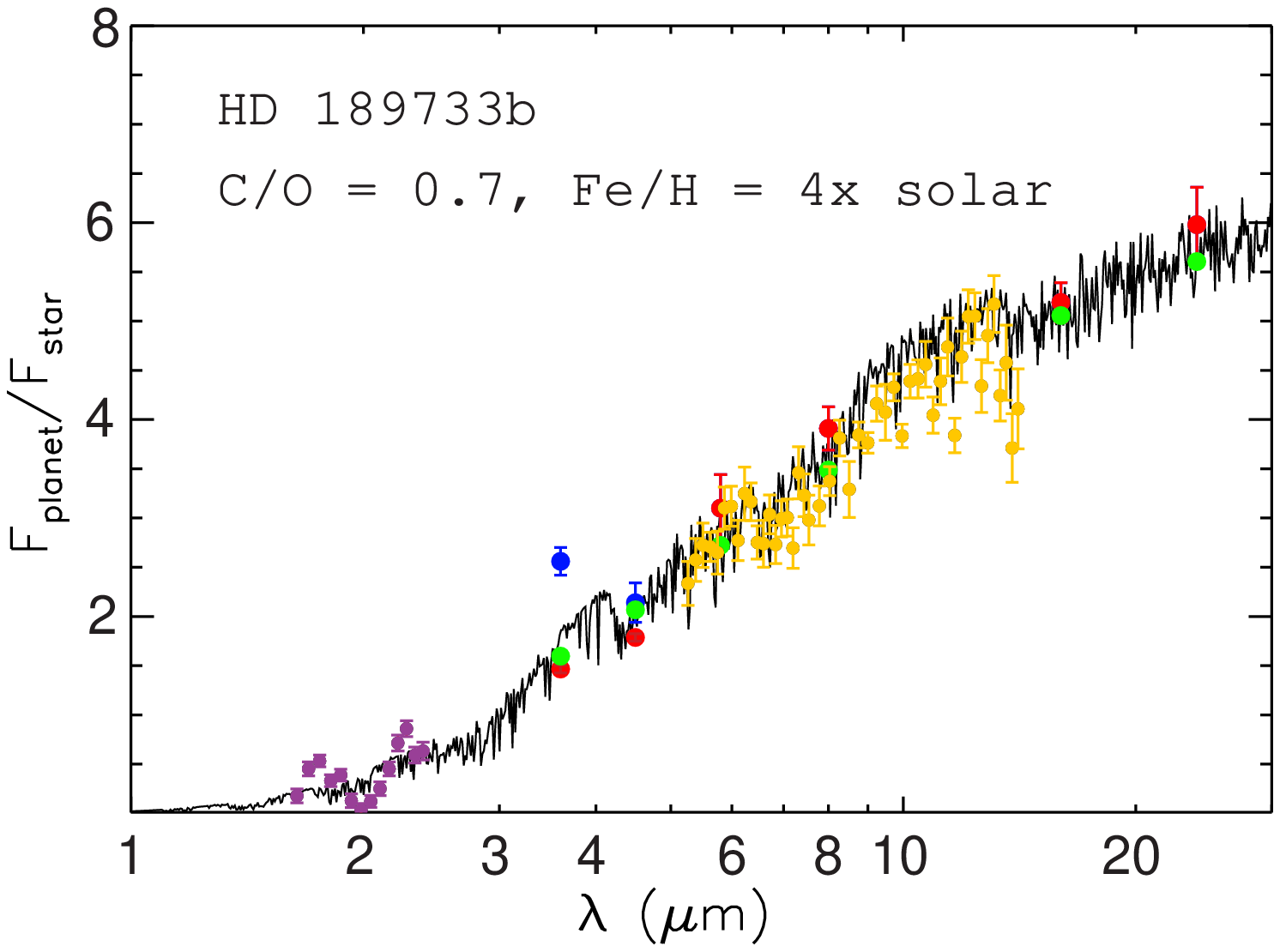}}
&
{\includegraphics[angle=0,clip=t,scale=0.54]{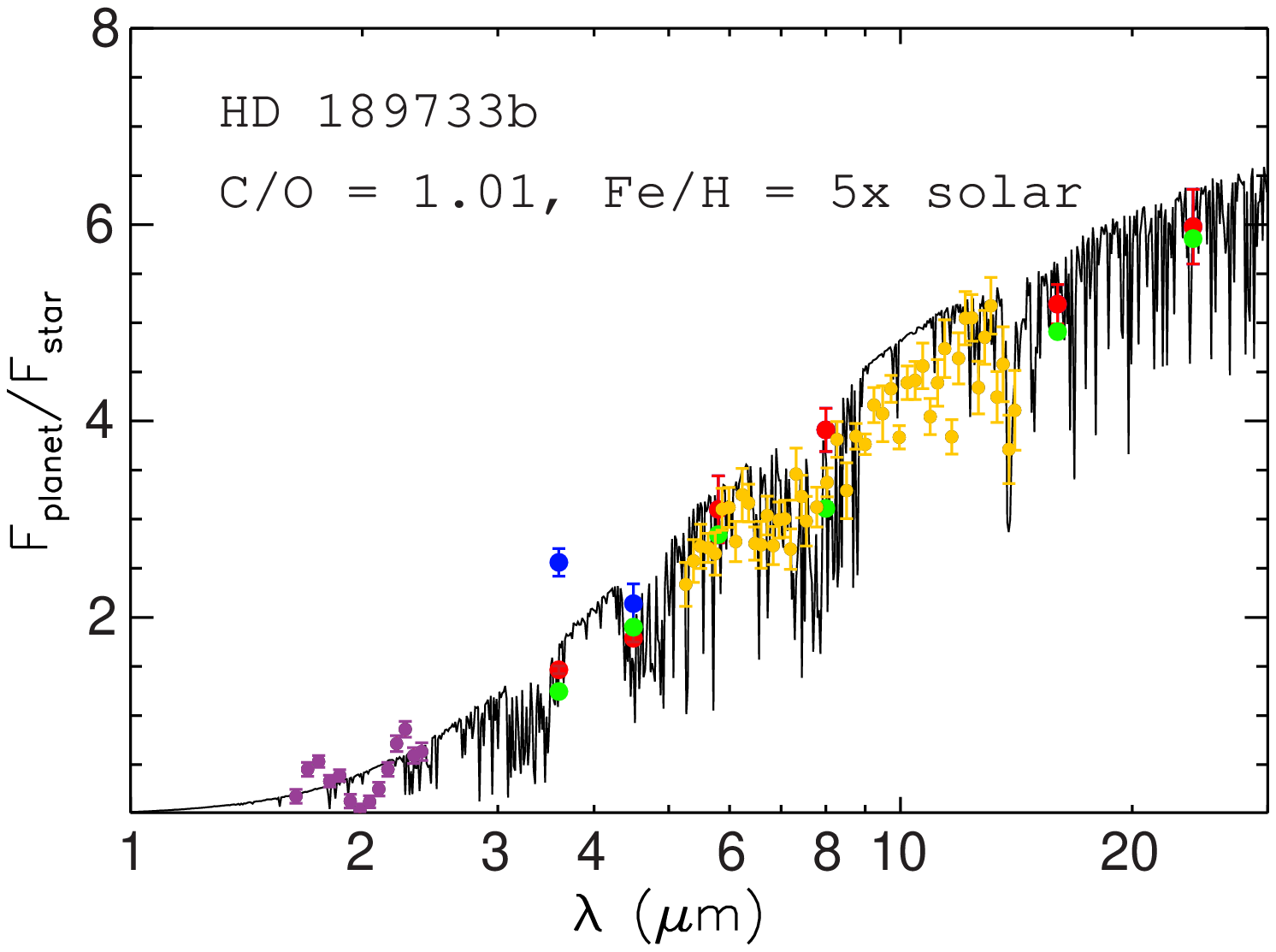}}
\\
\end{tabular}
\caption{Secondary-eclipse spectra for HD 189733b (in terms of the ratio of the flux 
of the planet to that of the host star) for models with (Left) a C/O ratio of 0.7 and 
a metallicity of 4 times solar, and (Right) a C/O ratio of 1.01 and a metallicity of 
5 times solar.  The solid black lines represent synthetic spectra from the results of 
our disequilibrium chemistry models.  The red circles with error bars represent the 
3.6 and 4.5 $\mu$m {\it Spitzer\/} photometric data from \citet{knut12} and the 5.8, 8, 
16, and 24 $\mu$m data from \citet{knut07,knut09hd189} and \citet{charb08}; 
the blue circles with error bars represent the older data in the 3.6 and 4.5 $\mu$m IRAC 
channels.  The purple circles with error bars represent the HST/NICMOS spectra
of \citet{swain09hd189}, and the yellow circles with error bars represent the {\it Spitzer\/} 
IRS spectra of \citet{grill08}.  The green circles without error bars represent our model 
results convolved over the {\it Spitzer\/} instrument response functions.  Note that the 
thermal contrast between the troposphere and stratosphere was reduced in our new models to 
better fit the new IRAC 3.6 and 4.5 $\mu$m points, and the resulting synthetic spectra compare 
better with the IRS spectra than the older models shown in Fig.~\ref{spechd189}.  A color version 
of this figure is available in the online journal.\label{figproof}}
\end{figure*}

\subsubsection{HD 189733b Recent Updates\label{hd189recent}}

The most recent HD 189733b {\it Spitzer\/} IRAC secondary 
eclispse data of \citet{knut12} imply significantly reduced fluxes in the 3.6 and 4.5 $\mu$m 
channels (see Fig.~\ref{figproof}).  With this new data, the derived $F_{planet}/F_{star}$ at 
3.6-$\mu$m no longer exceeds that at 4.5-$\mu$m, and in fact the entire spectrum more closely 
resembles a blackbody.  If the scattering hazes that are observed at ultraviolet and visible 
wavelengths \citep[e.g.,][]{sing11hd189,huitson12} are vertically optically thick at infrared 
wavelengths, then the spectrum at secondary eclipse might indeed be expected to resemble a 
blackbody, but a similar result could be obtained with optically thin hazes and less thermal 
contrast between the stratosphere and troposphere, or from a low-metallicity 
planet.  We have created new models to better fit the new \citet{knut12} 3.6 and 4.5 $\mu$m photometric 
points, along with the older 5.8, 8, 16, and 24 $\mu$m data from \citet{knut07,knut09hd189} and 
\citet{charb08}.  Some new results are shown in Fig.~\ref{figproof}.  When we consider models with a reduced 
$\Delta T$ between the troposphere and stratosphere, the predicted model excess-flux ``bulges'' at 
3.5-4 $\mu$m and 9-12 $\mu$m that were apparent in the previous models shown in Fig.~\ref{spechd189} 
are correspondingly reduced.  The new models then compare better 
with the IRS spectra of \citet{grill08} and have the appropriate relative fluxes in the 3.6 and 
4.5 $\mu$m IRAC channels, but the predicted amplitude in the near-IR absorption bands 
is smaller than is indicated by the HST/NICMOS data of \citet{swain09hd189}.  As with our previous 
results, we find that models with moderately high C/O ratios result in the low water abundances that 
are needed to provide better fits to the data than models with solar-like values of C/O $\sim$ 0.5; 
on the other hand, C/O ratios $\gta$ 1 provide too much absorption from methane and HCN at 3.6 and 
8 $\mu$m to be consistent with the IRAC data.  Our favored models have C/O ratios in the range 
$\sim$0.6-1.0.  In all tests, our models provide insufficient absorption at 4.5 $\mu$m to account for 
the \cite{knut12} flux in this channel, considering the exceedingly small statistical 
error bars cited for the 4.5 $\mu$m point.  \citet{knut12} had similiar difficulty with model fits 
at 4.5 $\mu$m.  Increasing the metallicity, with its corresponding increase in the CO$_2$ abundance, 
does not help the situation, as it is still difficult to fit the 4.5-$\mu$m point while keeping good 
fits at other wavelengths.  On the other hand, 
the potential systematic errors involved with the IRAC analyses, the observed 
variability of the star, and the uncertainties in the stellar models needed for calculating 
$F_{planet}/F_{star}$ for the synthetic spectra all come into play here, and fits such as are shown 
in Fig.~\ref{figproof} may be reasonable when all systematic observational and modeling uncertainties 
are considered.  In any case, spectral observations by the {\it James\ Webb\ Space\ Telescope\/} or 
dedicated missions like FINESSE and EChO \citep{swain12finesse,tinetti12} could help resolve the 
atmospheric composition and better constrain the C/O ratio on HD 189733b and other extrasolar panets.  

\subsection{XO-1b Results\label{xo1bsect}}

The transiting planet XO-1b was discovered by \citet{mccullough06}.
XO-1b, like HD 189733b, is a moderately-irradiated, moderately-hot transiting planet that is 
warm enough that CO should be the dominant carbon-bearing constituent in the middle atmosphere 
of the planet.  Whether this exoplanet falls into the O1 or C2 regime of \citet{madhu12coratio} 
depends on the C/O ratio.  \citet{madhu12coratio} suggests that the fit to the {\it 
Spitzer}/IRAC secondary-eclipse data of \citet{machalek08} is improved for models with C/O 
$\ge$ 1, although \citet{tinetti10} are able to find good fits to the same data (as well 
as to transit data) for more standard assumptions about the C/O ratio, depending on the 
CH$_4$ line parameters that are adopted.  The relative fluxes in the 3.6 vs.~4.5-$\mu$m 
channels and 5.8 vs.~8.0-$\mu$m channels led \citet{machalek08} to conclude that XO-1b has a 
thermal inversion, although their model fit to the data is not very good; \citet{madhu12coratio} 
demonstrates an improved fit for models with no thermal inversion and a C/O ratio $\ge$ 1.  
\citet{tinetti10} find a good fit to the secondary-eclipse data for models with and without 
thermal inversions, for various assumptions about species abundances.  For the case of no 
thermal inversion, the very high flux at 5.8 $\mu$m implies a very low water abundance and/or 
very high photospheric temperatures \citep{tinetti10,madhu12coratio}.  Degeneracies between the 
thermal structure and composition are clearly plaguing the interpretation of the XO-1b data 
\citep[e.g.,][]{tinetti10,madhu12coratio}, and we are unlikely to find a definitive solution 
to the problem.  We can, however, test whether the \citet{madhu12coratio} high-C/O-ratio scenarios 
for XO-1b can fit the secondary-eclipse data for chemically plausible species abundances that 
take into account disequilibrium processes.

\begin{figure*}
\begin{tabular}{ll}
{\includegraphics[angle=0,clip=t,scale=0.37]{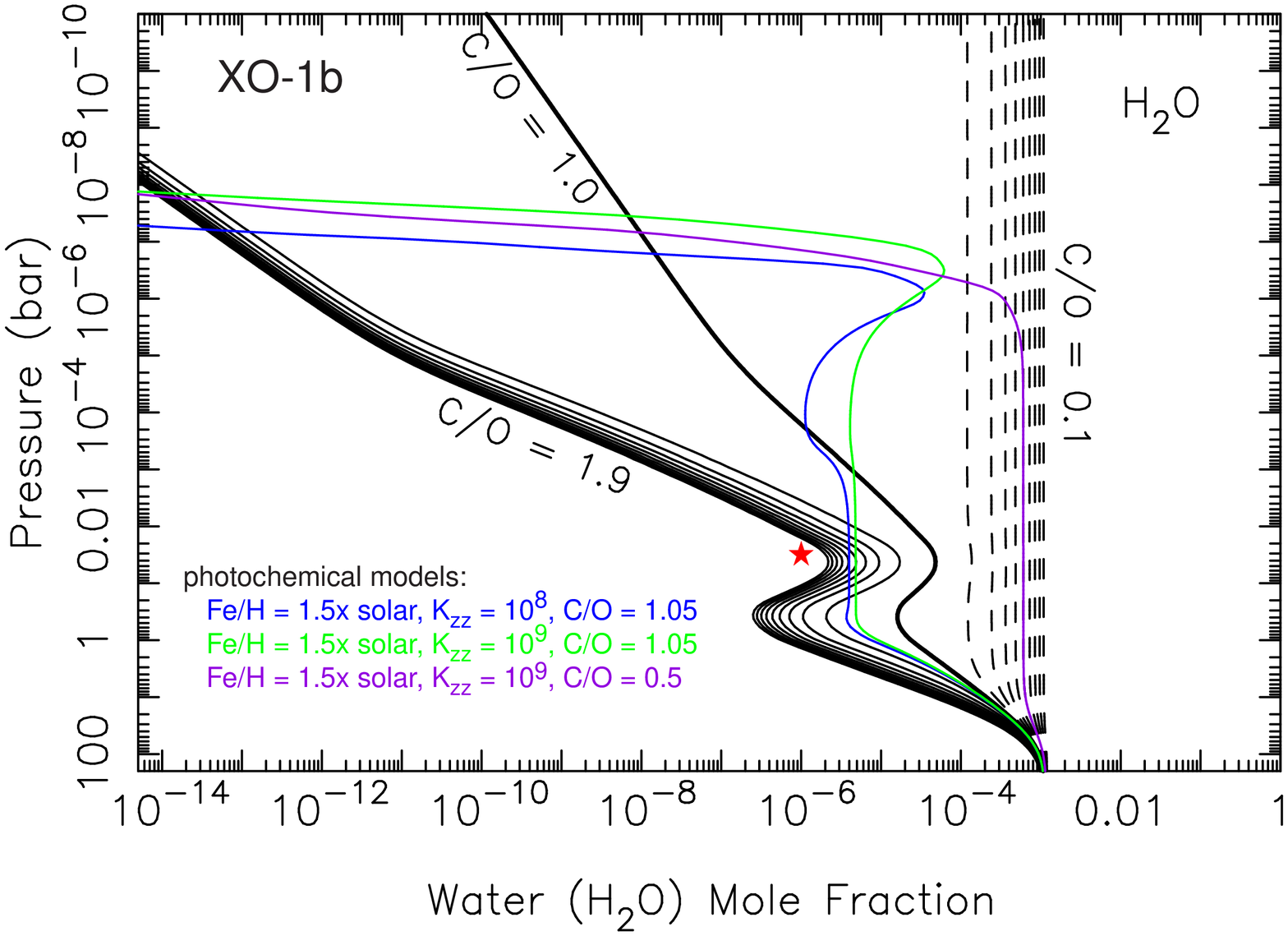}}
&
{\includegraphics[angle=0,clip=t,scale=0.37]{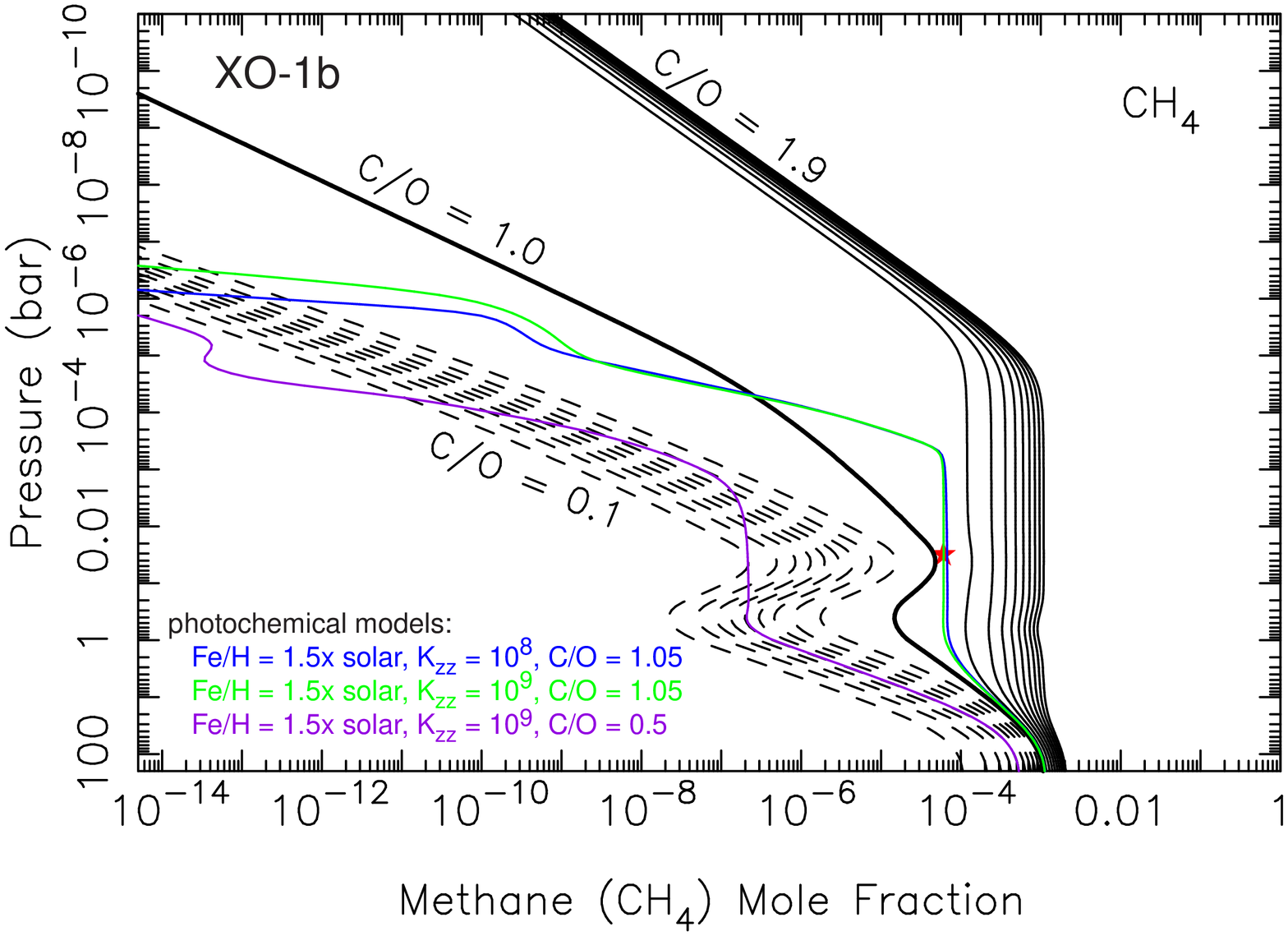}}
\\
{\includegraphics[angle=0,clip=t,scale=0.37]{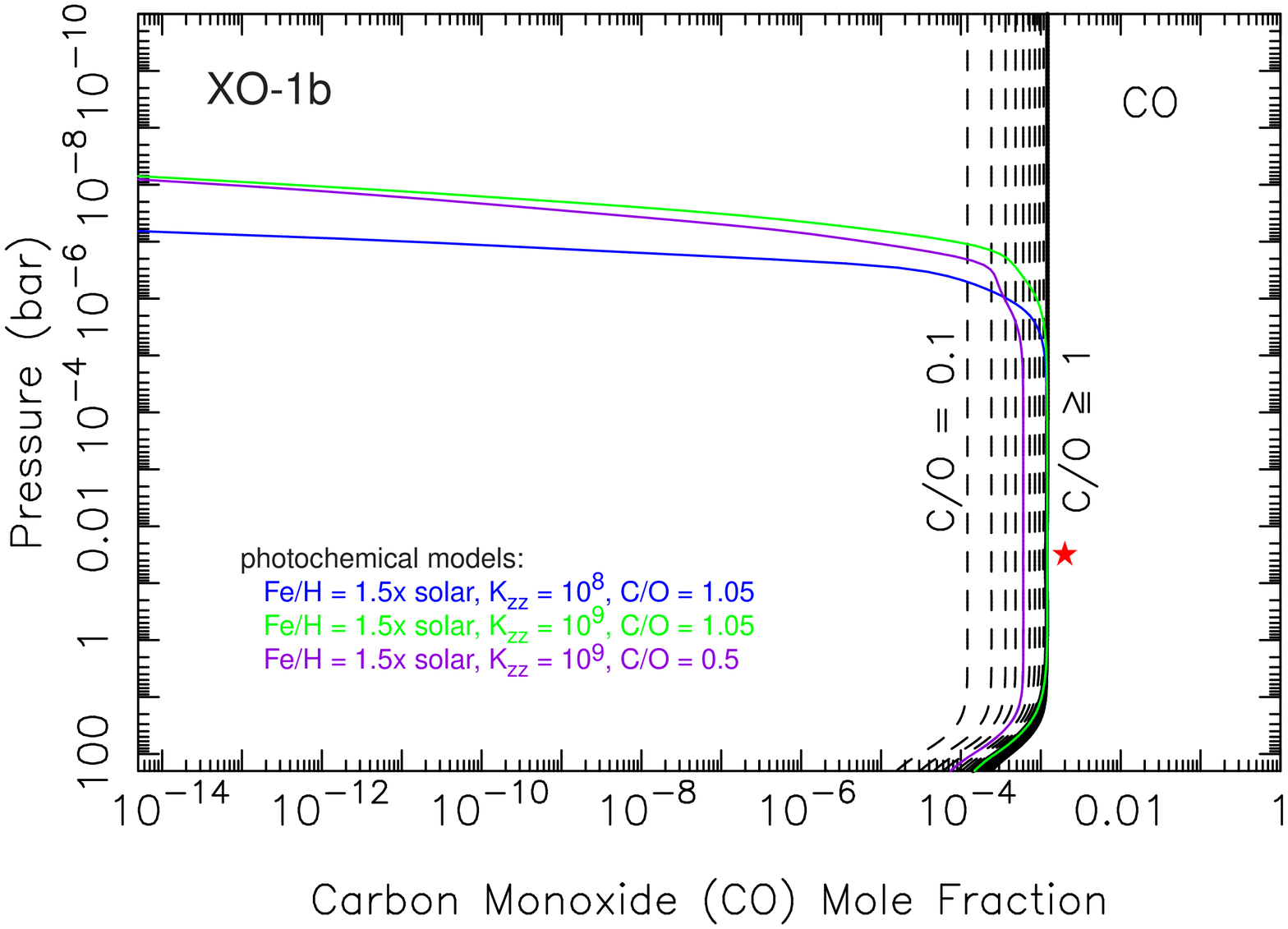}}
&
{\includegraphics[angle=0,clip=t,scale=0.37]{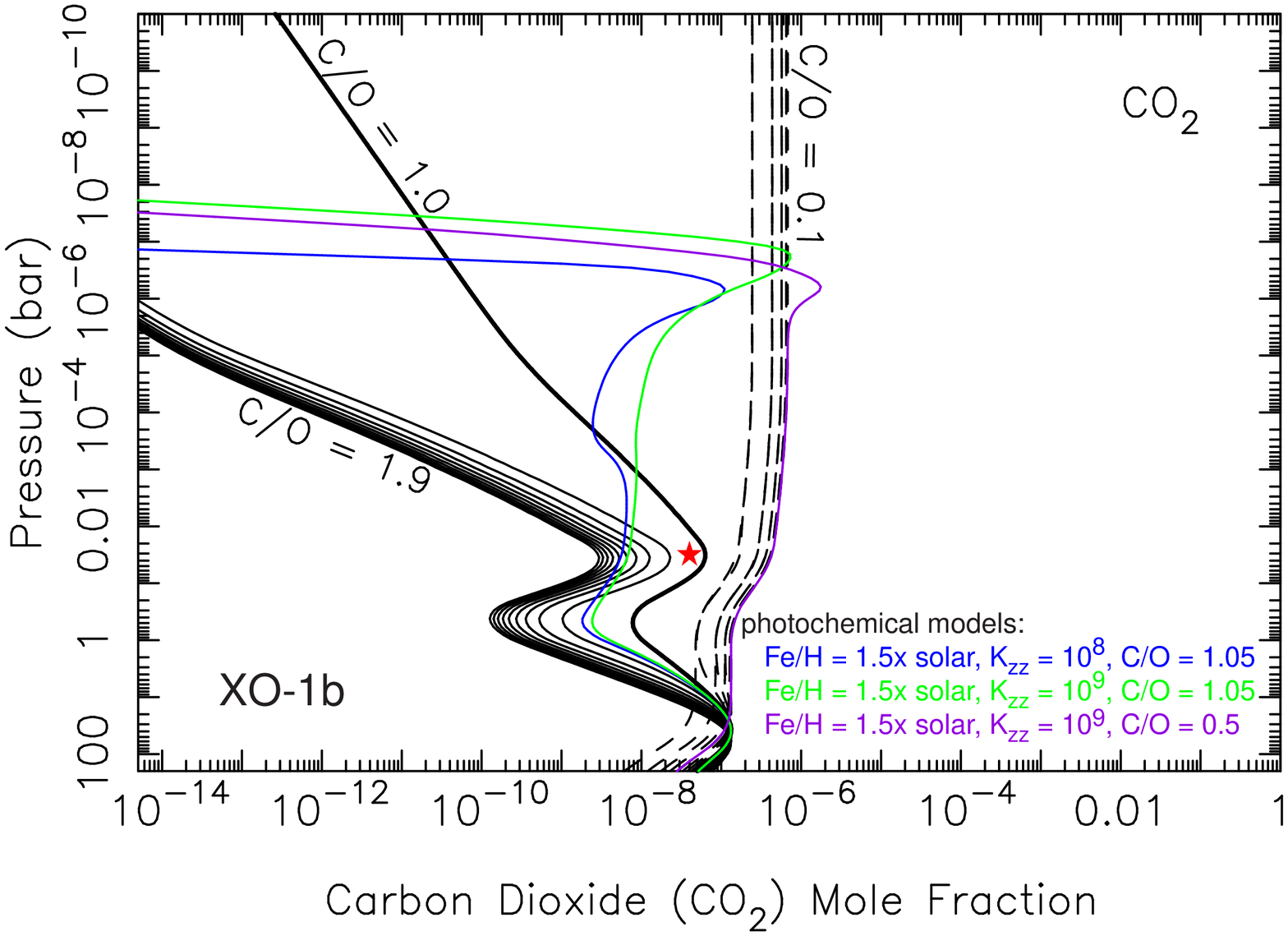}}
\\
{\includegraphics[angle=0,clip=t,scale=0.37]{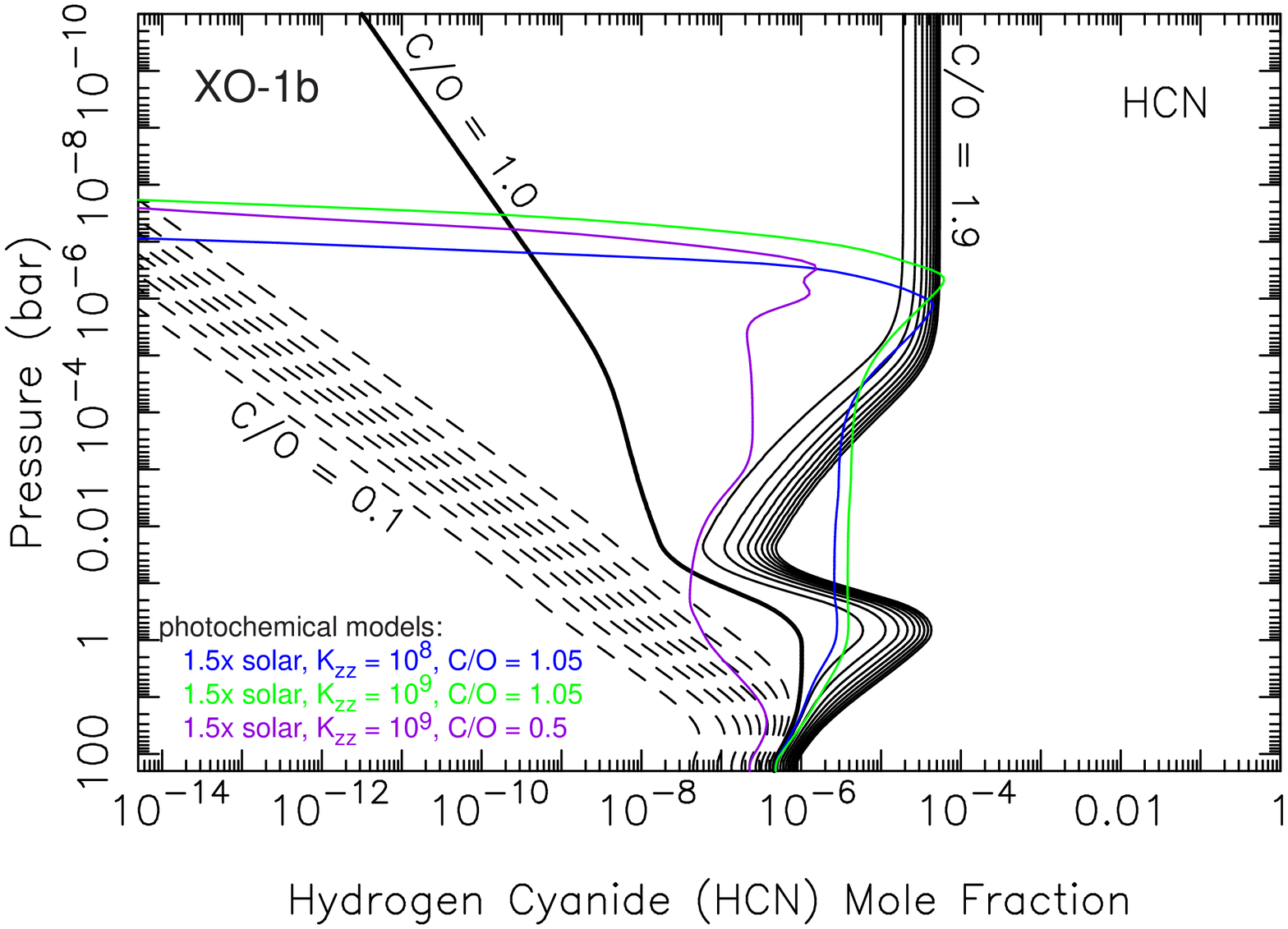}}
&
{\includegraphics[angle=0,clip=t,scale=0.37]{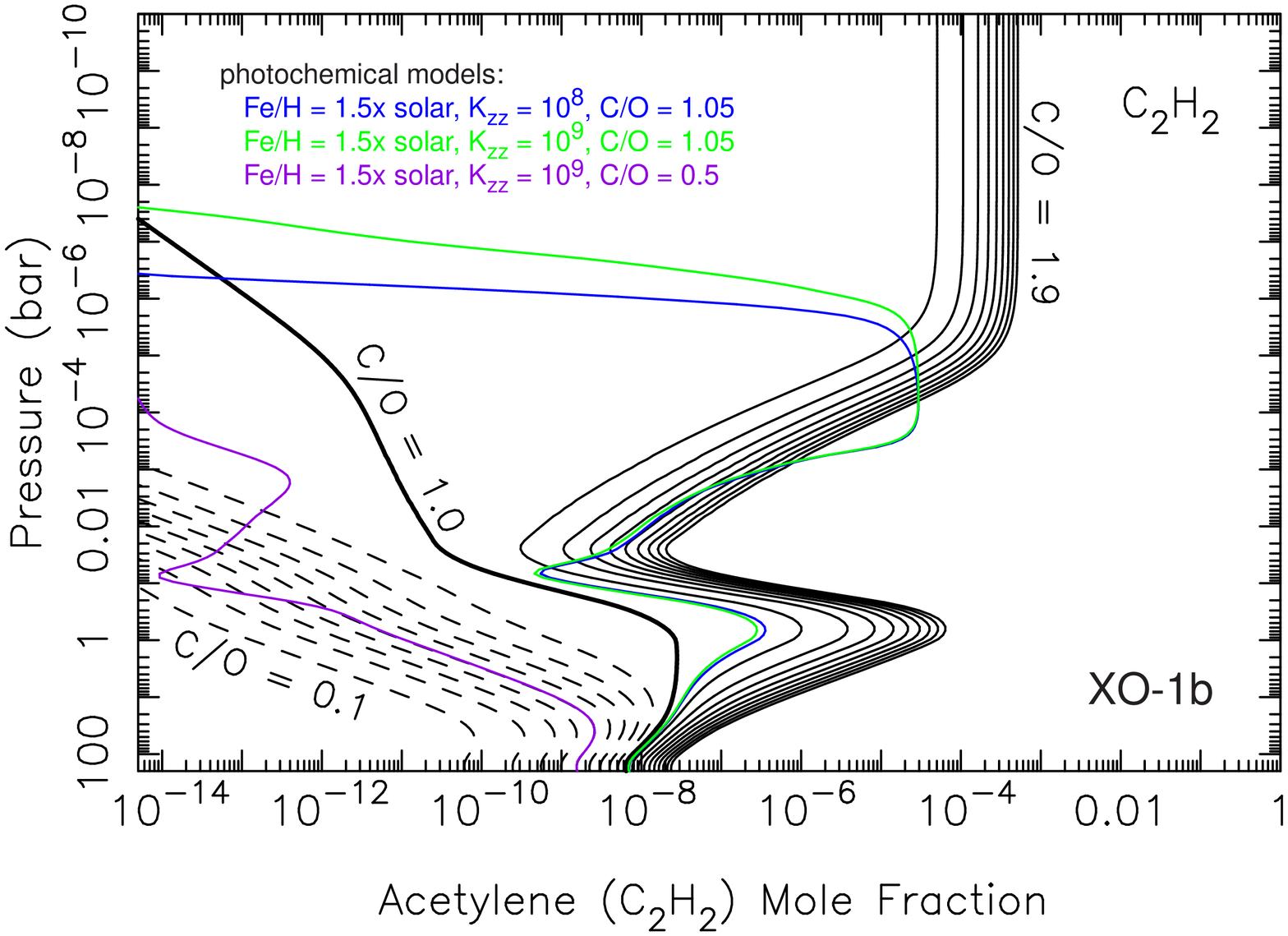}}
\\
\end{tabular}
\caption{Mixing-ratio profiles for H$_2$O, CH$_4$, CO, CO$_2$, HCN, and C$_2$H$_2$ on XO-1b 
from our thermochemical-equilibrium models with assumed 1.5$\times$ solar metallicity but different 
assumed C/O ratios (ranging from 0.1 to 1.9, incrementing by 0.1 --- dashed black lines for 
C/O $<$ 1, solid lines for C/O $\ge$ 1).  The colored lines represent disequilibrium-chemistry 
results for models with a solar-like C/O ratio of 0.5 and $K_{zz}$ = 10$^9$ $\cmtwo$ $\smone$ 
(purple), a model with a C/O ratio of 1.05 and $K_{zz}$ = 10$^9$ $\cmtwo$ $\smone$ (green), 
and a model with a C/O ratio of 1.05 and $K_{zz}$ = 10$^8$ $\cmtwo$ $\smone$ (blue).  All 
models have an assumed atmospheric metallicity of 1.5$\times$ solar.  The red stars represent 
abundance constraints from a model presented in \citet{madhu12coratio} based on the 
{\it Spitzer}/IRAC secondary-eclipse observations of \citet{machalek08}; HCN and C$_2$H$_2$ were 
not considered in that analysis.  A color version of this figure is presented in the online 
journal.\label{xo1bequil}}
\end{figure*}

\subsubsection{XO-1b Chemistry \label{xo1bchemsect}}

Figure~\ref{xo1bequil} illustrates how the thermochemical equili{\-}bri{\-}um composition of 
XO-1b varies as a function of the atmospheric C/O ratio for our assumed nominal 
dayside thermal structure (see Fig.~\ref{figtemp}) and an assumed 1.5$\times$ solar 
metallicity.  As with HD 189733b, CO is abundant for all assumptions of the C/O ratio, 
whereas H$_2$O and CO$_2$ are much more abundant for C/O $<$ 1, and conversely CH$_4$, HCN, and 
$\CtwoHtwo$ are much more abundant for C/O $>$ 1.  The spectral analysis of \citet{madhu12coratio} 
suggests that the C/O ratio on XO-1b could be greater than that on HD 189733b, perhaps even 
C/O $>$ 1.  
For a model with a thermal structure similar to our adopted nominal profile 
shown in Fig.~\ref{figtemp}, \citet{madhu12coratio} finds good fits to the \citet{machalek08} 
{\it Spitzer\/} IRAC secondary-eclipse spectra for the species mole fractions marked by 
a star in Fig.~\ref{xo1bequil}.  These constraints imply that CO is abundant, CO$_2$ is 
much less abundant, and CH$_4$ $\gg$ H$_2$O.  A C/O ratio near unity can satisfy all 
the these constraints for our adopted thermal profile, except the water abundance will be 
over-predicted for a C/O ratio $\sim$ 1.

\begin{figure}
\includegraphics[angle=-90,scale=0.37]{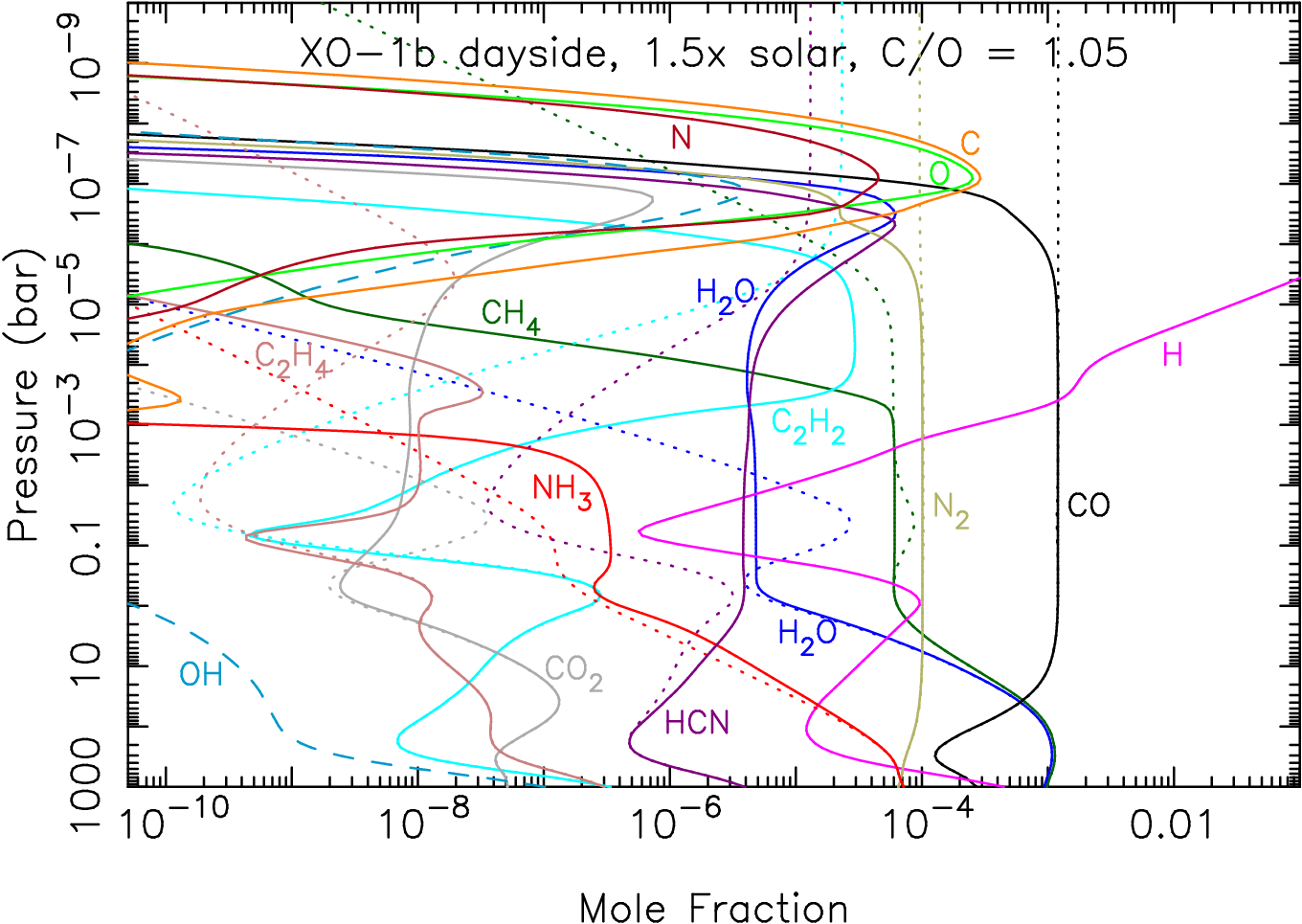}
\caption{Mole-fraction profiles for several important species (as labeled) in a 
thermochemical and photochemical kinetics and transport model for XO-1b.  Model assumptions 
include 1.5$\times$ solar metallicity, a C/O ratio of 1.05, and $K_{zz}$ = 10$^9$ $\cmtwo$ 
$\smone$ (see also the green curve in Fig.~\ref{xo1bequil}).  The dotted lines show the 
thermochemical-equilibrium abundances for the species, using the same color coding.  
The overall messiness of this figure indicates that disequilibrium chemistry is important 
on XO-1b.  A color version of this figure is available in the online journal.\label{xo1bpchem}}
\end{figure}

Disequilibrium processes do not affect this overall conclusion much, although the vertical 
profiles of the spectrally active molecules CH$_4$, H$_2$O, HCN, NH$_3$, and CO$_2$ are 
significantly altered by transport-induced quenching and photochemistry.  The results of three 
of our thermo/photochemical kinetics and diffusion models are shown in Fig.~\ref{xo1bequil}. 
Further results from one of these disequilibrium models are shown in Fig.~\ref{xo1bpchem}.   
The main reaction schemes controlling the abundances in the C/O = 0.5 model are the same as 
on HD 189733b; a full discussion can be found in \citet{moses11} and \citet{line10}. 
The disequilibrium processes affecting the molecules do change as C/O exceeds unity, 
although the main reaction schemes controlling the behavior typically remain the same.  The 
main differences pertain to the relative significance of quenching in controlling the 
mole fractions of different species in the middle and lower stratosphere, and the different 
relative abundances of the species at those quench points, which can lead to different ``parent'' 
molecules dominating the photochemistry in the upper atmosphere.  

For our favored model with 1.5$\times$ solar metallicity, C/O = 1.05, and $K_{zz}$ = 10$^9$ $\cmtwo$ 
$\smone$ (Fig.~\ref{xo1bpchem}), the only molecules relatively unaffected by disequilibrium 
processes (except at very high altitudes) are CO and N$_2$.  Compared with equilibrium, 
photochemistry strongly depletes CH$_4$ above $\sim$0.4 mbar and NH$_3$ above $\sim$1 mbar, 
although the NH$_3$ mole fraction in the 1-10$^{-3}$ bar region is enhanced by transport-induced 
quenching.  Scheme (2) of \citet{moses11} still dominates the CH$_4$ $\rightarrow$ CO quenching, 
but the very high equilibrium CH$_4$ abundance at the quench point allows methane to be a more 
important parent molecule for the subsequent photochemistry, with $\CtwoHtwo$ production in 
particular being enhanced in the C/O = 1.05 model as compared with the C/O = 0.5 model.  The 
large H abundance in the upper atmosphere drives the conversion of $\CHfour$ into $\CtwoHtwo$ 
through schemes such as
\begin{eqnarray}
2\, ( \H \, + \, \CHfour \, & \rightarrow & \, \CHthree \, + \, \Htwo \nonumber ) \\
2\, \CHthree \, + \, \M \, & \rightarrow & \, \CtwoHsix \, + \, \M \nonumber \\
\H \, + \, \CtwoHsix \, & \rightarrow & \, \CtwoHfive \, + \, \Htwo \nonumber \\
\CtwoHfive \, + \, \M \, & \rightarrow & \, \H \, + \, \CtwoHfour \, + \, \M \nonumber \\
\H \, + \, \CtwoHfour \, & \rightarrow & \, \CtwoHthree \, + \, \Htwo \nonumber \\
\H \, + \, \CtwoHthree \, & \rightarrow & \, \CtwoHtwo \, + \, \Htwo  \nonumber \\
\noalign{\vglue -10pt}
\multispan3\hrulefill \nonumber \cr
\Net \ \ 2\, \CHfour \, + \, 4\, \H \, & \rightarrow & \, \CtwoHtwo \, + \, 5\,\Htwo  , \\
\end{eqnarray}
where M represents any third body.  The key to this mechanism operating in the $\CHfour$ $\rightarrow$ 
$\CtwoHtwo$ direction is the reaction pair H + $\CHfour$ $\lrarrow$ $\CHthree$ + $\Htwo$, which can 
stay balanced in the lower stratosphere but operates with a net imbalance to the right in the upper 
stratosphere due to the large abundance of H atoms from $\HtwoO$ photolysis and subsequent catalytic 
destruction of $\Htwo$ \citep[see][]{moses11}.

The greater abundance of $\CHfour$ in the C/O = 1.05 model also leads to significantly enhanced 
disequilibrium production of HCN in the $\sim$0.1-to-10$^{-4}$ bar region through scheme (7) of 
\citet{moses11}.  At pressures less than 10$^{-4}$ bar, the HCN is derived from CO and $\Ntwo$ 
through schemes such as 
\begin{eqnarray}
\CO \, + \, h\nu \, & \rightarrow & \, \C \, + \, \O \nonumber  \\
\Ntwo \, + \, h\nu \, & \rightarrow & \, 2\, \N \nonumber  \\
\O \, + \, \Htwo \, & \rightarrow & \, \OH \, + \, \H \nonumber \\
\N \, + \, \OH \, & \rightarrow & \, \NO \, + \, \H \nonumber \\
\C \, + \, \NO \, & \rightarrow & \, \CN \, + \, \O \nonumber \\
\CN \, + \, \Htwo \, & \rightarrow & \, \HCN \, + \, \H \nonumber \\
\O \, + \, \Htwo \, & \rightarrow & \, \OH \, + \, \H  \nonumber \\
\OH \, + \, \Htwo \, & \rightarrow & \, \HtwoO \, + \, \H  \nonumber \\
2\, ( 2\, \H \, + \M \, & \rightarrow & \, \Htwo \, + \, \M \, ) \nonumber  \\
\noalign{\vglue -10pt}
\multispan3\hrulefill \nonumber \cr
\Net \ \ \CO \, + \, \Ntwo \, + \, 2\, \Htwo \, & \rightarrow & \, \HCN \, + \, \HtwoO \, + \, \N  \, + \, \H  , \\
\end{eqnarray}
where $h\nu$ represents an ultraviolet photon.

Note that photolysis of CO in schemes like the one above leads to the photochemical production of 
$\HtwoO$ at high altitudes, despite the low predicted equilibrium abundance of $\HtwoO$ in this 
region of the atmosphere in the C/O = 1.05 model. In fact, HCN and $\HtwoO$ have similar mole 
fractions at high altitudes in the C/O = 1.05 model because the O from CO photolysis ends up 
largely as $\HtwoO$, whereas the C ends up in HCN.  Both $\HtwoO$ and HCN are recycled efficiently 
at high altitudes (see \citealt{moses11}).  Transport-induced quenching is an important 
disequilibrium process 
for water in the C/O = 1.05 model, acting to smooth out the mole-fraction profile such that the 
H$_2$O abundance drops below equilibrium values near 0.1-0.01 bar but exceeds equilibrium values 
above $\sim$10$^{-2}$ mbar.  Quenching of water occurs through the same overall reaction scheme 
that affects the quenching of CO and $\CHfour$ (e.g., scheme (2) in \citealt{moses11}).  This 
scheme also operates in the C/O $<$ 1 models, but the quenching of water is more obvious in the 
C/O $>$ 1 models due to the lower equilibrium $\HtwoO$ abundance.  Carbon dioxide is also affected 
by the quenching of CO and $\HtwoO$, and the increased photochemical production of $\HtwoO$ and OH 
at high altitudes leads to a significantly increased CO$_2$ abundance in the C/O = 1.05 model as 
compared with equilibrium, but the overall abundance of CO$_2$ remains much lower than that of water.

As with the equilibrium model, the disequilibrium model has more H$_2$O than the spectral analysis 
of \citet{madhu12coratio} would indicate, but the resulting abundances of CO and CH$_4$ are 
still in line with the \citet{madhu12coratio} constraints.  The disequilibrium models 
underpredict the CO$_2$ abundance on XO-1b, as they did for HD 189733b,  but again, high 
CO$_2$ abundances are not required to fit the thermal-infrared data at secondary eclipse 
(see below).  The disequilibrium models also indicate that hydrogen cyanide is abundant 
enough that HCN opacity should be included in spectral models.

\subsubsection{XO-1b Spectra \label{xo1bspecsect}}

\begin{figure}
\includegraphics[angle=0,scale=0.48]{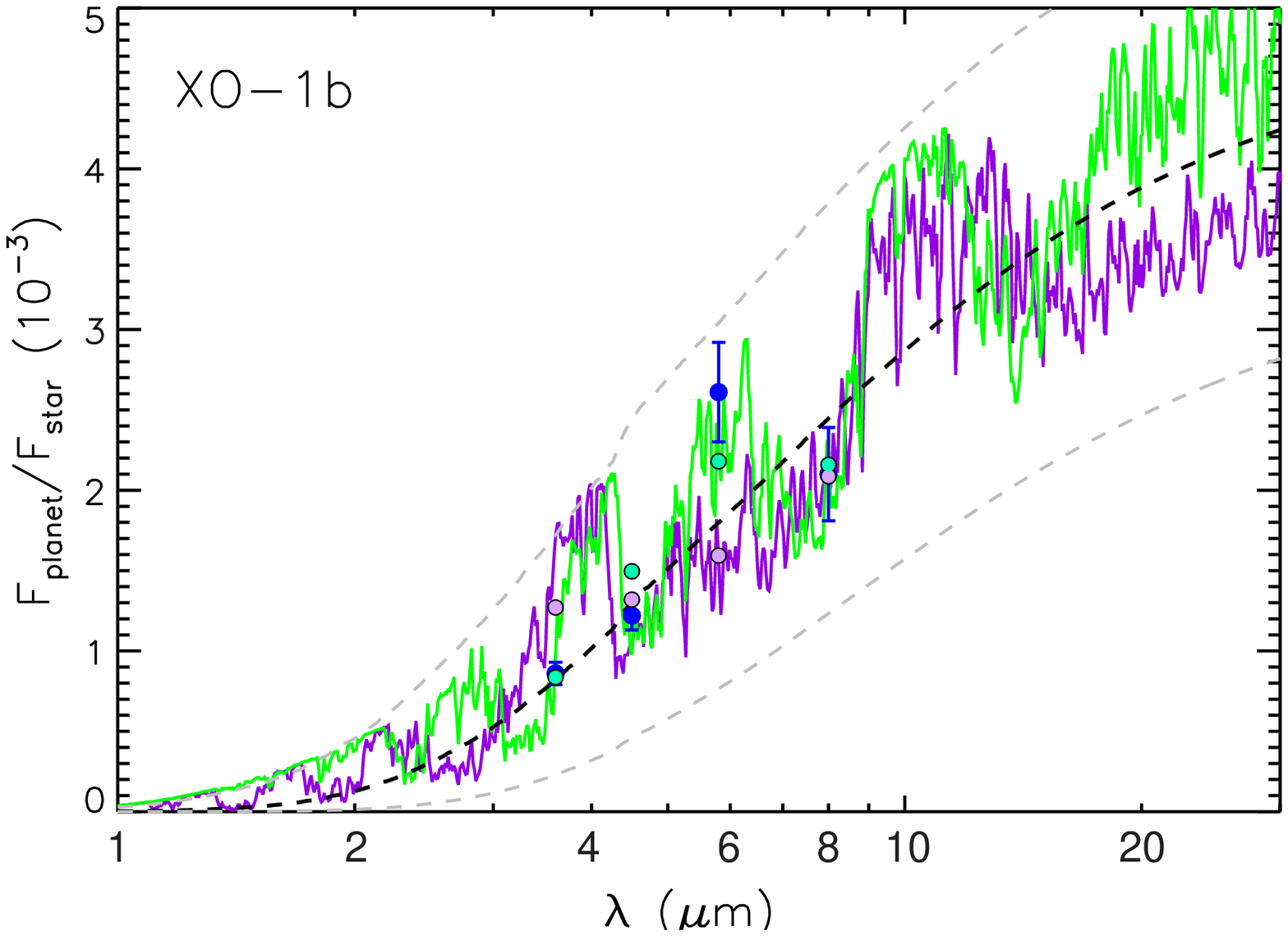}
\caption{Secondary-eclipse spectra for XO-1b.  The solid curves represent synthetic spectra 
from the results of our disequilibrium chemistry models with 1.5$\times$ solar metallicity, 
$K_{zz}$ = 10$^9$ $\cmtwo$ $\smone$, and C/O ratios of either 1.05 (green) or 0.5 (purple).
The blue circles with error bars represent the {\it Spitzer\/} IRAC observations of
\citet{machalek08}.  The circles without error bars represent the model results for 
C/O = 1.05 (green) or C/O = 0.5 (purple) convolved over the {\it Spitzer\/} bandpasses.  Note the 
much better fit to the data with the C/O = 1.05 model for this particular thermal structure and 
metallicity.  The dashed lines are blackbody curves for assumed temperatures of 900 K (bottom), 
1250 K (middle), and 1600 K (top).  A color version of this figure is available in the online 
journal.\label{xo1bspec}}
\end{figure}

In Fig.~\ref{xo1bspec}, we compare a synthetic spectrum from our nominal disequilibrium model with 
the XO-1b {\it Spitzer\/} data of \citet{machalek08}.  The photometric data indicate 
a flux at 4.5 $\mu$m that is larger than that at 3.6 $\mu$m and a flux at 5.8 $\mu$m that 
is larger than that at 8.0 $\mu$m.  In the absence of a thermal inversion, such flux ratios 
suggest that CH$_4$ (and/or HCN) are more abundant than $\HtwoO$, which is why \citet{madhu12coratio}  
suggested XO-1b could be carbon-rich in the first place.  Our spectral fit for this model with 
C/O = 1.05 is within 1.7-sigma of the data for all the {\it Spitzer\/} bands, with the biggest 
problem being an overestimation of the absorption at 5.8 $\mu$m.  Going to a higher C/O ratio, 
with its correspondingly lower water abundance, would improve the fit in this wavelength band, 
but the resulting larger CH$_4$ and HCN abundances would degrade the fit at 3.6 and 8.0 $\mu$m.
Our spectral comparison here demonstrates that carbon-rich models with plausible species 
abundances from disequilibrium-chemistry predictions can be consistent with the {\it Spitzer\/} 
secondary-eclipse data, validating the suggestion of \citet{madhu12coratio}.  

Given that XO-1b is cool enough that TiO is not expected to be present in significant quantities 
in the stratosphere, even for C/O ratios $<$ 1 (and thus TiO would not be available to produce a 
stratospheric thermal inversion; \citealt{hubeny03}, \citealt{fort06hd149,fort08unified}), the 
high C/O-ratio, non-inverted scenario for XO-1b has some theoretical advantages.  However, we 
must again emphasize that our models are not unique.  The results will be sensitive to the 
assumed metallicity and thermal structure, for example.  We note that the relative 
3.6-$\mu$m/4.5-$\mu$m and 5.8-$\mu$m/8.0-$\mu$m flux ratios from 
the XO-1b secondary-eclipse data cannot be explained with solar-like C/O ratios unless 
temperatures at $\sim$0.1-1-bar are much hotter ($\gta$2000 K) than is suggested by 
theoretical predictions or unless the planet has a stratospheric thermal inversion 
\citep[cf.][]{machalek08,tinetti10}.  In fact, the observed 5.8-$\mu$m/8.0-$\mu$m flux ratio 
is difficult to explain with any model --- whether thermally inverted, anomalously hot 
at depth, or possessing a high C/O ratio (see Fig.~\ref{xo1bspec}, \citealt{machalek08}, 
\citealt{tinetti10}, \citealt{madhu12coratio}) --- and additional high-precision photometric 
or spectral data from ground-based or space-based observations would be highly desirable in 
helping constrain chemical and physical properties of this unusual and intriguing exoplanet. 

\subsection{WASP-12b Results\label{waspsect}}

The transiting planet WASP-12b, discovered by \citet{hebb09}, is much hotter than either
HD 189733b or XO-1b (see Fig.~\ref{figtemp}).  As one of the most highly irradiated hot 
Jupiters discovered to date, WASP-12b would fall into the upper part of either 
the O2 or C2 regimes of the \citet{madhu12coratio} hot-Jupiter classification scheme, depending 
on the C/O ratio.   WASP-12b is hot enough that titanium would not be cold-trapped into 
condensed calcium-titanium oxides \citep[e.g.,][]{lodders02ti,lodders10} or other condensed 
phases.  Thus, for C/O ratios less than unity, TiO vapor would be a major 
titanium-bearing phase that would be expected to survive into the stratosphere of WASP-12b, 
where it could absorb at optical wavelengths, heat the atmosphere, and create a 
stratospheric thermal inversion \citep[e.g.,][]{hubeny03,fort06hd149,fort08unified,burrows08,spiegel09} 
such as has been suggested for HD 209458b, HAT-P-7b, and several other transiting planets based on 
{\it Spitzer\/} observations
\citep[e.g.,][]{burrows07,burrows08,knut08,knut09tres4,swain09hd209,madhu09,madhu10inv,spiegel10hatp7,christiansen10}.

However, \citet{madhu11wasp12b} conclude that a strong stratospheric thermal inversion in the 
0.01-1 bar region of WASP-12b can be statistically ruled out due to a poor fit to the 
secondary-eclipse {\it Spitzer}/IRAC data of \citet{campo11} and ground-based data of 
\citet{croll11wasp12b}.  Instead, \citet{madhu11wasp12b} suggest that the atmosphere of 
WASP-12b is enriched in carbon relative to solar elemental abundances, with WASP-12b spectra 
exhibiting strong absorption due to CO and CH$_4$, and much weaker absorption due to H$_2$O, as would 
be expected from an atmosphere with a C/O ratio greater than unity.  In fact, the model-data 
comparisons of \citet{madhu11wasp12b} suggest that a solar-like ratio of C/O = 0.54 for WASP-12b 
can be ruled out statistically at the 4.2$\sigma$-significance level, whereas chemical-equilibrium 
models assuming C/O = 1 yield abundances of CO, H$_2$O, and CH$_4$ that are consistent with 
secondary-eclipse observations.  Recently, \citet{madhu12coratio} updated the WASP-12b secondary-eclipse 
analysis to include HCN and $\CtwoHtwo$ in the spectral modeling, and their conclusions with 
respect to the high C/O ratio have not changed.  Moreover, {\it HST\/} Wide Field Camera 3 (WF3) 
spectral observations by \citet{swain12} confirm the suggestion of a very low water abundance on 
WASP-12b, as well as a low TiO abundance, furthering the inference of a high C/O ratio for this 
planet.  \citet{swain12} and \citet{madhu12coratio} demonstrate that the WFC3 secondary-eclipse 
observations are well reproduced by high C/O ratio models.  However, it is worth noting that 
\citet{swain12} favor a very-low-metallicity solution over a high-C/O-ratio solution in explaining 
the WFC3 observations because the high C/O ratio models provide a worse fit to the transit 
observations, especially at the short-wavelength end near 1.1-1.3 $\mu$m.  The degree to which 
Rayleigh scattering could be affecting the shorter wavelengths was not discussed.

In any case, the \citet{madhu11wasp12b} analysis provided the first strong evidence for a 
high C/O ratio on a transiting exoplanet. \citet{madhu11carbrich} further emphasized that TiO 
would not be abundant in giant-exoplanet atmospheres where C/O $\ge$ 1, thus providing a 
theoretically consistent explanation for the lack of a stratospheric thermal inversion on 
WASP-12b.  \citet{koppa12} followed up this analysis by developing photochemical models for WASP-12b for 
two different assumptions of the atmospheric C/O ratio; the molecules HCN and $\CtwoHtwo$ were 
included in these models based on a suggestion by \citet{moses11dps} that these species would 
be important for high C/O ratios \citep[see also][]{lodders10,madhu11carbrich}.  \citet{koppa12} 
found that $\CtwoHtwo$ and HCN were even more abundant than CH$_4$ in their photochemical model with 
C/O = 1.08, and they suggested that C$_2$H$_2$ rather than CH$_4$ is responsible for the strong 
8-$\mu$m absorption on WASP-12b.  We expand the \citet{koppa12} 
modeling by investigating the equilibrium and disequilibrium composition of WASP-12b for a 
variety of C/O ratios and by calculating the spectral consequences of abundant HCN and C$_2$H$_2$ 
for the carbon-rich cases.

\begin{figure*}
\begin{tabular}{ll}
{\includegraphics[angle=0,clip=t,scale=0.37]{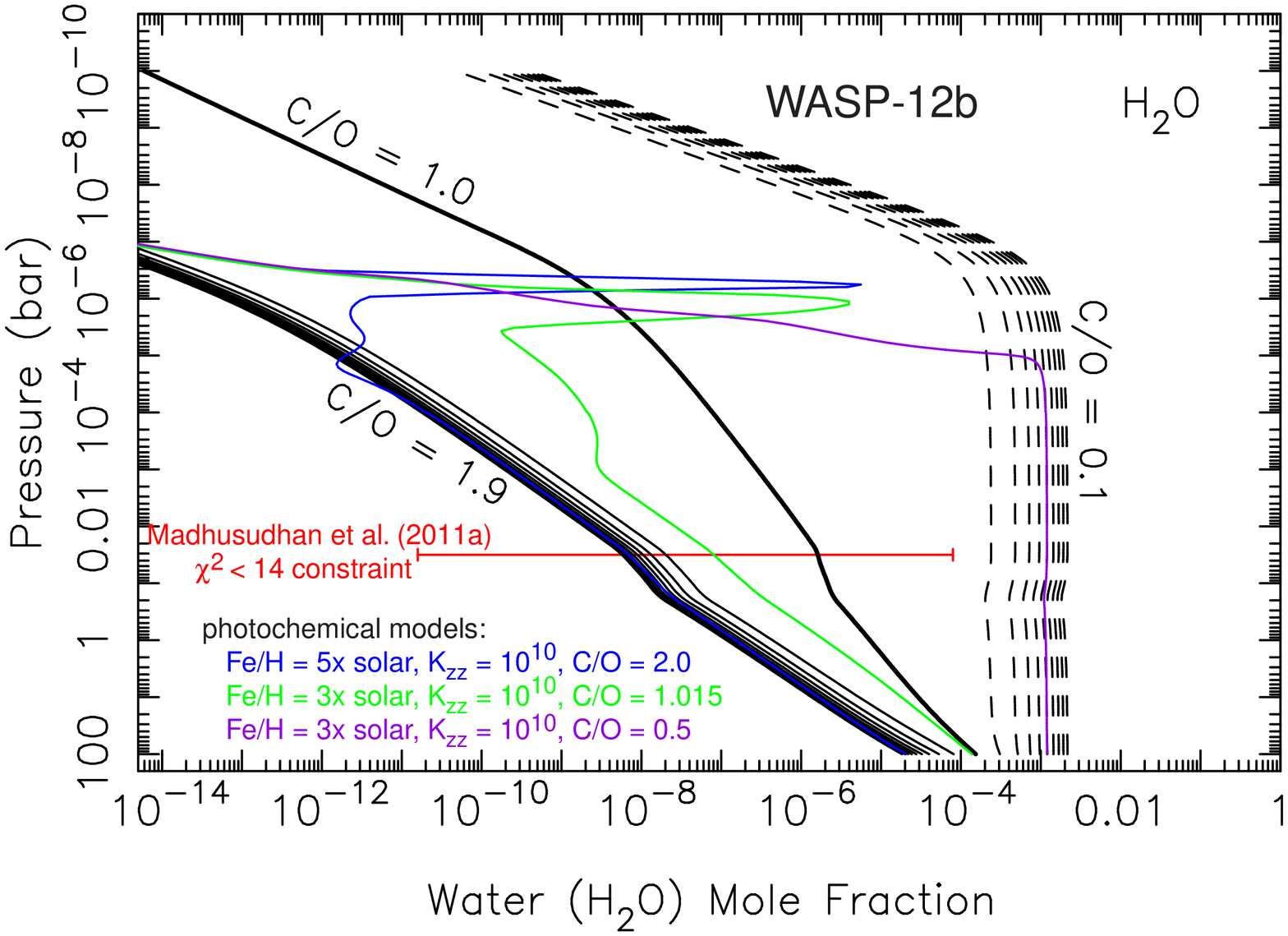}}
&
{\includegraphics[angle=0,clip=t,scale=0.37]{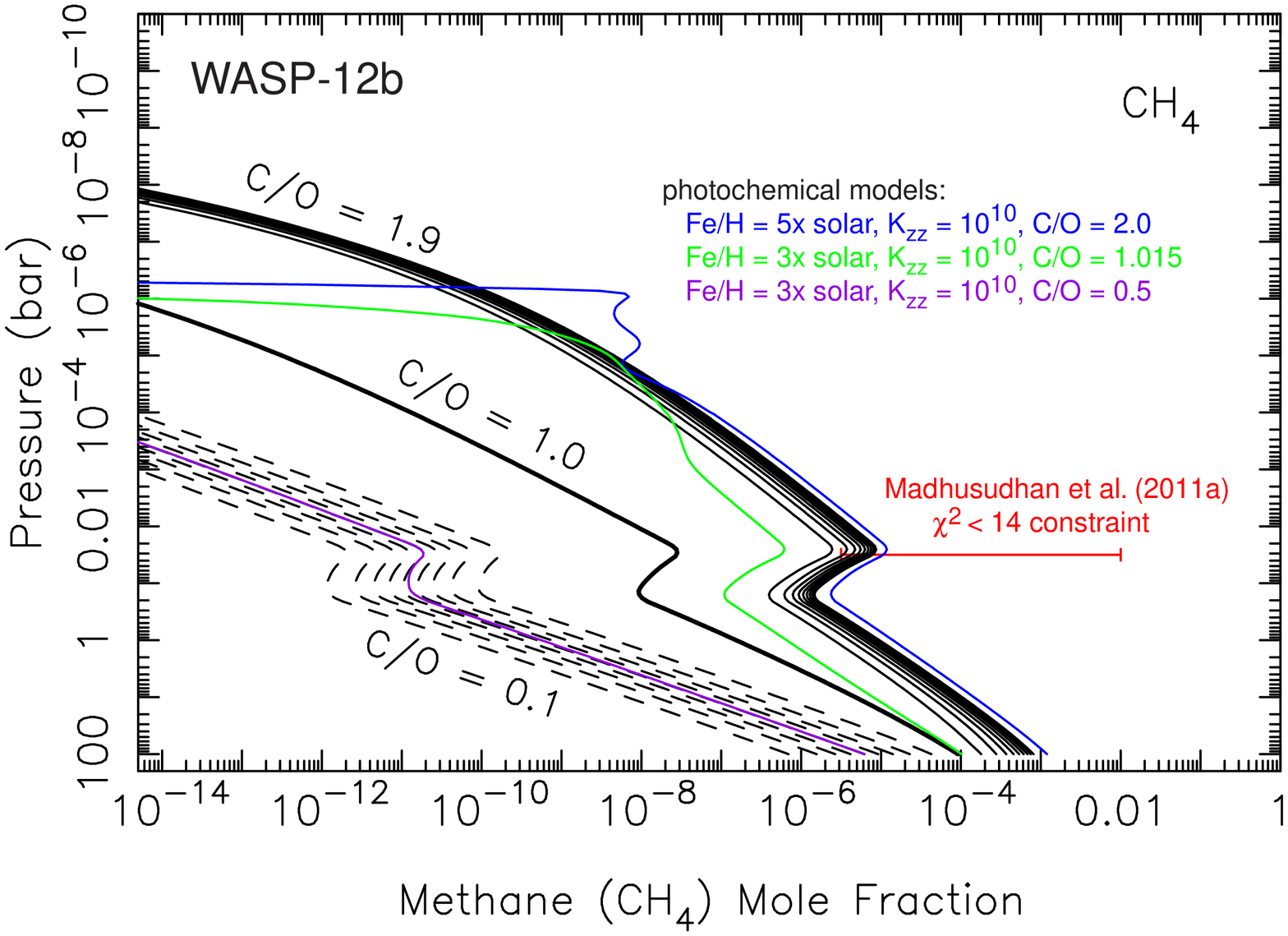}}
\\
{\includegraphics[angle=0,clip=t,scale=0.37]{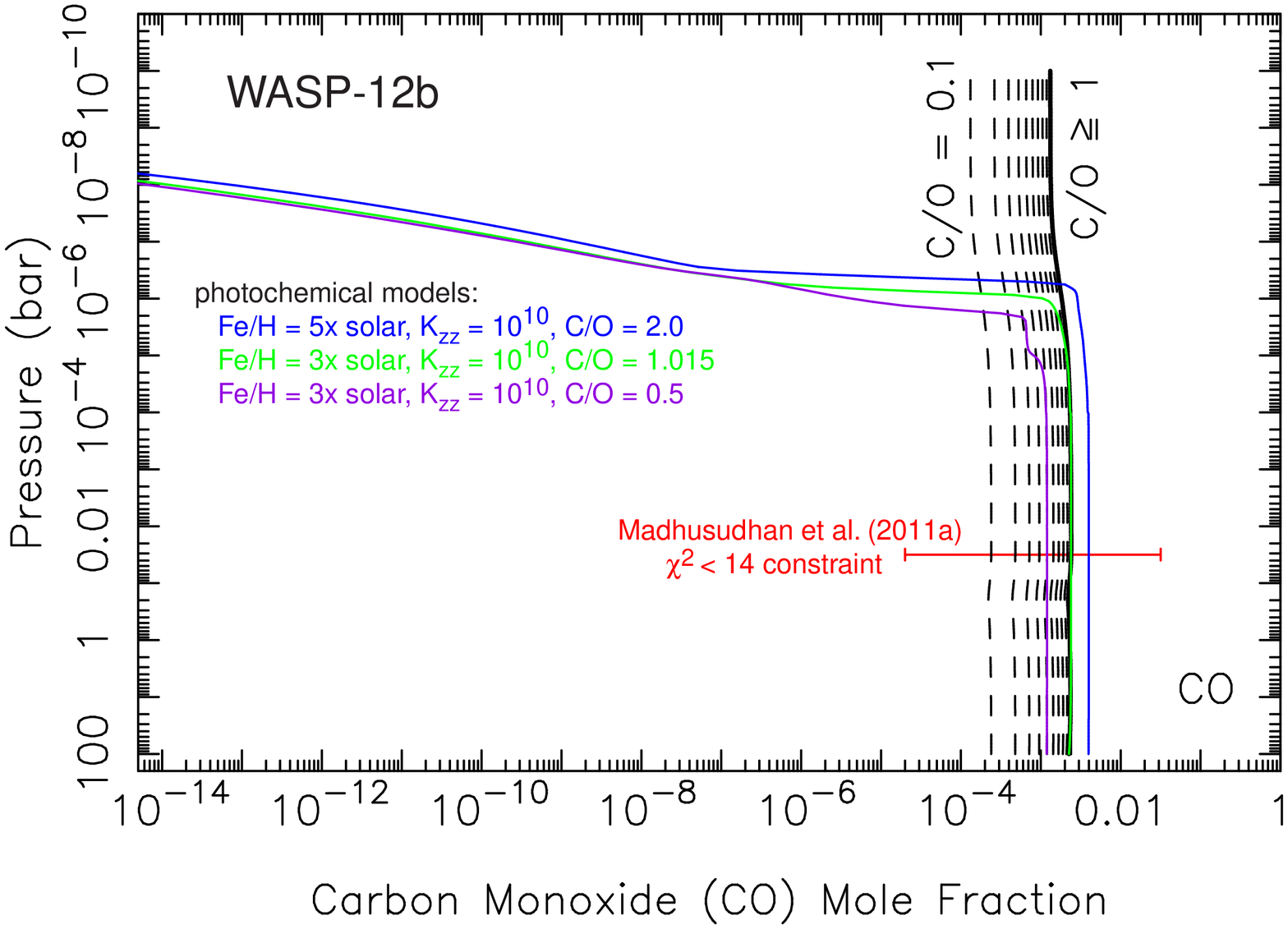}}
&
{\includegraphics[angle=0,clip=t,scale=0.37]{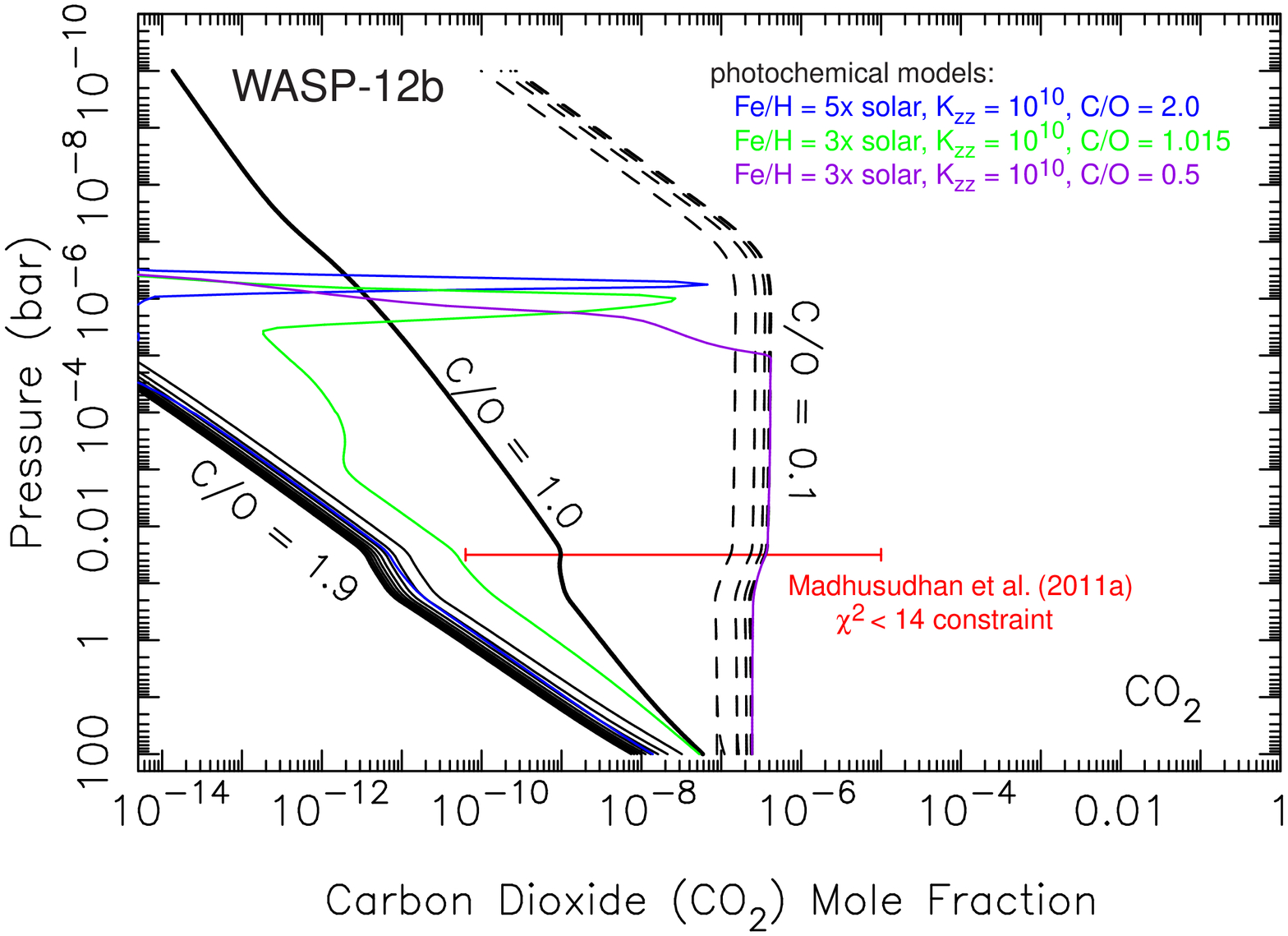}}
\\
{\includegraphics[angle=0,clip=t,scale=0.37]{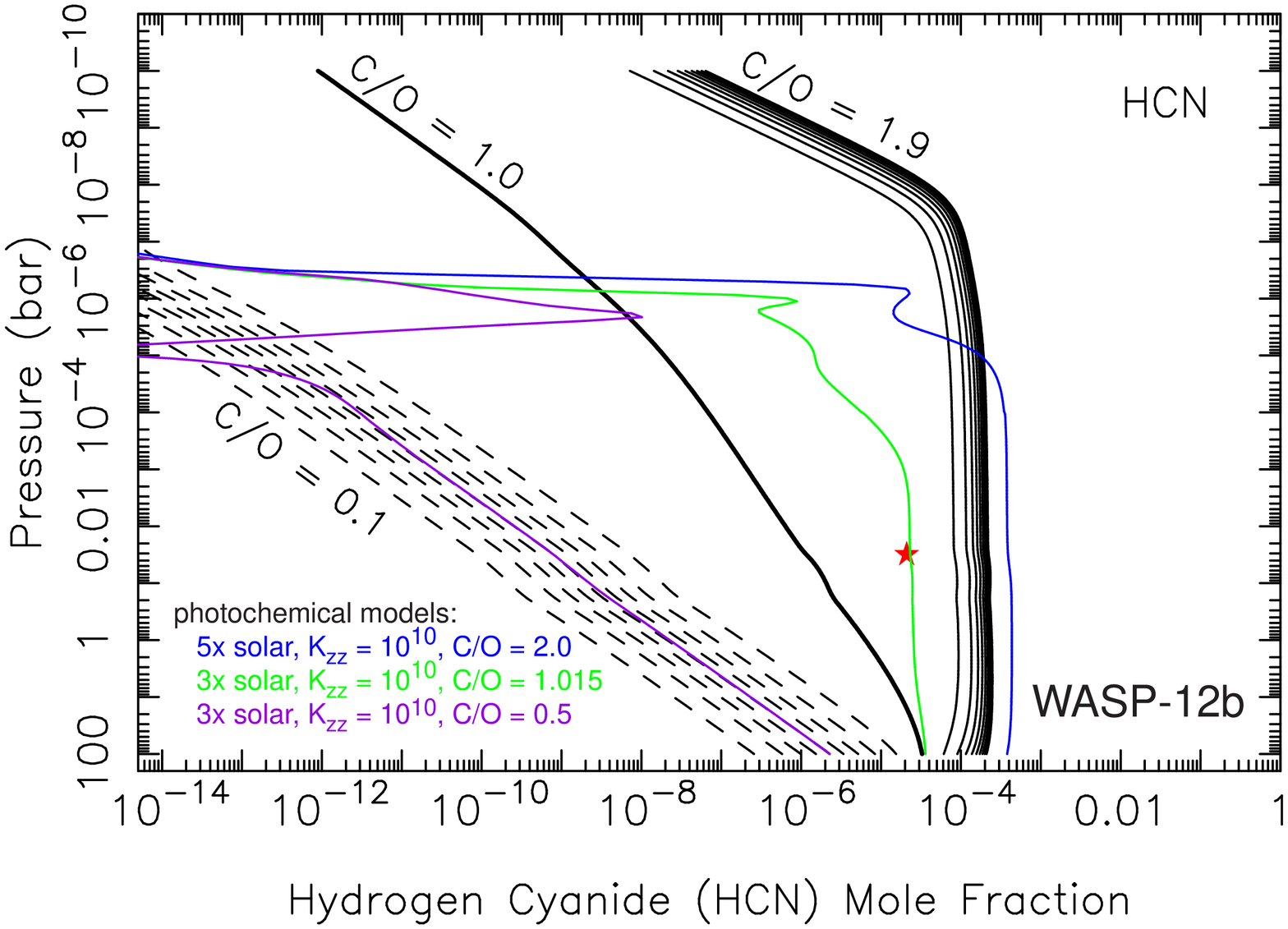}}
&
{\includegraphics[angle=0,clip=t,scale=0.37]{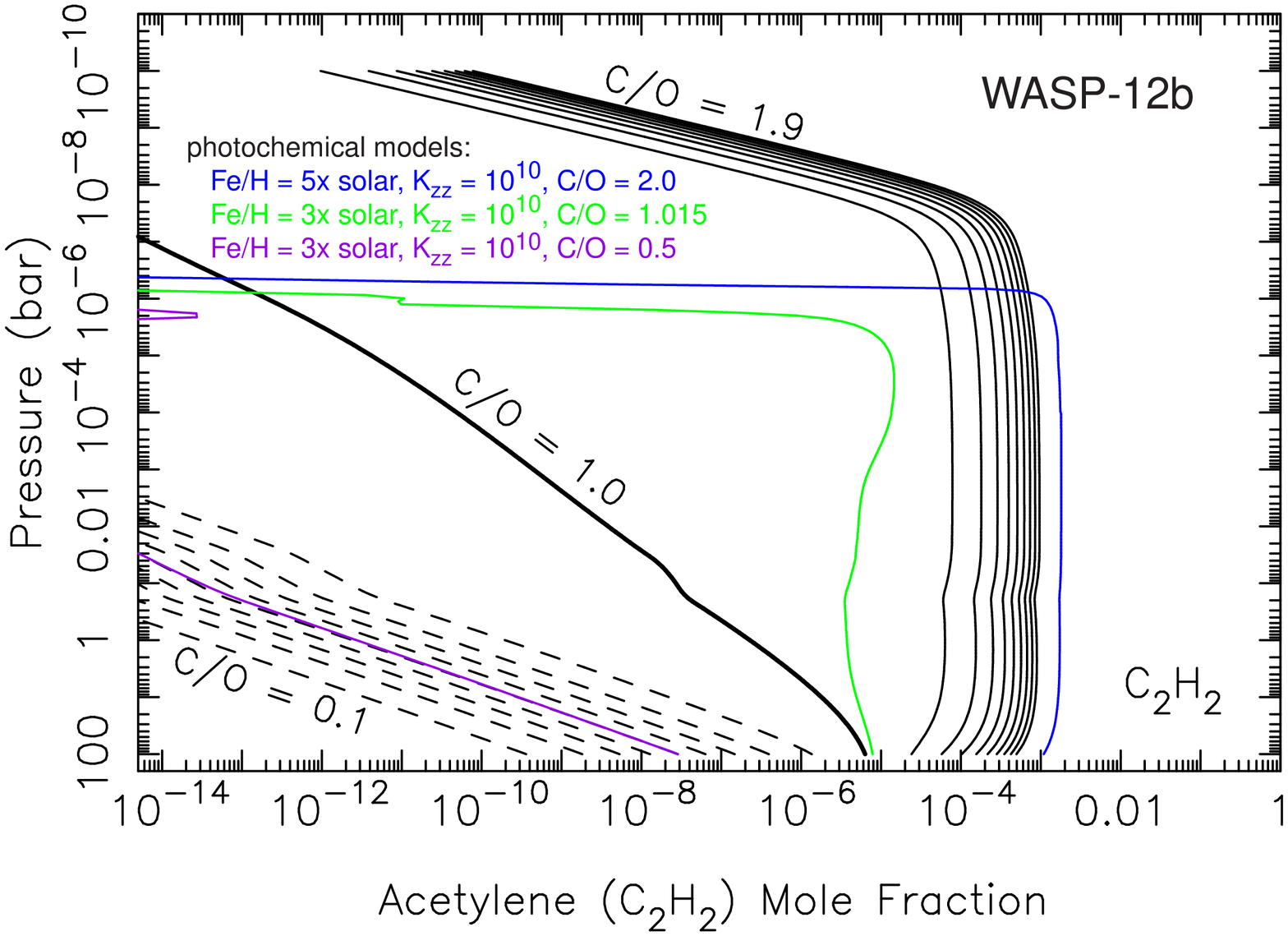}}
\\
\end{tabular}
\caption{Mixing-ratio profiles for H$_2$O, CH$_4$, CO, CO$_2$, HCN, and C$_2$H$_2$ on WASP-12b 
from our thermochemical-equilibrium models with assumed 3$\times$ solar metallicity but 
different assumed C/O ratios (ranging from 0.1 to 1.9, incrementing by 0.1 --- dashed black 
lines for C/O $<$ 1, solid lines for C/O $\ge$ 1).  The colored lines represent disequilibrium-chemistry 
results for models with 3$\times$ solar metallicity and a solar-like C/O ratio of 0.5 (purple), 
a model with 3$\times$ solar metallicity and a C/O ratio of 1.015 (green), and a model with 
5$\times$ solar metallicity and a C/O ratio of 2.0 (blue).  All models have an assumed vigorous 
$K_{zz}$ = 10$^{10}$ $\cmtwo$ $\smone$, which might be expected for such a highly irradiated planet.  
The red horizontal line segments represent abundance constraints from models presented in 
\citet{madhu11wasp12b} based on the {\it Spitzer}/IRAC secondary-eclipse observations of 
\citet{campo11} and the ground-based $Ks$, $H$, and $K$-band secondary-eclipse observations of 
\citet{croll11wasp12b}.  Note that HCN and C$_2$H$_2$ were not considered in the \citet{madhu11wasp12b}
analysis, and the red star in the HCN figure is from the present analysis and indicates the HCN 
abundance that best reproduces the secondary-eclipse data for our adopted nominal thermal structure 
shown in Fig.~\ref{figtemp}.  
A color version of this figure is presented in the online journal.\label{waspequil}}
\end{figure*}

\subsubsection{WASP-12b Chemistry \label{waspchemsect}}

Figure~\ref{waspequil} illustrates how the thermochemical equili{\-}bri{\-}um composition of 
WASP-12b varies as a function of the atmospheric C/O ratio for our assumed nominal 
dayside thermal structure (see Fig.~\ref{figtemp}) and an assumed 3$\times$ solar 
metallicity.  Because of WASP-12b's high temperatures, the chemical-equilibrium abundance 
of methane in the photosphere of WASP-12b is predicted to be smaller than that on the 
cooler HD 189733b or XO-1b, for similar assumptions of atmospheric metallicity and C/O ratio.
Methane, with its relatively weak C-H bond, is less stable at high temperatures and low 
pressures than CO, which has a very strong carbon-oxygen bond.  As with the other planets 
we have investigated, CO is again abundant for all assumptions of the C/O ratio, whereas 
H$_2$O and CO$_2$ are only important for C/O $<$ 1, and HCN and $\CtwoHtwo$ become more 
important for C/O $>$ 1.  At the high temperatures predicted for WASP-12b, HCN --- which also 
has a strong carbon-nitrogen bond --- quickly replaces CH$_4$ as the second-most abundant 
carbon-bearing molecule (behind CO) as the pressure decreases, even for C/O $<$ 1.  With 
increasing C/O ratio, $\CtwoHtwo$ also becomes a major equilibrium constituent whose 
abundance can eventually exceed that of CO when C/O $\gg$ 1.  The implied large methane abundance 
required to fit the secondary-eclipse data in the \citet{madhu11wasp12b} statistical analysis 
suggests a C/O ratio of $\sim$2 or larger for WASP-12b (see Fig.~\ref{waspequil}); however, 
this conclusion changes when HCN and $\CtwoHtwo$ opacities are included in the spectral models. 
For example, as is discussed in Section \ref{waspspecsect}, at least part of the opacity hitherto 
attributed to CH$_4$ on WASP-12b could instead result from HCN and/or C$_2$H$_2$, which are 
abundant species in carbon-rich atmospheres with C/O ratios $\ge$ 1 \citep[see 
also][]{koppa12,madhu12coratio}.

\begin{figure}
\includegraphics[angle=-90,scale=0.37]{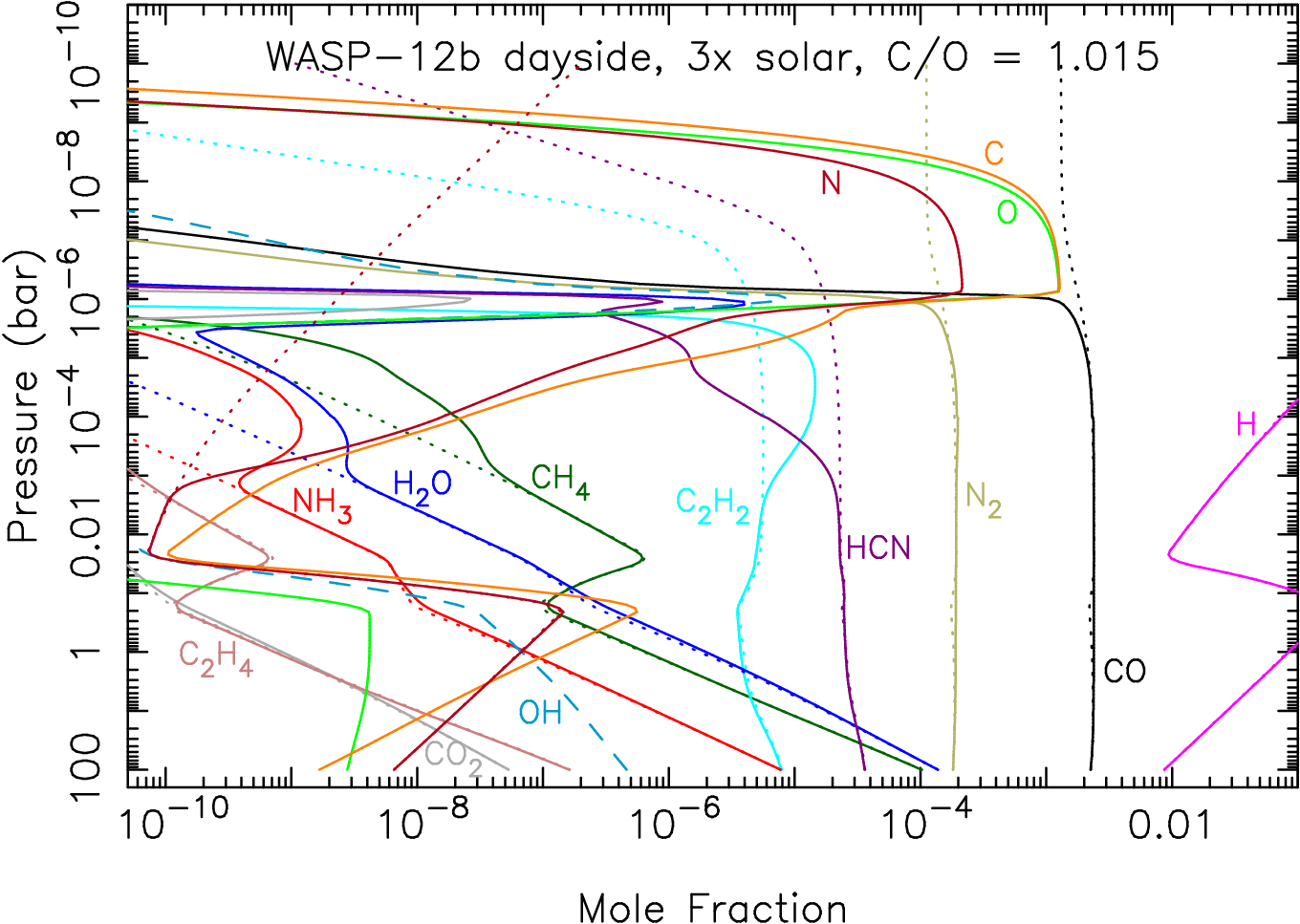}
\caption{Mole-fraction profiles for several important species (as labeled) in a 
thermochemical and photochemical kinetics and transport model for WASP-12b.  Model assumptions 
include 3$\times$ solar metallicity, a C/O ratio of 1.015, and $K_{zz}$ = 10$^{10}$ $\cmtwo$ 
$\smone$ (see also the green curves in Fig.~\ref{waspequil}).  The dotted lines show the 
thermochemical-equilibrium abundances for the species, using the same color coding.  
Note that both HCN and $\CtwoHtwo$ are more abundant than $\CHfour$ throughout the 
middle atmosphere.  
A color version of this figure is available in the online journal.\label{wasppchem}}
\end{figure}

Disequilibrium processes like photochemistry and transport-induced quenching result in relatively 
minor changes in the profiles of the dominant species in our C/O = 1.015 model, except for HCN, 
$\CtwoHtwo$, and atomic species (see Fig.~\ref{wasppchem}).  
Hydrogen cyanide becomes a parent molecule of other photochemical products in much 
the same way as $\NHthree$ and CH$_4$ do in cooler atmospheres.  Due to the high temperatures and 
large overall H abundance at high altitudes, kinetic loss of HCN via H + HCN $\rightarrow$ CN + 
$\Htwo$ dominates over HCN photolysis as the main photochemical loss process.  The CN radicals 
produced can recycle the HCN or become photolyzed themselves, leading to C + N.  Atomic N can then 
react either with NO to form $\Ntwo$ + O or with CN to form $\Ntwo$ + C, and the atomic carbon can 
react further to eventually produce $\CtwoHtwo$.  The dominant $\CtwoHtwo$ photochemical production 
scheme in our C/O = 1.015 model is
\begin{eqnarray}
\Htwo \, + \, \M \, & \rightarrow & \, 2\, \H \, + \, \M \nonumber  \\
2\, ( \H \, + \, \HCN \, & \rightarrow & \, \CN \, + \, \Htwo ) \nonumber  \\
2\, ( \CN \, + \, h\nu \, & \rightarrow & \, \C \, + \, \N ) \nonumber  \\
2\, ( \C \, + \, \Htwo \, & \rightarrow & \, \CH \, + \, \H ) \nonumber  \\
\CH \, + \, \Htwo \, & \rightarrow & \, \threeCHtwo \, + \, \H \nonumber  \\
\threeCHtwo \, + \, \Htwo \, & \rightarrow & \, \CHthree \, + \, \H \nonumber  \\
\CH \, + \, \CHthree \, & \rightarrow & \, \CtwoHthree \, + \, \H \nonumber  \\
\H \, + \, \CtwoHthree \, & \rightarrow & \, \CtwoHtwo \, + \, \Htwo \nonumber \\
\N \, + \, \Htwo \, & \rightarrow & \, \NH \, + \, \H \nonumber \\
\NH \, + \, \O \, & \rightarrow & \, \NO \, + \, \H \nonumber \\
\N \, + \, \NO \, & \rightarrow & \, \Ntwo \, + \, \O \nonumber \\
3\, ( 2\, \H \, + \M \, & \rightarrow & \, \Htwo \, + \, \M \, ) \nonumber  \\
\noalign{\vglue -10pt}
\multispan3\hrulefill \nonumber \cr
\Net \ \ 2\, \HCN \, & \rightarrow & \, \CtwoHtwo \, + \, \Ntwo  . \\
\end{eqnarray}
The acetylene abundance increases with altitude in the $\sim$2-mbar to 2-$\mu$bar region due 
to this HCN photochemistry, while the HCN abundance decreases.  
The nitrogen from HCN loss largely ends up as $\Ntwo$ at pressures greater than $\sim$1 $\mu$bar and 
as atomic N at pressures less than $\sim$1 $\mu$bar.  In fact, no molecules (including $\Htwo$) survive 
at altitudes much above $\sim$1 $\mu$bar, as the intense ultraviolet flux and large H abundance attack most 
molecular bonds and favor atomic species.  Some photochemical production of $\HtwoO$, $\CHfour$, and 
$\NHthree$ occurs from the CO and HCN photochemistry, but these species remain below ppm levels at 
pressures less than $\sim$0.1 bar and only become significant at high pressures and in a narrow 
region centered near $\sim$1 $\mu$bar where photolysis of CO and $\Ntwo$ is significant.  In all, 
disequilibrium processes have less of an effect on hotter planets like WASP-12b than they 
do on cooler planets like HD 189733b and XO-1b \citep[see
also][]{liang03,liang04,zahnle09sulf,line10,line11gj436,moses11,venot12}.

\begin{figure}
\includegraphics[angle=0,scale=0.48]{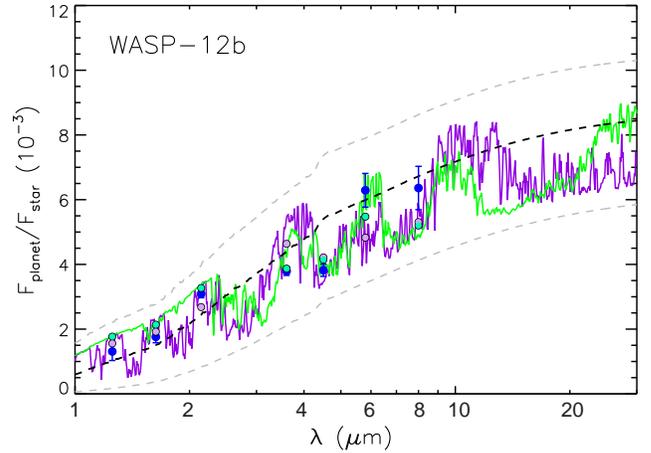}
\caption{Secondary-eclipse spectra for WASP-12b. The solid lines represent synthetic spectra 
from the results of our disequilibrium chemistry models with 3$\times$ solar metallicity, 
$K_{zz}$ = 10$^{10}$ $\cmtwo$ $\smone$, and C/O ratios of either 1.015 (green) or 0.5 (purple).
The blue circles with error bars represent the {\it Spitzer}/IRAC data from \citet{campo11} 
and ground-based data from \citet{croll11wasp12b}.  The circles without error bars represent 
the model results for C/O = 1.015 (green) or 0.5 (purple) convolved over the {\it Spitzer\/} 
bandpasses.  Note the better fit with the C/O = 1.015 model for this assumed metallicity 
and thermal structure, particularly at 2.1, 3.6, and 5.8 $\mu$m.  
The dashed curves represent blackbodies with temperatures of 1800 K (bottom), 2500 K (middle), 
and 3000 K (top).  A color version of this figure is available in the online journal.\label{specwasp}}
\end{figure}

Our photochemical models for a high C/O ratio on WASP-12b differ somewhat from those of \citet{koppa12}.  
Their upper atmosphere is cooler, which tends to reduce reaction rates.  Transport-induced quenching 
of $\CHfour$ and $\HtwoO$ therefore become correspondingly more important, as expected, leading to 
increased abundances of $\HtwoO$ and $\CHfour$.  Their adopted temperature-pressure profile and 
C/O ratio of 1.08 also result in $\CtwoHtwo$ being slightly more abundant than HCN in equilibrium 
throughout the photosphere, which can affect the photochemistry.  However, the HCN and $\CtwoHtwo$ 
vertical profiles of \citet{koppa12} are almost completely unaffected by photochemistry, in 
contrast to our models, for reasons that are unclear but likely have to do with different 
reaction mechanisms.  Their main conclusion that HCN and $\CtwoHtwo$ are important on WASP-12b for 
the case of C/O $\ge$ 1 under both equilibrium and disequilibrium conditions is qualitatively 
supported by our models \citep[see also][]{lodders10,moses11dps,madhu11carbrich,madhu12coratio}. 
 
\subsubsection{WASP-12b Spectra \label{waspspecsect}}

Like methane, both HCN and $\CtwoHtwo$ have bands that fall within the {\it Spitzer}/IRAC 
8.0-$\mu$m spectral 
bandpass, and to a lesser extent within the short-wavelength end of the 3.6-$\mu$m spectral 
bandpass \citep[e.g.,][]{shabram11}.  We find that all three molecules contribute to the 
calculated flux in the 3.6 and 8-$\mu$m channels, but for our nominal thermal structure and 
C/O ratio $\sim$ 1, the channel-integrated flux at these wavelengths is particularly sensitive 
to the HCN mole fraction.  
For a 3$\times$ solar metallicity and our adopted thermal structure, C/O ratios just over unity 
result in an HCN abundance that provides a good fit to the {\it Spitzer}/IRAC 3.6-$\mu$m band 
depth, while providing slightly too much absorption in the lower-signal-to-noise 8.0-$\mu$m 
band.  This result will be model-dependent, particularly in terms of the relative contributions 
of different molecules for different thermal structures, but Fig.~\ref{specwasp} demonstrates 
that our model with C/O = 1.015 provides a significantly better fit (with $\xi ^2$ $<$ 2.6) to the 
{\it Spitzer\/} data than the model with C/O = 0.5 (with $\xi ^2$ $<$ 10.2); this result is 
consistent with the detailed statistical analysis of 
\citet{madhu11wasp12b}.  The C/O = 1.015 model has much less absorption at 5.8 $\mu$m and much 
more absorption at 3.6 $\mu$m than the C/O = 0.5 model, and the predicted relative strengths 
in all four {\it Spitzer}/IRAC bands are much more consistent with observations for the higher 
C/O ratio model.  We therefore concur with the suggestion of \citet{madhu11wasp12b} that WASP-12b 
could be a carbon-enriched planet, although in our model it is HCN rather than $\CHfour$ that 
provides the dominant opacity source in the 3.6 and 8.0 $\mu$m channels (see also Section
4 and \citealt{madhu12coratio}).  Note that a high C/O ratio model for WASP-12b is also more 
consistent with the recent {\it HST\/} WFC3 secondary-eclipse observations of \citet{swain12} 
in which a very low water abundance is needed to explain the nearly featureless 1.0-1.7 micron 
spectrum of WASP-12b \citep[e.g.,][]{swain12,madhu12coratio}.

The large predicted HCN abundance on a putative carbon-rich WASP-12b is also likely to affect its 
near-infrared transit spectrum, where \citet{cowan12} observe a larger transit depth in the 
3.6-$\mu$m {\it Spitzer\/} channel than in the 4.5 $\mu$m channel.  \citet{cowan12} did not 
consider the possible influence of HCN on the 3.6-$\mu$m absorption, and in fact only 
considered enhanced CO abundances as a consequence of a potential carbon-rich scenario.  
Future transit simulations should consider the influence of HCN and $\CtwoHtwo$.  At 1-2 $\mu$m, 
the spectrum in the high C/O ratio case is much more featureless than the solar C/O ratio case, 
where water absorption bands are prominent.

\subsection{CoRoT-2b Results\label{corotsect}}

The transiting planet CoRoT-2b, discovered by \citet{alonso08}, has often been called a 
``misfit'' because of its unusually large radius \citep[e.g.,][]{alonso08,gillon10,guillot11} 
and because traditional models that assume solar elemental abundances cannot explain its 
unusual flux ratios from secondary-eclipse observations in the {\it Spitzer}/IRAC channels 
at 3.6, 4.5, and 8.0 $\mu$m, regardless of whether the planet is assumed to have a stratospheric 
thermal inversion or not \citep{gillon10,deming11corot}.  The main problem is an anomalously low flux 
at 8.0 $\mu$m and an unusually large 4.5-$\mu$m/8.0-$\mu$m flux ratio. \citet{madhu12coratio}
suggests that all the secondary-eclipse observations for CoRoT-2b can be reproduced to within their 
1-$\sigma$ uncertainties for chemically plausible models with C/O $\ge$ 1; \citet{madhu12coratio} 
also finds good solutions for oxygen-rich models, but those models tend to have implausibly large 
$\CHfour$/CO abundance ratios.  Because there is no {\it Spitzer}/IRAC data available for CoRoT-2b 
from the 5.8-$\mu$m IRAC channel,  model solutions for CoRoT-2b will tend to be more degenerate, but 
if \citet{madhu12coratio} is correct in his suggestion of a carbon-rich atmosphere, then 
CoRoT-2b would fall within the C2 regime in their classification scheme.

\begin{figure*}
\begin{tabular}{ll}
{\includegraphics[angle=0,clip=t,scale=0.37]{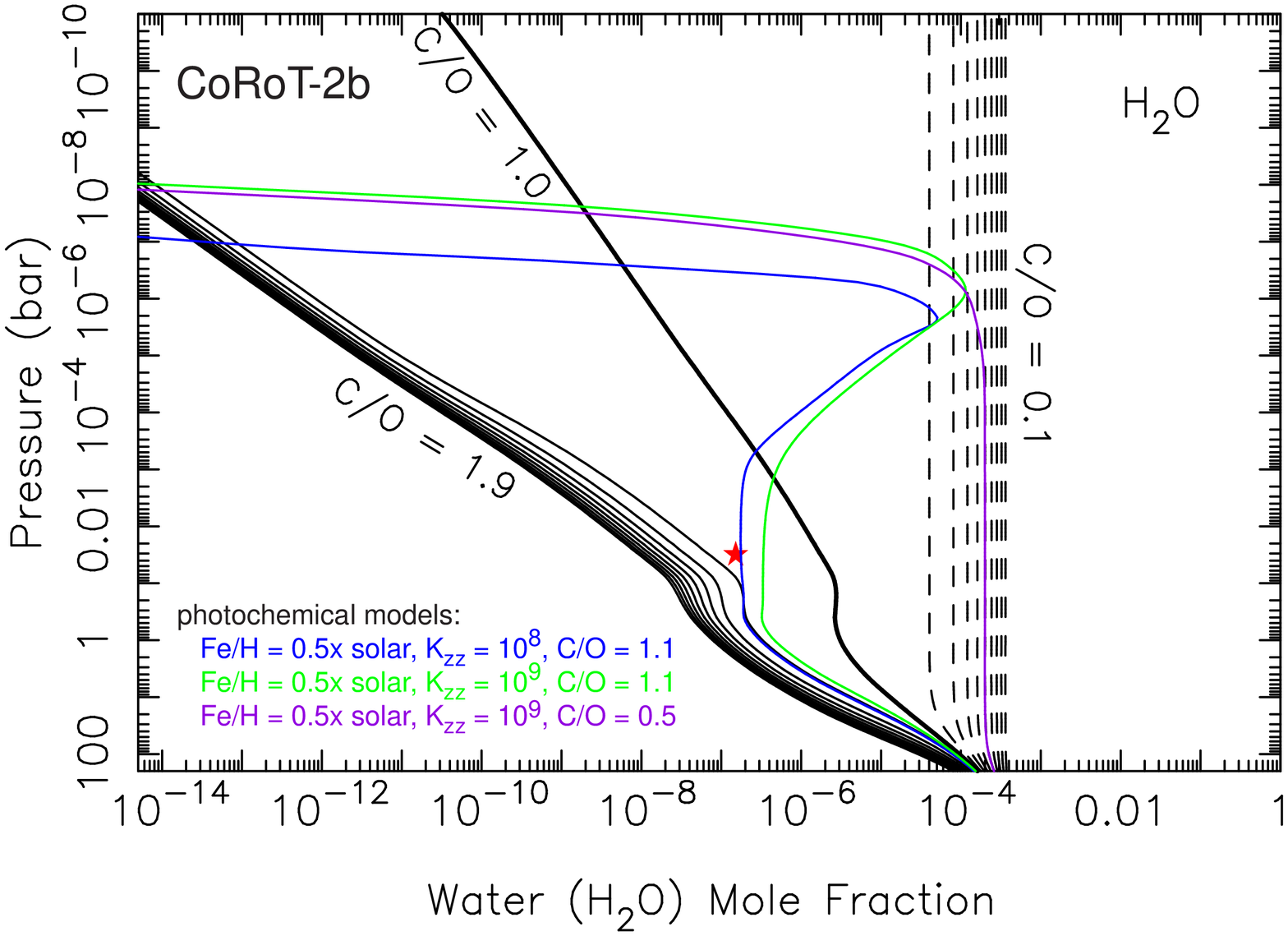}}
&
{\includegraphics[angle=0,clip=t,scale=0.37]{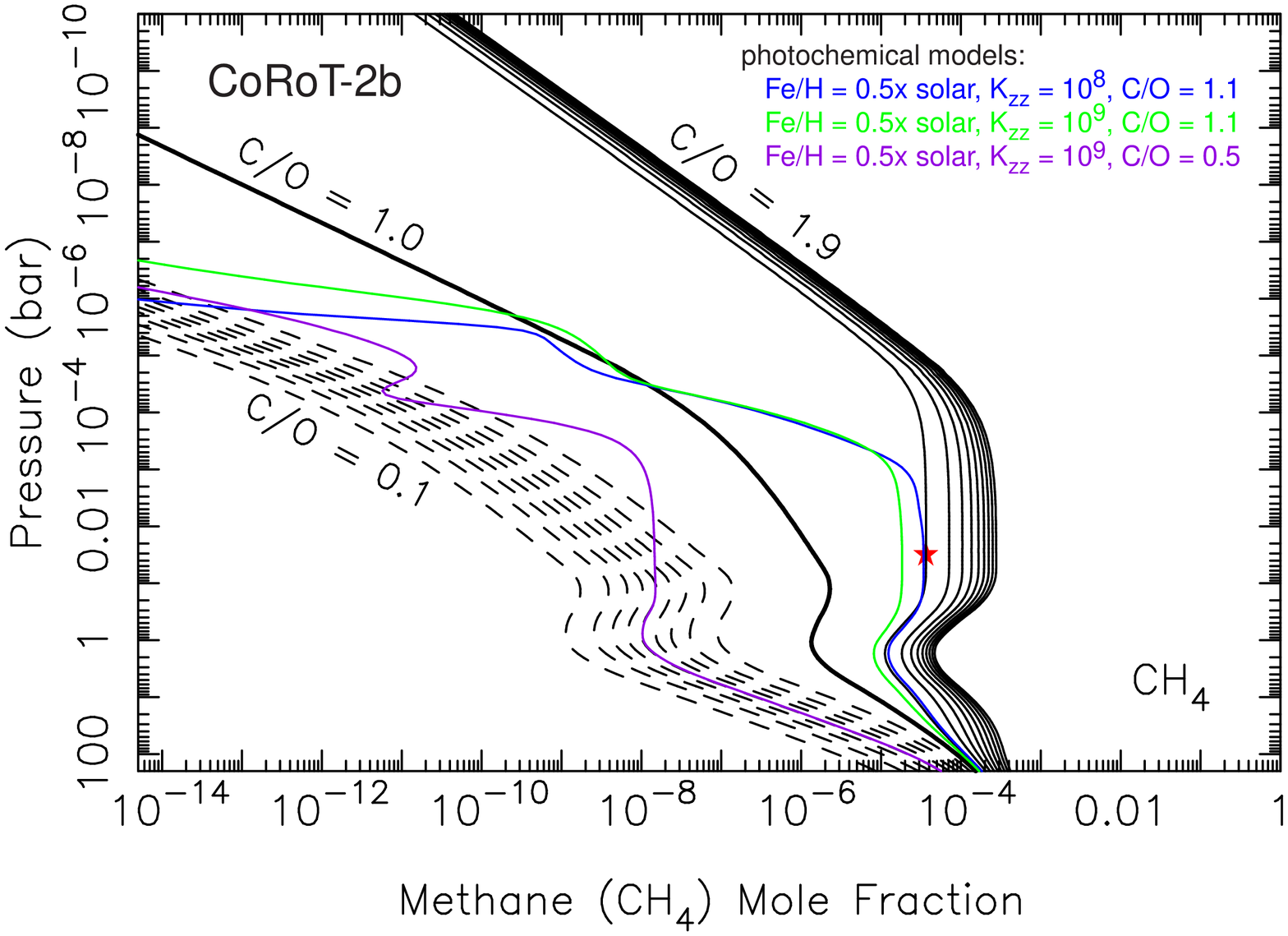}}
\\
{\includegraphics[angle=0,clip=t,scale=0.37]{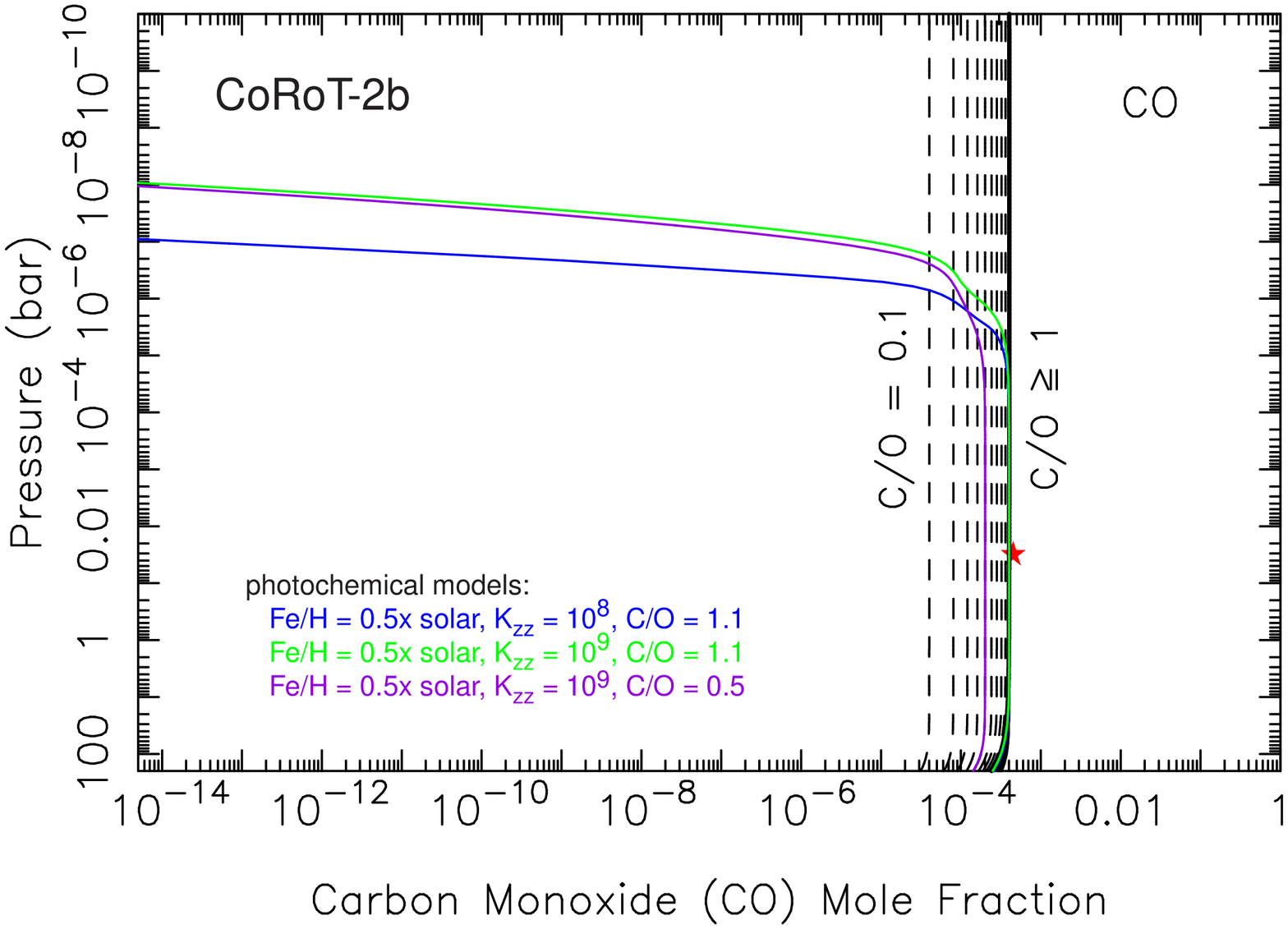}}
&
{\includegraphics[angle=0,clip=t,scale=0.37]{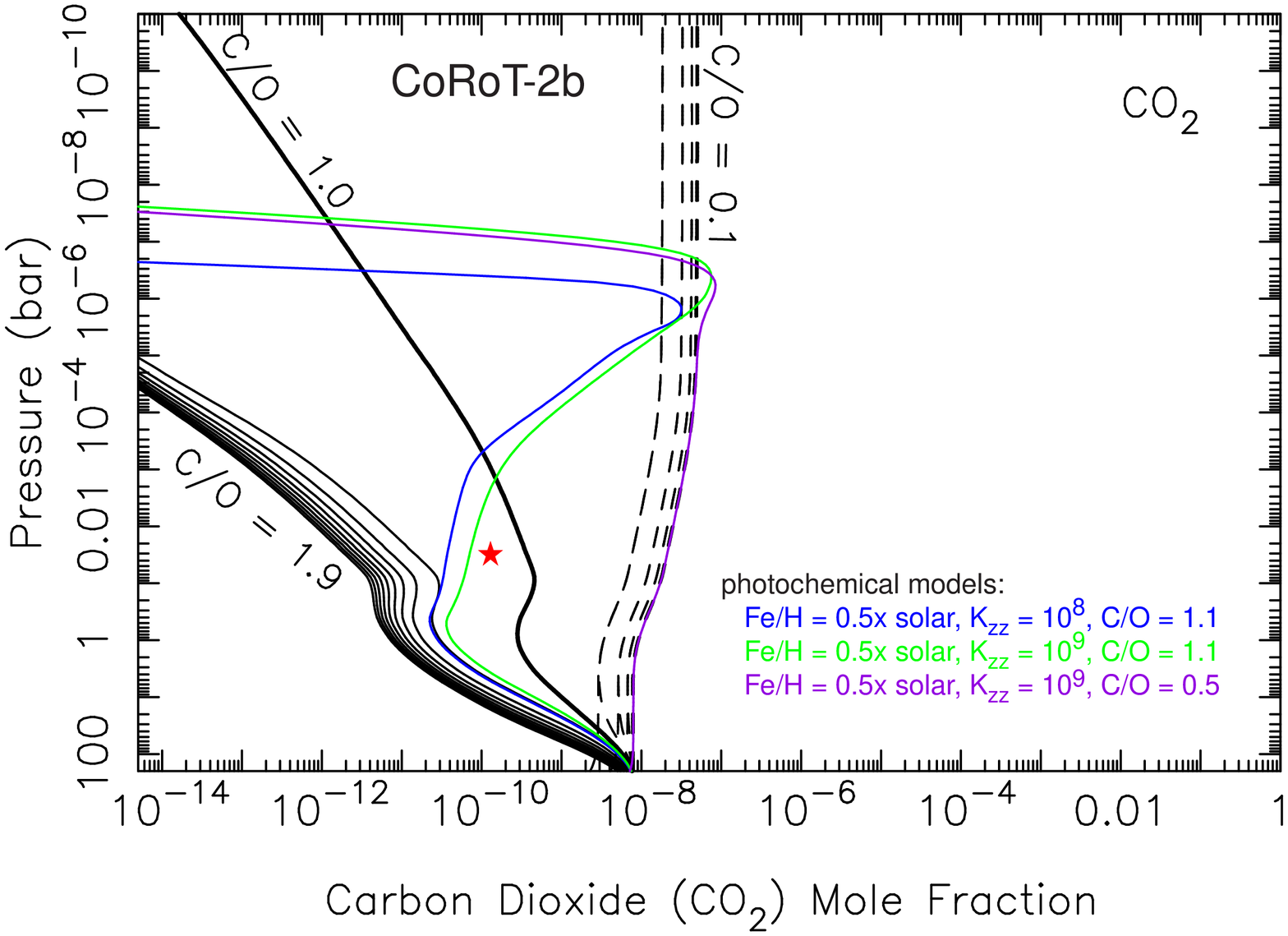}}
\\
{\includegraphics[angle=0,clip=t,scale=0.37]{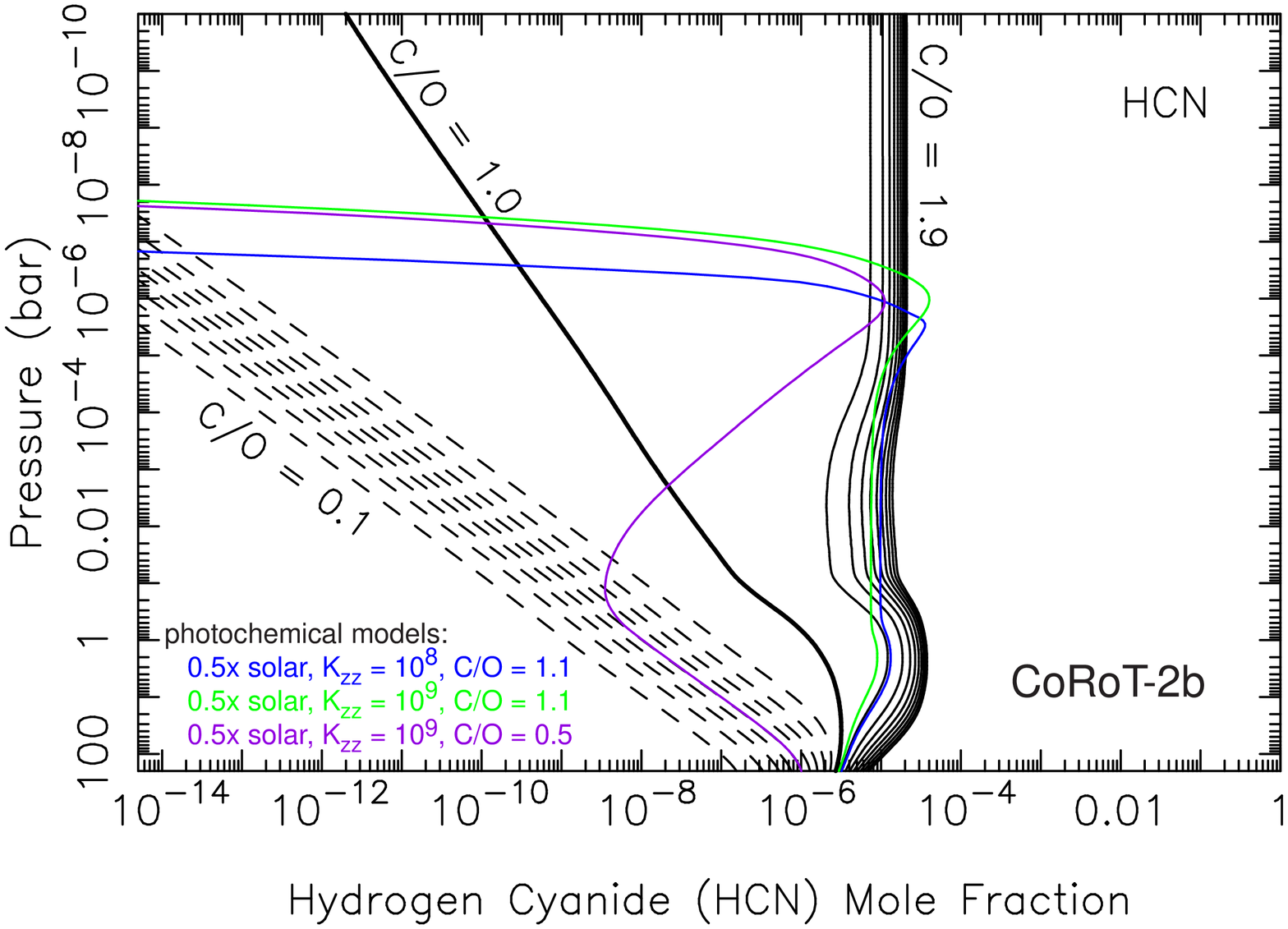}}
&
{\includegraphics[angle=0,clip=t,scale=0.37]{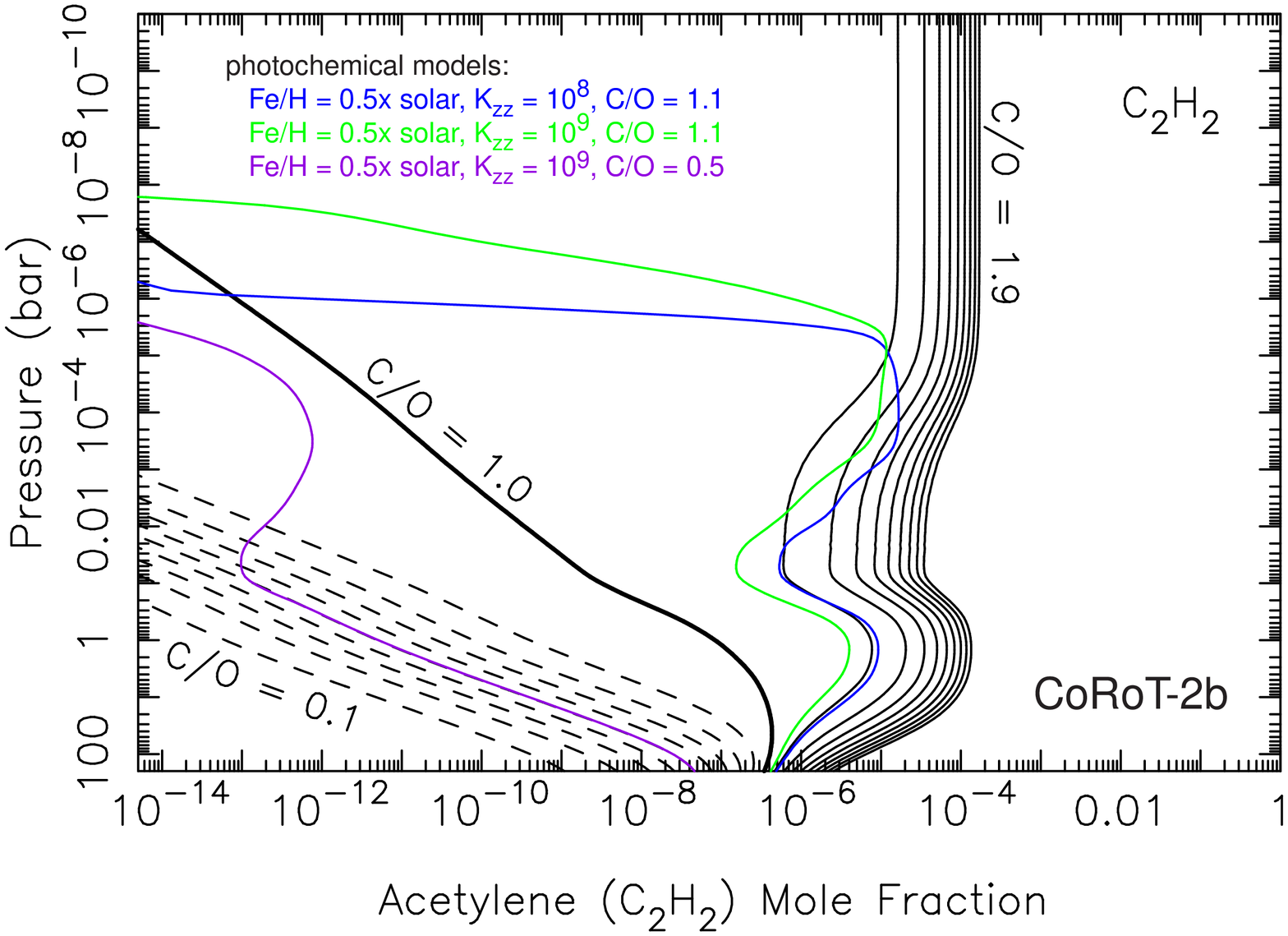}}
\\
\end{tabular}
\caption{Mixing-ratio profiles for H$_2$O, CH$_4$, CO, CO$_2$, HCN, and C$_2$H$_2$ on CoRoT-2b
from our thermochemical-equilibrium models with assumed 0.5$\times$ solar metallicity but different 
assumed C/O ratios (ranging from 0.1 to 1.9, incrementing by 0.1 --- dashed black lines for 
C/O $<$ 1, solid lines for C/O $\ge$ 1).  The colored lines represent disequilibrium-chemistry 
results for models with a solar-like C/O ratio of 0.5 and $K_{zz}$ = 10$^9$ $\cmtwo$ $\smone$ 
(purple), a model with a C/O ratio of 1.1 and $K_{zz}$ = 10$^9$ $\cmtwo$ $\smone$ (green), 
and a model with a C/O ratio of 1.1 and $K_{zz}$ = 10$^8$ $\cmtwo$ $\smone$ (blue).  All 
models have an assumed atmospheric metallicity of 0.5$\times$ solar.  The red stars represent 
abundance constraints from a preliminary model \citep{madhu12coratio} based on the 
{\it Spitzer}/IRAC secondary-eclipse observations of \citet{gillon10} and \citet{deming11corot} 
and the ground-based secondary-eclipse $Ks$-band observations of \citet{alonso10}, in which HCN 
and C$_2$H$_2$ were not considered in the analysis.  A color version of this figure is presented 
in the online journal.\label{corotequil}}
\end{figure*}

\subsubsection{CoRoT-2b Chemistry \label{corotchemsect}}

Figure~\ref{corotequil} illustrates how the thermochemical-equilibrium composition of 
CoRoT-2b varies as a function of the atmospheric C/O ratio for our assumed nominal 
dayside thermal structure (see Fig.~\ref{figtemp}) and an assumed 0.5$\times$ solar 
metallicity.  The low metallicity we are considering here is required to prevent too 
much absorption from CO in the {\it Spitzer}/IRAC channel at 4.5 $\mu$m for our assumed 
thermal structure; a warmer atmosphere would allow higher metallicities while still being 
consistent with the data.  As with all the other planets we have considered, CO is abundant 
in chemical equilibrium for all assumptions of the C/O ratio, whereas H$_2$O and CO$_2$ are 
only abundant for C/O $<$ 1, and CH$_4$, HCN, and $\CtwoHtwo$ become more abundant for C/O $>$ 1.  
For a model with a thermal structure similar to our adopted nominal profile shown in 
Fig.~\ref{figtemp}, \citet{madhu12coratio} finds good fits to the \citet{alonso10}, 
\citet{gillon10}, and \citet{deming11corot} secondary-eclipse data for the species mole fractions 
marked by a star in Fig.~\ref{corotequil}.  These constraints imply that CO and $\CHfour$ are 
abundant, whereas $\HtwoO$ and $\COtwo$ remain below ppm levels.  It appears that a C/O ratio 
near 1.1 can readily satisfy all the these constraints.  The relatively low upper atmospheric 
temperatures adopted for our nominal model suggest that photochemistry will be important on 
CoRoT-2b, as we found for XO-1b and HD 189733b.

\begin{figure}
\includegraphics[angle=-90,scale=0.37]{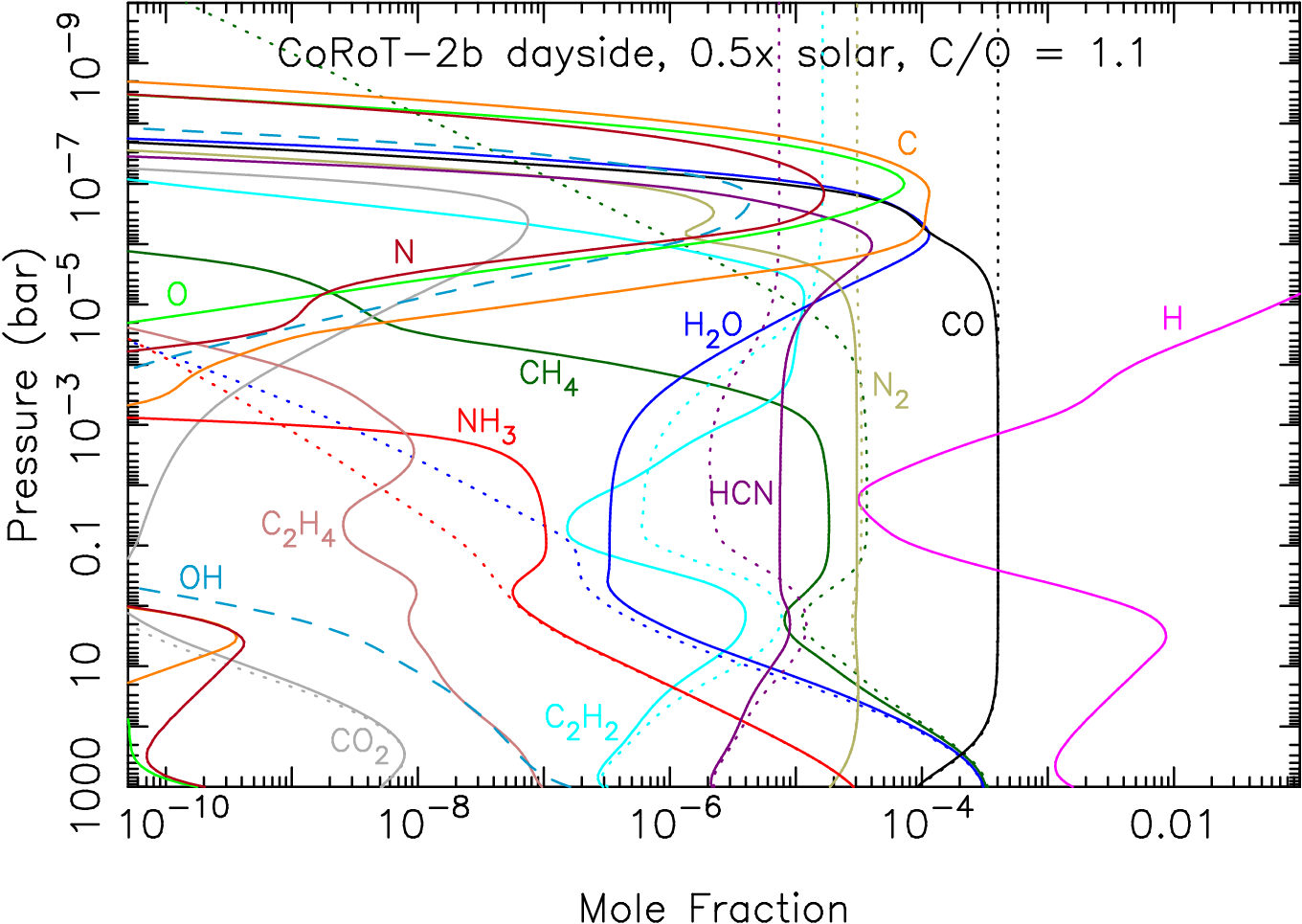}
\caption{Mole-fraction profiles for several important species (as labeled) in a 
thermochemical and photochemical kinetics and transport model for CoRoT-2b.  Model assumptions 
include 0.5$\times$ solar metallicity, a C/O ratio of 1.1, and $K_{zz}$ = 10$^9$ $\cmtwo$ 
$\smone$ (see also the green curves in Fig.~\ref{corotequil}).  The dotted lines show the 
thermochemical-equilibrium abundances for some of the species, using the same color coding.  
Note the enhanced disequilibrium abundances of HCN, $\HtwoO$, NH$_3$, and CO$_2$ over equilibrium 
values.  A color version of this figure is available in the online journal.\label{corotpchem}}
\end{figure}

Indeed, Fig.~\ref{corotpchem} shows that disequilibrium chemistry is important for our CoRoT-2b 
model with an assumed C/O ratio of 1.1, a subsolar metallicity of 0.5$\times$ solar, and 
$K_{zz}$ = 10$^9$ $\cmtwo$ $\smone$.  Transport-induced quenching of water is very important, 
enhancing the $\HtwoO$ photospheric column abundance by about an order of magnitude, and 
CO photochemistry further enhances the H$_2$O abundance in the upper atmosphere by many orders 
of magnitude.  Photochemical production of water occurs via CO photolysis in the upper 
stratosphere to produce C + O, followed by reactions of O with $\Htwo$ to form OH, which reacts 
with $\Htwo$ to form water.  The atomic C can react with NO to form CN and eventually HCN (see 
scheme (2) above) or the C can react with $\Htwo$ to form CH, which eventually goes 
on to produce $\CtwoHtwo$.  We are even seeing a non-trivial formation rate of small 
C$_3$H$_x$ compounds at high altitudes in the model from reaction of C and/or CH with $\CtwoHtwo$; 
these C$_3$H$_x$ species could be soot precursors, although 
benzene and complex nitriles themselves never exceed ppb levels in this constantly-illuminated model.  
Transport-induced quenching also affects $\CHfour$, HCN, $\NHthree$, $\COtwo$, and many other 
species in the middle atmosphere, whereas photochemistry affects profiles in the upper atmosphere.  
Although photochemical production of $\COtwo$ greatly enhances its abundance in the upper atmosphere, 
the column abundance in the photosphere is still quite small, and $\COtwo$ should have little 
influence on spectra.  The chemistry of a carbon-rich CoRoT-2b is similar to that of HD 189733b 
and XO-1b, which is covered in more detail in Sections~\ref{hd189chemsect} \& \ref{xo1bchemsect}; 
the chemistry of a solar-like-composition CoRoT-2b is similar to that of other 
``warm'' transiting planets (see \citealt{moses11} and \citealt{line10} for further details).

\begin{figure}
\includegraphics[angle=0,scale=0.48]{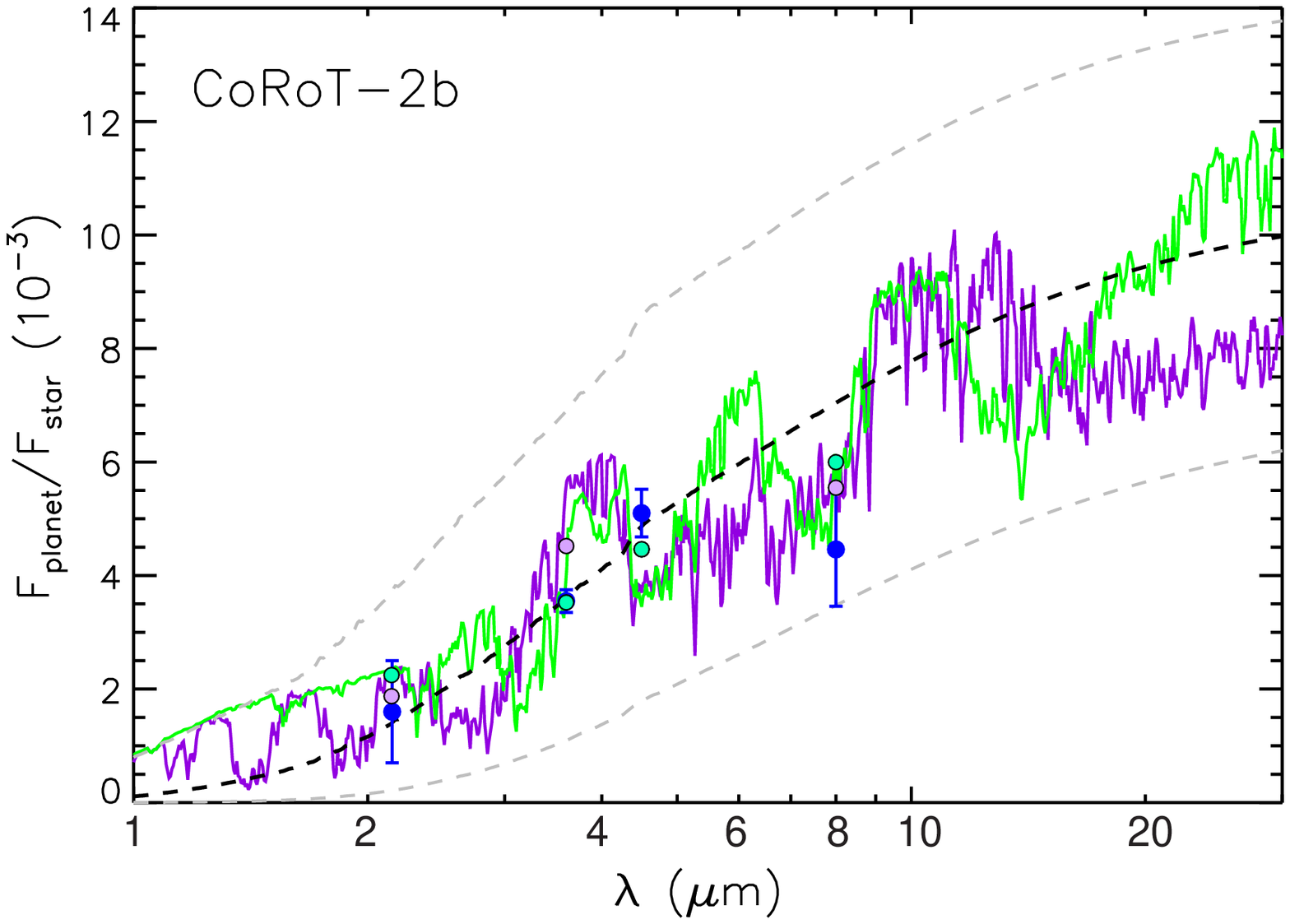}
\caption{Secondary-eclipse spectra for CoRoT-2b. The solid lines represent synthetic spectra 
from the results of our disequilibrium chemistry models with 0.5$\times$ solar metallicity, 
$K_{zz}$ = 10$^9$ $\cmtwo$ $\smone$, and C/O ratios of either 1.1 (green) or 0.5 (purple).
The blue circles with error bars represent the {\it Spitzer}/IRAC data from \citet{gillon10} 
and \citet{deming11corot} and ground-based data from \citet{alonso10}.  The circles without 
error bars represent the model results for C/O = 1.1 (green) or 0.5 (purple) convolved over 
the spectral bandpasses.  Note the better fit with the C/O = 1.1 model at 3.6 $\mu$m for this 
particular assumed metallicity and thermal structure.  
A color version of this figure is available in the online journal.\label{speccorot}}
\end{figure}

\subsubsection{CoRoT-2b Spectra \label{corotspecsect}}

Figure \ref{speccorot} shows how two models with different C/O ratios compare with the 
{\it Spitzer}/IRAC data of \citet{gillon10} and \citet{deming11corot} and the ground-based 
data of \citet{alonso10} at secondary eclipse.  Both models exhibit too much absorption 
at 4.5 $\mu$m, which demonstrates why we did not investigate models with higher metallicities 
for this particular thermal structure.  Both models exhibit less flux in the 8-$\mu$m 
channel than in the 4.5 $\mu$m channel, in contrast to observations, which further emphasizes 
the unusual observed 4.5-$\mu$m/8-$\mu$m flux ratio for this planet \citep[see][]{deming11corot,madhu12coratio}, 
although both models are within $\sim$1.5-$\sigma$ of the uncertainties at these wavelengths.  The 
main difference between the models lies in the 3.6-$\mu$m/4.5-$\mu$m flux ratio and in the 
overall fit in the 3.6-$\mu$m channel, for which the C/O = 1.1 model provides a significantly 
better fit to the existing secondary-eclipse data.  In all, this carbon-rich model provides 
a statistically better fit ($\xi ^2$ = 1.3) than the solar-composition models presented here 
($\xi ^2$ = 7.2) and in \citet{madhu12coratio} and \citet{deming11corot}, while still having 
theoretically consistent species abundances.  The suggestion of \citet{madhu12coratio} that 
CoRoT-2b could have a carbon-rich atmosphere is therefore viable, although further observations 
at additional wavelengths would be required to fully test the suggestion.

As an aside, we note that unlike the case for the very hot WASP-12b or the much cooler XO-1b, 
our nominal thermal profile for CoRoT-2b first crosses the enstatite and forsterite condensation 
curves right within the planet's photosphere (for solar-like elemental compositions), which 
suggests that silicate clouds could have more spectral consequences on CoRoT-2b than they do 
on much hotter or cooler planets.  Opacity from thick clouds can reduce the depth (and broadness) 
of absorption bands and make the observed planetary spectrum appear more like a blackbody.  We 
note, however, that the observed secondary-eclipse data for CoRoT-2b do not much resemble a blackbody, 
particularly in terms of the deep absorption in the 8-$\mu$m channel.  If clouds were present 
near $\sim$0.1 bar or deeper, they would have relatively little effect on the photochemistry 
of C, N, and O-bearing species.  High-altitude clouds located at 10$^{-3}$ mbar or above, on 
the other hand, could shield species from photolysis, with some interesting consequences.  We 
also note that the host star for CoRoT-2b is very young and spectrally active \citep{bouchy08,guillot11}, 
which would result in a high x-ray and EUV flux and enhanced charged particle fluxes that 
would be conducive for ion chemistry in the atmosphere of CoRoT-2b.  As mentioned by \citet{moses11}, 
Titan-like ion chemistry at high altitudes could enhance the destruction of $\Ntwo$ (and perhaps 
CO) and enhance the production of complex hydrocarbons, which could further influence spectral 
behavior. 

\section{Implications With Respect to Spectra\label{implicatesect}}

\begin{figure}
\includegraphics[angle=0,scale=0.48]{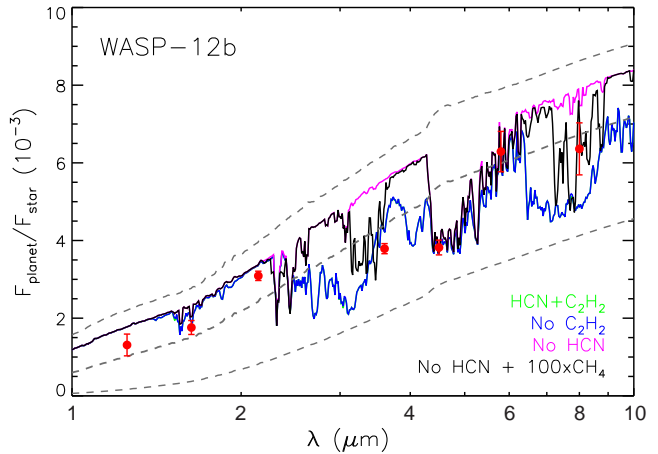}
\caption{The contribution of various molecules in our 3$\times$ solar metallicity, C/O = 1.015 
disequilibrium-chemistry model to the predicted secondary-eclipse spectra for WASP-12b.
The green curve represents synthetic spectra from our nominal model (see Figs.~\ref{wasppchem} \& 
\ref{specwasp}), whereas the blue curve is the same model with $\CtwoHtwo$ eliminated.  The 
two curves overlap almost exactly, illustrating the relative insignificance of $\CtwoHtwo$ as an 
opacity source in this model.  For the magenta curve, the contribution of HCN has been eliminated 
from the nominal model; a comparison of the magenta curve with the blue and/or green curves 
illustrates the dominance of HCN opacity in controlling the spectra behavior in many wavelength 
regions in our model.  The black curve is a synthetic spectrum without the contribution of HCN, 
but in which the $\CHfour$ abundance has been increased by a factor of 100.  The red circles 
with error bars represent the {\it Spitzer\/} and ground-based secondary-eclipse data of 
\citet{campo11} and \citet{croll11wasp12b}.  The dashed lines are blackbody curves for assumed 
temperatures of 1800 K (bottom), 2500 K (middle), and 3000 K (top).  A color version of this 
figure is available in the online journal.\label{wasphcn}}
\end{figure}

The atmospheric composition of carbon-rich planets differs considerably from that of solar-composition 
planets, and those differences have spectral implications, as is obvious from Figs.~\ref{spechd189}, 
\ref{figproof}, \ref{xo1bspec}, \ref{specwasp}, and \ref{speccorot}.  
Photochemical production of HCN and $\CtwoHtwo$ 
is efficient enough on cooler exoplanets (even those that are only mildly carbon-enhanced) that 
these species could affect spectra in certain wavelength regions.  Acetylene would provide opacity 
mainly in the 13-15 $\mu$m region, and to a lesser extent in the $\sim$3 and 7-8 $\mu$m regions 
(with the latter bands falling within the {\it Spitzer\/} 3.6 and 8.0-$\mu$m channels).  Hydrogen 
cyanide would provide opacity mainly in the $\sim$3-$\mu$m, 6.5-7.8 $\mu$m, and 12-16 $\mu$m regions 
(with broader wavelength ranges for higher temperatures). 
Even on the hottest transiting exoplanets, species like HCN can have a major influence on the total 
atmospheric opacity if the atmospheric C/O ratio exceeds unity.  For example, Fig.~\ref{wasphcn} 
shows the major influence that HCN has on the infrared spectra of our nominal carbon-rich WASP-12b 
model with C/O = 1.015.  Although $\CtwoHtwo$ has relatively little influence on the spectrum, only 
becoming noticeable in the $\sim$3-$\mu$m region, HCN opacity completely dominates in several 
wavelength regions, even surpassing the contribution of CH$_4$, CO, H$_2$O, and other molecules at 
$\sim$2.4-4.1 $\mu$m and in the 6-10 $\mu$m region.  Note that CH$_4$ opacities at high temperatures 
are poorly known and may be underestimated by as much as two orders of magnitude or more in our spectral
calculations.  If we approximate this effect by increasing the CH$_4$ abundance by a factor of 100, we 
see that CH$_4$ opacity could still be important on WASP-12b, but that does not change the fact that 
HCN is likely a major opacity source in the {\it Spitzer\/} 3.6 and 8.0 
$\mu$m channels.  Clearly, HCN is a potentially important molecule that can contribute to exoplanet 
spectral behavior and should be included in analyses of exoplanet spectra; $\CtwoHtwo$ and 
NH$_3$ are important disequilibrium species under some conditions and should also be considered if 
those conditions are met (e.g., potentially high C/O ratio or cooler atmosphere; see also 
\citealt{moses11}, \citealt{koppa12}).

Note, however, that to truly determine relative molecular abundances from exoplanet spectra, more 
detailed information on the line parameters (including intensities) from hot bands is needed in the molecular 
spectroscopy databases like GEISA and HITRAN \citep{jacquinet08,rothman09}.  There is currently 
insufficient published high-temperature information on $\CHfour$, HCN, and $\CtwoHtwo$ to reliably 
evaluate their potential quantitative contributions to the spectra.

Carbon dioxide can influence the spectra, particularly near $\sim$2 $\mu$m, $\sim$4.3 $\mu$m, and 
in the 14-16 $\mu$m region.  Our carbon-enhanced nominal disequilibrium models that provide good 
fits to the {\it Spitzer}/IRAC data tend to have CO$_2$ abundances lower than those derived from 
the analyses of \citet{madhu09}, \citet{madhu11wasp12b}, and \citet{madhu12coratio} for these 
four exoplanets (see Figs.~\ref{hd189equil1x}, \ref{xo1bequil}, \ref{waspequil}, \& \ref{corotequil}).  
Of all the {\it Spitzer}/IRAC channels, carbon dioxide has the most influence on the flux in the 
4.5-$\mu$m channel, and Figs.~\ref{spechd189}, \ref{figproof}, \ref{xo1bspec}, \& \ref{specwasp} 
demonstrate that our band-integrated fluxes at 4.5 $\mu$m do 
indeed fall slightly above the IRAC-derived fluxes at this wavelength for HD 189733b, 
XO-1b, and WASP-12b (CoRoT-2b being the exception).  Because other molecules such as CO contribute to the 
flux in this IRAC channel, an improved fit would be obtained for slight increases in the assumed 
metallicity of these planets; therefore, our low derived CO$_2$ abundances do not seem to be 
a major problem for the {\it Spitzer\/} IRAC fits.

\begin{figure*}
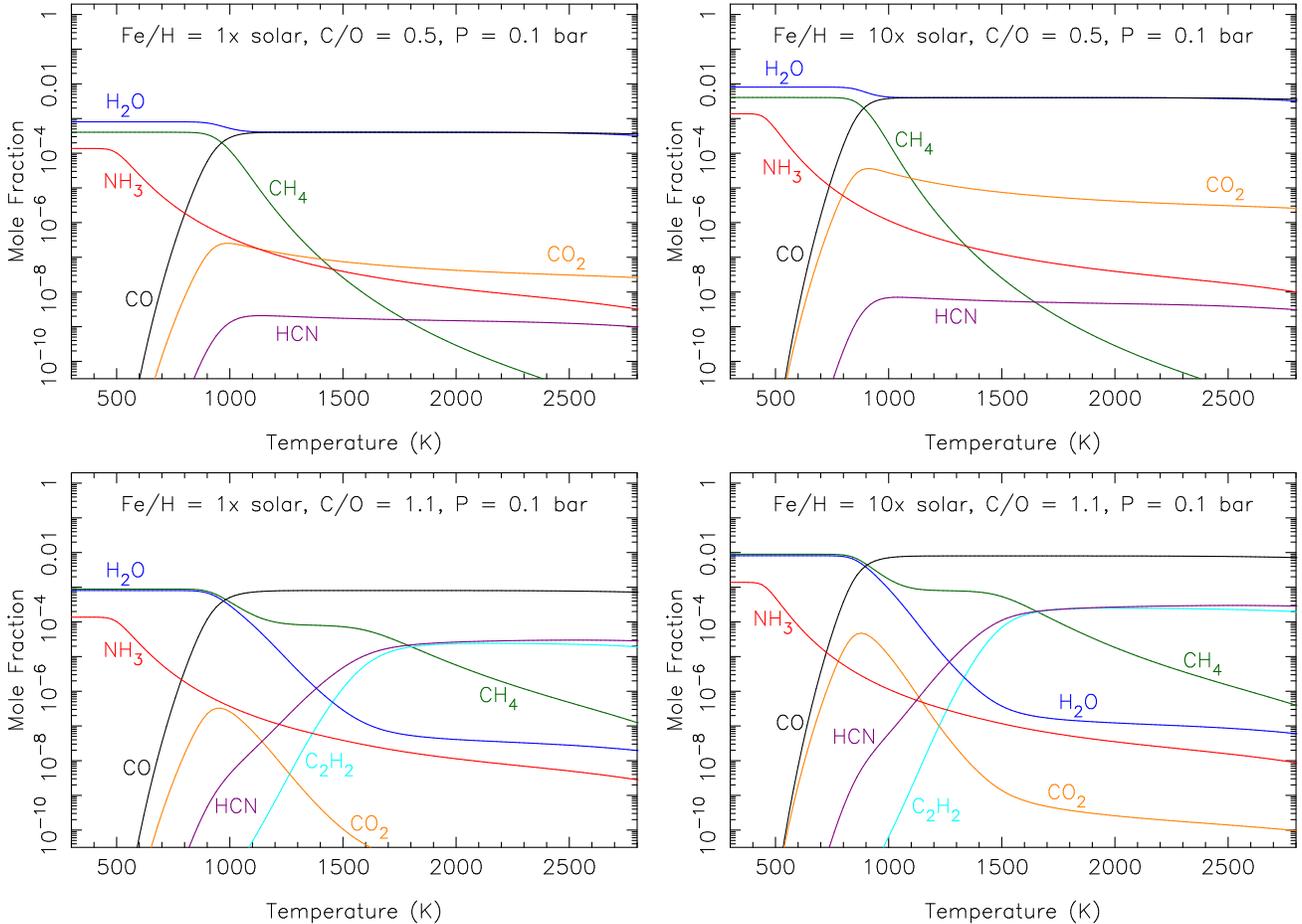

\begin{tabular}{ll}
{\includegraphics[angle=-90,clip=t,scale=0.37]{fig17a_color.ps}}
&
{\includegraphics[angle=-90,clip=t,scale=0.37]{fig17b_color.ps}}
\\
{\includegraphics[angle=-90,clip=t,scale=0.37]{fig17c_color.ps}}
&
{\includegraphics[angle=-90,clip=t,scale=0.37]{fig17d_color.ps}}
\\
\end{tabular}
\caption{The variation in the thermochemical-equilibrium abundances of key spectrally  
active species as a function of temperature at 0.1 bar, which is a typical pressure 
predicted to be within the infrared photospheres of hot Jupiters, for 1$\times$ solar metallicity 
and C/O = 0.5 (top left), for 10$\times$ solar metallicity and C/O = 0.5 (top right), for 
1$\times$ solar metallicity and C/O = 1.1 (bottom left), and for 10$\times$ solar metallicity 
and C/O = 1.1 (bottom right).  Note the relatively low CO$_2$ abundance in relation to H$_2$O 
for all temperatures, even for the more favorable case of 10$\times$ solar metallicity at a 
C/O ratio of 0.5.  A color version of this figure is available in the online journal.\label{figco2test}}
\end{figure*}

That is not the case at near-IR wavelengths, however, at least for HD 189733b.  Note from 
Figs.~\ref{hd189equil1x}, \ref{xo1bequil}, \ref{waspequil}, \& \ref{corotequil} that our predicted 
CO$_2$ mole fraction for all four of the exoplanets investigated never exceeds ppm levels, 
regardless of the C/O ratio, for both equilibrium and disequilibrium chemistry.  Even for 
10$\times$ solar metallicities, Fig.~\ref{figco2test} demonstrates that carbon dioxide is not 
expected to be a major product on hot Jupiters, and photochemistry does not change this 
conclusion.  

The relative unimportance of CO$_2$ in our chemical models is in stark contrast 
to the conclusions based on secondary-eclipse observations of HD 189733b in the near-infrared 
obtained with HST/NICMOS \citep{swain09hd189}.  The constraints for CO$_2$ based on this dataset 
range from a derived mole fraction of 10$^{-6}$-10$^{-1}$ \citep{swain09hd189}, to a best fit 
of $\sim$7$\scinot-4.$ \citep{madhu09}, to a best fit of $\sim$2$\scinot-3.$ \citep{lee12}, to 
derived values of (1.7--6.7)$\, \times \, 10^{-3}$ \citep{line11opti}.  Such large 
CO$_2$ abundances are inconsistent with the {\it Spitzer}/IRAC data in the 4.5-$\mu$m channel, 
which has led to the suggestion of atmospheric variability \citep{madhu09}.  However, 
note that the retrievals from this data set suggest CO$_2$/H$_2$O ratios of $\sim$3 
\citep{lee12} or even $\sim$30 \citep{line11opti}, which is not likely for a hydrogen-dominated 
planet.  In fact, to get CO$_2$ abundances comparable to those of H$_2$O under 
thermochemical-equilibrium conditions for our nominal thermal structure, 
HD 189733b would have to have a metallicity of order $\sim$2000, at which point H$_2$ would no 
longer be the dominant constituent by number --- a suggestion seemingly inconsistent with the 
planet's overall density and mass-radius relationship \citep{marley07}. 

Disequilibrium chemistry cannot help resolve this problem.  The dominant production mechanism 
for CO$_2$ is CO + OH $\rightarrow$ CO$_2$ + H, which is indeed effective on the dayside of 
HD 189733b due to OH production from H$_2$O photolysis (and other kinetic processes) and due 
to the large overall abundance of CO.  The main problem is that the reverse of this reaction 
is also effective at atmospheric conditions on HD 189733b, due to the large available 
H abundance and the relatively high atmospheric temperatures.  The forward and reverse reaction 
rates balance each other to a large degree, and although there is a small positive net production 
of CO$_2$ in our dayside atmospheric models, the CO$_2$ column abundance never achieves the 
levels needed to explain the HST/NICMOS secondary-eclipse observations of \citet{swain09hd189}.  
It is possible that we are missing some key photochemical mechanism that effectively converts 
water to CO$_2$ on transiting exoplanets, which would be needed to explain a high derived 
CO$_2$/H$_2$O ratio \citep[see][]{lee12,line11opti}, but given 
the efficiency of $\HtwoO$ recycling in a hydrogen-dominated atmosphere \citep[see 
also][]{moses11,line10}, it is hard to imagine that any chemical mechanism could produce this 
result while hydrogen is still the dominant atmospheric constituent.

The problem could potentially be resolved if the absorption attributed to CO$_2$ in the 
\citet{swain09hd189} NICMOS data actually resulted from some other atmospheric constituent.   
Species like $\NHthree$, HCN, and C$_2$H$_2$ come readily to mind as candidates, as they are 
all relatively abundant in our disequilibrium models, particularly for super-solar C/O ratios 
(for the case of HCN and $\CtwoHtwo$) or rapid atmospheric mixing at depth (for the case of 
$\NHthree$).  Hydrogen cyanide has strong bands in the $\sim$3-$\mu$m region and much weaker 
bands in the $\sim$1.85 $\mu$m and $\sim$1.53 $\mu$m regions; as such, HCN seems unlikely to 
cause significant absorption in the needed 1.9-2.2 $\mu$m range, despite the relatively large 
predicted column abundance of $\sim$1$\scinot19.$ $\cmmtwo$ above 1 bar in our nominal 
C/O = 0.88 disequilibrium model.   For the same model, $\NHthree$ and $\CtwoHtwo$ have column 
abundances of $\sim$2.5$\scinot19.$ and $\sim$2.5$\scinot16.$ $\cmmtwo$, respectively.  
Acetylene (and HCN) could be more abundant than we derive if the upper atmosphere is cooler 
than we have assumed \citep[e.g.,][]{moses11} or if ion chemistry enhances its abundance, but 
the $\CtwoHtwo$ bands near $\sim$1.53 and $\sim$2.45 $\mu$m do not seem good spectral matches 
to the NICMOS data.  Ammonia has bands centered bear $\sim$1.97 and 2.2-2.4 $\mu$m that could be 
of interest to the NICMOS data, particularly given that models with different assumed thermal 
structures and $K_{zz}$ values \citep{moses11} predict a significantly larger NH$_3$ abundance 
than the models shown here.  Although the NH$_3$ band at 1.97 $\mu$m lines up well with the 
minimum in flux seen in the NICMOS spectra, the continued absorption in the $\sim$2.1 $\mu$m region 
seen in the NICMOS data does not seem consistent with NH$_3$; however, due to the likely 
presence of NH$_3$ on HD 189733b and other transiting planets, further investigation is 
warranted.  At the very least, NH$_3$ and HCN should be considered in both near-IR and thermal-IR 
spectral models for hot Jupiters, as these molecules are important disequilibrium products for 
a variety of C/O ratios, as well as important equilibrium products for cooler planets and/or for 
planets with high C/O ratios.  

Models that contain more ammonia (e.g., because of stronger atmospheric mixing or a cooler 
thermal structure at depth) could also help improve the discrepancy between our best-fit 
solutions to the secondary-eclipse {\it Spitzer\/} IRAC data as compared with the IRS 
spectra for HD 189733b \citep[see also][]{madhu09}.   A hot lower atmosphere was found necessary
to fit the high flux in the 3.6 $\mu$m-channel IRAC data in both our analysis and that of 
\citet{madhu09}, but that solution also caused excess flux in the 9-12 $\mu$m region that is 
inconsistent with the IRS data (see Fig.~\ref{spechd189}).  \citet{moses11} demonstrate that 
the flux in the spectral models at $\sim$9-12 $\mu$m is considerably reduced when opacity from 
a moderate amount of quenched NH$_3$ is included in the calculations (see Fig.~13 of 
\citealt{moses11}).  However, the very recent {\it Spitzer\/} IRAC analysis of \citet{knut12} 
suggests a very large 7.5$\sigma$ reduction in the 3.6-$\mu$m flux at secondary eclipse, 
eliminiating the need for such a warm lower atmosphere to explain the 3.6-$\mu$m data 
and reducing the model flux in the 9--12 $\mu$m region to be more in line with the IRS 
observations (see Section \ref{hd189recent} and Fig.~\ref{figproof} for more details).  

Although we have focused on secondary-eclipse observations in this paper, transit observations 
can also provide compositional information.  The C/O ratio is a bulk atmospheric property that 
does not change on observational time scales.  Atmospheric temperatures, on the other hand, 
change relative quickly through much of the photospheric region due to changes in the incidence 
angle of the radiation coming from the host star and to the relatively short radiative heating 
and cooling time scales \citep[e.g.,][]{barman05}.  Dynamical time scales are of order 
$\sim$10$^5$ s, which is typically longer than radiative time scales, allowing horizontal 
thermal gradients to be maintained \citep[e.g.,][]{showman09}.  The average temperature structure 
therefore changes with viewing geometry \citep[e.g.,][]{knut07}.  If chemical time scales are 
shorter than dynamical time scales, the composition will also change with viewing geometry.  

Chemical time scales range from very short ($\ll$ 10$^5$ s) in the deepest, hottest regions 
below the photosphere and in the uppermost regions of the photosphere where disequilibrium 
photochemistry is active, to $>$ 10$^5$ s in the mid-to-lower photosphere.  In the mid-to-upper 
photosphere, chemical time scales can either be $\lta$10$^5$ s or $\gta$10$^5$ s, depending on 
stratospheric temperatures.  When chemical lifetimes are longer than transport time scales, which 
is likely true for a good portion of the photosphere on the cooler transiting exoplanets, then
transport-induced quenching can operate effectively to homogenize composition horizontally as 
well as vertically \citep[e.g.,][]{coop06}, significantly reducing the temperature-related 
equilibrium compositional variations shown in plots like Fig.~\ref{figco2test}.  However, 
chemical lifetimes will be shorter on warmer planets and for certain altitude regions on cooler 
planets, making compositional variations possible between transit observations (which are 
sensitive to the limb atmosphere at the terminators) and eclipse observations (which are 
sensitive to the fully illuminated dayside atmosphere).  Accurate predictions of the composition 
as a function of planetary viewing angle will therefore require more sophisticated time-variable 
disequilibrium models, which we delegate to future investigations.

\section{The Case for Planetary C/O Ratios Greater than the Host Star\label{coratiosect}}

According to our current understanding of star formation, the elemental ratios within a stellar 
photosphere reflect to a large degree the elemental ratios of the protostellar nebula in which the 
star formed, except for a slight depletion in heavy elements relative hydrogen due to later diffusive 
settling in the stellar atmosphere \citep[e.g.,][and references therein]{lodders09}.  Since planets 
also formed within that same nebula (i.e., protoplanetary disk), one might naively expect giant 
planets that efficiently accrete nebular gas to have elemental ratios that match that of 
the host star and bulk protoplanetary disk.  In that naive view, giant planets with high C/O ratios 
would then be expected to derive from a protoplanetary disk with high bulk C/O ratios, and their 
host stars should be similarly carbon-enriched.  Many factors can complicate this naive picture 
(see below), but detailed evolution models for WASP-12b suggest that a protoplanetary disk with a 
supersolar C/O ratio is required to produce WASP-12b with C/O $\ge$ 1 \citep{madhu11carbrich}.  

Although most stars appear to have near-solar relative abundances of carbon and oxygen, there is 
some observational evidence for a C/O-rich tail of the stellar population distribution, indicating a
non-trivial percentage of stars with C/O ratios greater than 1 (\citealt{ecuvillon04,ecuvillon06}, 
\citealt{delgado10}, \citealt{petigura11}; see also \citealt{bond10}).  \citet{gaidos00} points out 
the C/O ratio of the gas in the galactic disk is likely increasing with time, leading to a C/O ratio 
that is larger for younger planetary systems in our galaxy and for those farther away from the 
galactic bulge, and \citet{petigura11} provide 
observational evidence that stars with planetary systems have statistically higher C/O ratios than 
those without planets.  Both these factors could potentially explain observations of carbon-rich 
giant planets.  On the other hand, \citet{fort12} notes that stars with C/O ratios greater than 
1 have likely been overestimated in observational studies 
\citep{ecuvillon04,ecuvillon06,delgado10,petigura11} such that carbon-rich stars (and by correlation,
carbon-rich protoplanetary disks) are more likely to be very rare in the galaxy.  In this vein, it is 
notable that the host star WASP-12 appears to have a subsolar C/O ratio of 0.40$^{+0.11}_{-0.07}$ 
\citep{petigura11}, in contrast to the suggestion of \citet{madhu11wasp12b} and the work presented 
here that WASP-12b's atmosphere could be carbon rich, and in conflict with the naive view of 
similar elemental ratios in a giant planet and its host star.  How then could a giant planet 
develop a C/O ratio different from its host star?

Several processes --- some better understood than others --- can conspire to produce this 
result.  Disk chemistry is not uniform with location and time, and physical and chemical 
fractionations can lead to different C/O ratios in the nebular gas and/or solids 
\cite[e.g.,][]{krot00,cuzzi05,petaev05,lodders10}.  Evidence from meteorites in our own solar 
system suggests that although the bulk nebula had a solar C/O ratio, there were pockets that must
have been both oxygen-rich and oxygen-poor (at different locations and/or times) to account for 
certain mineral assemblages in different meteorite classes, with enstatite chondrites in particular 
suggesting some regions in the solar nebula had C $\approx$ O at the time and location in which 
the enstatite chondrites were formed \citep[cf.][]{larimer75,krot00,hutson00,pasek05}.  

One obvious fractionation process is condensation.  Condensation of water ice at the ``snow line'' in 
a diffusive or turbulently mixed nebula progressively depletes water vapor from the inner nebula 
\citep{stevenson88}, and this process plus inward drift of ice grains from the outer nebula 
\citep{stepinski97} leads to a pile up of water-ice solids near the snow line that can greatly 
facilitate the formation of giant planets (assuming the core-accretion 
model for giant-planet formation; e.g., \citealt{mizuno78}; \citealt{pollack96}).  As more water 
becomes trapped in proto-giant-planet cores and other planetesimals, the overall C/O ratio of the 
remaining vapor increases, and water-vapor transport to this cold trap can eventually leave the 
inner nebula depleted in water \citep{stevenson88,cyr98}, although there may be temporary regions 
of enhanced water vapor where inwardly drifting icy grains evaporate \citep{cuzzi04,ciesla06}.  The 
outer nebula will also be depleted in water vapor wherever water ice condenses.  Several 
such condensation/evaporation fronts for different ices occur at various radial distances 
in the nebula, leading to solids with different degrees of carbon and oxygen enrichments/depletions 
and vapor with complementary elemental depletions/enrichments, all of which can influence the 
composition of giant-planet atmospheres \citep{lodders04,dodson09,oberg11}.  Solids of sufficient 
size can decouple from the gas and evolve differently in the disk such that although the solar 
nebula had a bulk solar C/O ratio initially, the relative abundances of carbon and oxygen in the 
solids and gas varies with location and evolves with time \citep[e.g.,][]{ciesla06}.  The C/O 
ratio in the atmosphere of a giant planet then depends on the location and timing of its formation 
and the subsequent evolutionary history.

If one assumes that a planet's carbon enrichment is largely due to the accretion of solid material, 
which could incidentally also result in planetary metallicities enhanced over the host star's, then 
there is a limited region within a protoplanetary disk from which the planet could have accreted 
these carbon-enriched solids \citep[e.g.,][]{lodders04,lodders10}.  Water ice condenses beyond the 
$\HtwoO$ snow line (which forms initially near $\sim$5 AU in the solar nebula but moves inward with 
time; see \citealt{ciesla06}), 
whereas carbon can remain predominantly as gas-phase CO out to its condensation front (at $\gta$ 
10-30 AU; \citealt{bergin09}; \citealt{dodson09}), resulting in oxygen-enriched solids and 
carbon-enriched gas in the region between the $\HtwoO$ and CO condensation fronts.  Some additional 
volatile elements can be trapped in clathrate hydrates within this region of the disk \citep{lunine85}, 
but as these materials are also water-rich, the solids between the $\HtwoO$ and CO condensation fronts 
are expected to be enriched in oxygen and depleted in carbon relative to bulk nebular values 
\citep{hersant01}.  Therefore, the addition of accreted planetesimals from this region of the disk 
would be expected to {\it decrease\/} the C/O ratio of a giant planet's atmosphere below the stellar 
ratio \citep[e.g.,][]{gautier01,hersant04}.  In contrast, \citet{lodders04} 
suggests that disks also contain a ``tar line'' inward of the snow line, where refractory organics 
are stable but water ice has evaporated.  If a carbon-rich giant planet gets its extra carbon from 
solid material, then it is likely from Lodders' tarry solids or from some kind of carbon-rich 
refractory phase that remains in the inner solar system after the ice has evaporated 
\citep[e.g.,][]{ebel11}.  This scenario suggests that the carbon-rich giant planet would have to 
have accreted these solids during or after migration into the inner solar system.

Note that if the planet were to accrete greatly enhanced silicate or metal abundances, a potentially greater 
fraction of the oxygen could be tied up in condensed silicates and oxides in these planets, leaving 
the atmosphere above these condensate clouds to be oxygen-depleted.  However, silicate and metal 
condensates in the protoplanetary disk are typically already saturated with oxygen (i.e., metal oxides 
and silicates like MgSiO$_3$ and Mg$_2$SiO$_4$), making it unlikely that the accretion of 
metals and silicates in excess of solar proportions would be a mechanism for depleting oxygen in 
the planets.

In another scenario, the carbon enrichment in a giant planet could result from accreted gas rather 
than from solids.  \citet{oberg11} suggest that planets like WASP-12b could have formed beyond 
the $\HtwoO$ snow line, with the C/O enhancement simply resulting from the enhanced gas-phase 
nebular C/O ratio that results from $\HtwoO$ condensation.  For that scenario to work, there cannot 
be much mixing between the protoplanetary core (which is presumably water-rich due the planet's 
formation region beyond the snow line) and the atmosphere after the planet undergoes its fast 
runaway gas-accretion phase.  Moreover, the planet cannot continue to accrete large quantities 
of more oxygen-rich gas as it migrates to its present position close to its host star, and it 
cannot accrete a late veneer of oxygen-rich solids that would bring its C/O ratio back to the 
bulk nebular value.  If the planet becomes massive enough early on to create a gap in the disk 
around its orbit, then Type II migration could ensue \citep{ward97}; because the creation of 
the gap can substantially slow down the rate at which the planet accretes gas
\citep{dangelo08}, continued gas accretion may not be much of a factor for these planets once 
they begin migrating.  
Accretion of solids could still be important, but the current orbital positions of the transiting 
planets lie well within the condensation fronts for things like silicate and metals within the 
protoplanetary disk, and evaporation of most condensates might prevent these planets from 
accreting significant quantities of solids once the planets become parked at their present orbits.  
One observational 
consequence of this scenario would be a planetary metallicity the same order as the host star 
\citep{oberg11}, as superstellar metallicites would imply a larger contribution from solids 
(unless photoevaporation of the disk is responsible for enhancing virtually all heavy elements in 
the protoplanetary disk, see \citealt{guillot06}).  With this scenario, one would expect C/O 
ratios near unity, as CO is by far the dominant carbon- and oxygen-bearing gas in the region 
between the $\HtwoO$ and CO condensation fronts; note that CO$_2$ seems to be less important in 
a column-integrated sense than \citet{oberg11} have assumed \citep[see, e.g.,][]{willacy00,willacy09,walsh12}.

There is one other region of the protoplanetary disk in which the gas-phase abundance of CO 
is expected to significantly exceed that of water (even assuming bulk solar C/O ratios), and that is 
in the very inner regions of the disk ($\lta$0.1 AU) where high temperatures ($>$ 1800 K) begin 
to affect the stability of water.  Water thermally dissociates at about 2500 K, whereas CO 
can survive to $\sim$4000 K \citep[e.g.,][]{najita07}.  Excess oxygen not tied up in CO would 
be present as O and O$^+$ in these regions and might be expected to have different vertical 
distributions than CO due to molecular and ambipolar diffusion, with O and O$^{+}$ being 
concentrated at higher altitudes in the disk atmosphere than CO.  At these high altitudes, the 
O and O$^{+}$ might be more vulnerable to loss processes like stellar-wind stripping 
\citep{matsuyama09} or accelerated accretion onto the star \citep{elmegreen78}.  As such, the 
gas near the mid-plane could be carbon-rich in these warm inner-disk regions, with CO the 
dominant carbon and oxygen constituent.  Disk-chemistry models do not generally extend to regions so near the 
star, but observations of CO overtone emission in young low-mass stellar objects suggests that 
the CO emission derives from the $\sim$0.05--0.3 AU region \citep{chandler93,najita00}. 
Observations of CO and H$_2$O emission obtained for the same disk are consistent with the CO being 
located inward of, and at higher temperatures than, the H$_2$O \citep{carr04,thi05}.  Since hot 
Jupiters end up in these inner-disk regions after whatever migration process gets them there, 
local accretion of gas may also account for high atmospheric C/O ratios.  Again, since gas 
accretion is the culprit for the high C/O ratio in this scenario, one might expect the planet's 
metallicity to be similar to the host star's, unless H has been preferentially lost in the 
nebula \citep{guillot06}.  Another potential observational test for this scenario would be 
the N/O ratio, as $\Ntwo$ would be the only other stable heavy molecule abundant under these 
conditions.

The fact that we derive C/O ratios near unity for three out of four of the hot Jupiters 
investigated in this paper (and the fourth might also be just below 1) seems to suggest that 
the CO-rich gas accretion scenarios from either the hot inner nebula or from beyond the 
$\HtwoO$ snow line \citep[e.g.,][]{oberg11} represent the most likely explanation for the 
atmospheric carbon enrichment, as the carbon-rich-solid accretion scenario could potentially 
lead to C/O $\gg$ 1.  However, for carbon-rich atmospheres, graphite is 
expected to be a major condensate under temperature-pressure conditions in extrasolar giant 
planets, and graphite would thus commandeer the excess available carbon, leaving C/O = 1 
above the graphite clouds \citep[e.g.,][]{lodders97}.  The carbon-rich-solid scenario is 
therefore still a viable option, even for planets with an apparent C $\approx$ O.  Note, 
however, that WASP-12b is too hot for graphite to form, and if the atmosphere had a C/O ratio 
much greater than 1, we would expect to see spectral evidence to that effect.

Our own giant planets do not shed much light on the likelihood of these scenarios.  The 
overall enhanced metallicity for Jupiter and the other giant planets suggests that solid 
planetesimals were instrumental in supplying heavy elements to our own solar-system giant 
planets \citep[e.g.,][]{owen99}.  However, the ultimate cause of the near-equal enrichments 
of most heavy elements measured by the {\it Galileo\/} probe is under debate 
\citep{niemann98,wong04,owen99,gautier01,lodders04,owen06,hersant04,alibert05,guillot06,mousis09min}.  
The fact that the probe entered a dry hot-spot region on Jupiter and was not able to measure 
the deep water abundance is unfortunate, as the O/H enhancement over solar (or lack thereof) 
could have helped distinguish between different formation scenarios \citep[e.g.,][]{lunine04,atreya05}.  
Indirect evidence based on the observed CO abundance and disequilibrium-chemistry arguments 
is uncertain enough that a range of bulk Jovian C/O ratios is possible from these calculations 
\citep[cf.][]{fegley94,lodders02,bezard02,viss10co}, although the latest models seem to 
favor C/O ratios $\gta$ solar on Jupiter \citep{viss11}, seeming to select against some of the 
high-water-ice and high-clathrate accretion scenarios.  Given that Jupiter and transiting planets 
had very different migration histories, it is not clear that their carbon and oxygen inventories 
would have come from a similar source.

In any case, the C/O ratio and bulk metallicity within a giant planet's atmosphere provide 
important clues that may help us unravel the puzzle of giant planet formation and evolution. 
Further investigations that can constrain these properties from transit and eclipse observations 
are warranted.  Ultimately, it will be important to collect a statistically significant sample 
of exoplanet atmospheric properties to determine how common carbon-rich giant planets are 
within the exoplanet population.

\section{Summary\label{summarysect}}

Analyses of transit and eclipse observations suggest that some transiting extrasolar giant 
planets have unexpectedly low $\HtwoO$ abundances 
\citep[e.g.,][]{seager05,richardson07,grill07,swain09hd189,swain09hd209,madhu09,madhu10inv,desert09,sing09,line11opti,lee12,cowan12,madhu12coratio}.
Atmospheres with solar-like elemental abundances in 
thermochemical equilibrium are expected to have abundant water \citep[e.g.,][]{lodders02}, and 
disequilibrium processes like photochemistry cannot seem to deplete water sufficiently in the 
infrared photospheres of these planets to explain the observations \citep[e.g., see 
Section 3;][]{line10,moses11}.  
While aerosol extinction might help explain the lack of water-absorption signatures in 
some of the near-infrared transit observations \citep{lecave08hd189,pont08,sing09,sing11hd189,gibson12}, 
scattering and absorption from clouds or hazes should be less of a factor for the mid-infrared 
secondary-eclipse observations, unless the clouds are optically thick in the vertical, contain 
relatively large particles, and are located at pressure levels within the infrared photosphere
\citep[e.g.,][]{liou02}.  Such conditions may not be met for the cooler transiting planets like 
HD 189733b, TrES-1, or XO-1b, where major cloud-forming species like iron or magnesium silicates 
will condense too deep to affect the secondary-eclipse spectra \citep{fort05tres1,fort06hd149,fort10}, 
or for the hottest hot Jupiters like WASP-12b, where temperatures remain so hot throughout the column 
that iron and silicates will not condense.  
Extinction from clouds is therefore not expected to be the culprit for the low 
derived water abundances from thermal-infrared secondary-eclipse observations of many exoplanets, and 
if the interpretation is robust, we must look to other potential sources for the observed behavior.

A low water abundance is a 
natural consequence of an atmosphere with a C/O ratio of 1 or greater.  Motivated by the recent 
derivation of a carbon-rich atmosphere for WASP-12b based on secondary-eclipse observations 
\citep{madhu11wasp12b}, we have examined the influence of the C/O ratio on the composition of 
extrasolar giant planet atmospheres, both from a thermochemical equilibrium standpoint and as a 
result of disequilibrium processes like transport-induced quenching and photochemistry.  
We find that the equilibrium composition of hot Jupiters is very sensitive to the C/O ratio \citep[see
also][]{seager05,kuchner05,fort05,lodders10,madhu11wasp12b,madhu11carbrich,madhu12coratio,koppa12}.
Carbon monoxide is a major constituent on all hot Jupiters whose atmospheres reside on the CO side 
of the $\CHfour$-vs.-CO stability field, regardless of the C/O ratio (at least for the range 0.1 $<$ 
C/O $<$ 2), whereas H$_2$O and CO$_2$ become significant constituents for C/O $<$ 1, and 
CH$_4$, HCN, and $\CtwoHtwo$ become significant constituents for C/O $>$ 1.  Disequilibrium 
processes do not change this conclusion, although photochemistry can greatly enhance the HCN and 
$\CtwoHtwo$ abundances, and transport-induced quenching can enhance the lower-stratospheric 
abundances of CH$_4$, NH$_3$, and HCN, for a variety of assumed C/O ratios 
\citep[see also][]{moses11,viss11}.  Despite rapid photolytic destruction by the intense 
incident UV radiation and despite kinetic attack by atomic H, water survives in hot-Jupiter atmospheres 
due to the efficient recycling processes that occur in an $\Htwo$-dominated atmosphere.  A low 
derived water abundance for a transiting hot Jupiter may therefore be indicative of an atmosphere 
with a C/O ratio greater than solar \citep[see also][]{madhu11wasp12b,madhu12coratio}.  

We compare the results of our thermochemical and photochemical kinetics and transport models with 
{\it Spitzer\/} secondary-eclipse data from WASP-12b, XO-1b, CoRoT-2b, and HD 189733b --- four 
exoplanets whose atmospheres have been described as having potentially low water abundances
\citep[e.g.,][]{madhu09,madhu11wasp12b,madhu12coratio}.  We find that disequilibrium models with 
C/O $\sim$ 1 are consistent with photometric data from WASP-12b, XO-1b, and CoRoT-2b, confirming 
the possible carbon-rich nature of these planets.  In contrast, spectra from HD 189733b are more 
consistent with C/O $\lta$ 1.  In particular, our synthetic spectra for HD 189733b compare well 
with {\it Spitzer\/} secondary-eclipse photometric data for models with metallicities between 
1-5$\times$ solar for moderately enhanced C/O ratios between 0.88-1.0.  These fits are not unique: 
the specific derived C/O ratio that provides the best fit to the data will be dependent on the 
assumed thermal structure and atmospheric metallicity, with lower metallicites and/or higher 
atmospheric temperatures allowing lower (i.e., more solar-like) C/O ratios to remain consistent 
with the low inferred water absorption from these planets.  In those cases, however, the relative 
band ratios in the {\it Spitzer\/} IRAC channels are not always well reproduced.  
 
Our models indicate that hydrogen cyanide is an important atmospheric opacity source when C/O 
$\gta$ 1, such that HCN can even dominate the opacity in the 3.6 and 8.0 $\mu$m {\it Spitzer\/} 
channels for a variety of conditions.  For example, we find that HCN is the main absorber in 
the 3.6 and 8.0 $\mu$m bands on the very hot WASP-12b, rather than CH$_4$ being responsible
(as was suggested by \citealt{madhu11wasp12b}) or $\CtwoHtwo$  
being responsible (as was suggested by \citealt{koppa12}); however, this result could change for 
different assumptions about the thermal structure and C/O ratio.  Acetylene is less important 
overall for the {\it Spitzer\/} bandpasses, but $\CtwoHtwo$ is an important photochemical product 
in cooler atmospheres for a variety of C/O ratios and can become a significant equilibrium 
constituent in hotter atmospheres when C/O $>$ 1.  

Carbon dioxide is not a dominant constituent in our equilibrium or disequilibrium models 
regardless of the C/O ratio (again, for the range 0.1 $<$ C/O $<$ 2 and near-solar metallicities).  
A CO$_2$/H$_2$O ratio 
greater than unity for HD 189733b \citep[see][]{swain09hd189,madhu09,line11opti,lee12} would require 
a metallicity greater than 2000$\times$ solar or a C/O ratio $\ll$ 0.1, with both scenarios 
seeming to be inconsistent with the planet's bulk density and secondary-eclipse observations.
We therefore cannot explain the relatively large $\COtwo$ abundance inferred from {\it HST}/NICMOS 
observations of \citep{swain09hd189}.  In general, we do not expect CO$_2$ to be a major 
constituent in the atmospheres of near-solar-composition hot Jupiters.  Therefore, spectral modelers 
may want to consider the effects of species like HCN and NH$_3$ instead of (or in addition to) 
CO$_2$ and/or photochemical modelers will need to come up with new mechanisms to explain enhanced 
CO$_2$ abundances if spectral features on hot Jupiters can be tied uniquely to CO$_2$.

Although an atmospheric C/O ratio $\gta$ 1 on a giant planet could result from a carbon-enriched 
protoplanetary disk \citep[e.g.,][]{madhu11carbrich}, there are also ways in which such a planet 
could form in a disk that possesses more solar-like bulk elemental ratios.  Vapor condensation within 
the protoplanetary disk is one significant way in which elements can be fractionated, with the solids 
and vapor ending up with very different C/O ratios.  Carbon-rich giant planets could have accreted 
water-ice-poor, carbon-rich solids during their migration through the disk regions inward of the 
snow line \citep[e.g.,][]{lodders04,ebel11}, or they could get their carbon enrichment from the 
accretion of CO-rich, $\HtwoO$-poor gas, either during the runaway accretion phase as they formed 
in the region between the water and CO condensation fronts in the outer disk \citep[e.g.,][]{oberg11} 
or {\it in situ\/} from the inner disk once their migration stopped very close to the host star.  
Regardless of whether these specific scenarios are viable or not, the atmospheric C/O ratio and 
bulk metallicity provide important clues regarding the formation and evolution of giant planets, 
and it is hoped that future spectral observations can reveal how common carbon-rich giant planets 
are within the exoplanet population.



\acknowledgments

We thank Mark Marley and Jonathan Fortney for interesting exoplanet science discussions.  
The first author (JM) gratefully acknowledges support from the NASA Planetary Atmospheres 
Program grant number NNX11AD64G, CV acknowledges support from NNX10AF64G, and NM acknowledges 
support from the Yale Center of Astronomy and Astrophysics through a YCAA Postdoctoral 
Fellowship.





\clearpage

\end{document}